\documentclass[pra,aps,preprint,amsmath,amssymb,showpacs]{revtex4}
\usepackage{graphicx}
\usepackage{epsfig}
\usepackage{epstopdf}
\DeclareGraphicsExtensions{.png,.pdf}

\include{graphics}
\begin{document}


\title{Enhancement of transmission quality in soliton-based optical waveguide systems 
by frequency dependent linear gain-loss and the Raman self-frequency shift}

\author{Avner Peleg$^{1}$, Debananda Chakraborty$^{2}$}

\affiliation{$^{1}$ Department of Exact Sciences, Afeka College of Engineering, 
Tel Aviv 69988, Israel 
\\
$^{2}$ Department of Mathematics, New Jersey City University, 
Jersey City, New Jersey 07305, USA}

\date{\today}

\begin{abstract}
We study transmission stabilization against radiation emission in soliton-based nonlinear 
optical waveguides with weak linear gain-loss, cubic loss, and delayed Raman response. 
We show by numerical simulations with perturbed nonlinear Schr\"odinger propagation models 
that transmission quality in waveguides with frequency independent linear gain and cubic loss 
is not improved by the presence of delayed Raman response 
due to the lack of an efficient mechanism for suppression of radiation emission. 
In contrast, we find that the presence of delayed Raman response leads to significant enhancement 
of transmission quality in waveguides with frequency dependent linear gain-loss and cubic loss. 
Enhancement of transmission quality in the latter waveguides is enabled by the separation of the 
soliton's spectrum from the radiation's spectrum due to the Raman-induced self-frequency shift 
and by efficient suppression of radiation emission due to the frequency dependent linear gain-loss. 
Further numerical simulations demonstrate that the enhancement 
of transmission quality in waveguides with frequency dependent linear gain-loss, cubic loss, 
and delayed Raman response is similar to transmission quality enhancement in waveguides 
with linear gain, cubic loss, and guiding filters with a varying central frequency.   
\end{abstract}


\maketitle


\section{Introduction}
\label{Introduction}
Transmission of solitons in nonlinear optical waveguide systems has been the subject of intensive research 
in the last several decades due to the stability and shape preserving properties of the solitons 
\cite{Agrawal2001,Mollenauer2006,Hasegawa95,Iannone98}. 
In addition, since Kerr nonlinearity does not cause any pulse distortion in single-soliton 
propagation, soliton-based transmission can be used to realize higher transmission rates 
and larger error-free transmission distances compared with other transmission methods 
\cite{Mollenauer2006,Iannone98,Mollenauer97,MM98,Nakazawa2000}. 
This is true for example for transmission in optical fibers. 
Indeed, in Ref. \cite{Nakazawa91}, error-free optical fiber transmission of a single sequence 
of optical solitons at a bit rate of 10 Gb/s over $10^{6}$ km was experimentally demonstrated  
by using synchronous modulation. In another experiment, error-free transmission of seven soliton 
sequences at 10 Gb/s per sequence over transoceanic distances was realized, 
using dispersion-tapered optical fibers and guiding filters with a varying central frequency \cite{MMN96}. 
Even larger transmission rates were experimentally demonstrated with dispersion-managed solitons. 
In particular, in Ref. \cite{Nakazawa2000}, transmission of 25 sequences of dispersion-managed solitons 
at 40 Gb/s per sequence over 1500 km was achieved. Furthermore, transmission of 109 dispersion-managed 
soliton sequences at 10 Gb/s per sequence over $2 \times 10^{4}$ km was demonstrated in Ref. \cite{Mollenauer2003}. 
    

In the current paper, we study transmission stabilization of conventional optical solitons, 
that is, of solitons of the cubic nonlinear Schr\"odinger (NLS) equation without dispersion management. 
Our reasons for considering conventional optical solitons are the following. 
First, as stated in the first paragraph, because of the stability and shape-preserving properties 
of the solitons, soliton-based transmission is advantageous compared with other transmission methods. 
Second, due to the integrability of the unperturbed cubic NLS equation, derivation of the equations for 
dynamics of the soliton parameters in the presence of perturbations can be done in a rigorous manner. 
Third, the simpler properties of conventional solitons compared with dispersion-managed solitons 
make them more suitable for usage in optical networks and in other optical systems, 
where simplicity and scalability are important. 
Fourth, even though the details of pulse dynamics in other transmission systems might be different, 
analysis of transmission stabilization of conventional optical solitons can still provide a rough idea 
on how to realize transmission stabilization in other waveguide setups.

In several earlier works, we developed a general method for stabilizing the dynamics of 
optical soliton amplitudes in multisequence nonlinear optical waveguide systems 
with weak nonlinear dissipation \cite{NP2010,PNC2010,PC2012,CPJ2013,NPT2015,PNT2016,PNH2017,PC2016}. 
The method is based on showing that amplitude dynamics induced by the dissipation 
in $N$-sequence optical waveguide systems can be approximately  
described by $N$-dimensional Lotka-Volterra (LV) models.  
Stability analysis of the equilibrium states of the LV models can then be used 
for realizing stable amplitude dynamics along ultra-long distances.  
However, due to the instability of multisequence soliton-based transmission 
against resonant and non-resonant emission of radiation, 
the distances along which stable amplitude dynamics was observed   
in numerical simulations with the perturbed NLS equation were initially limited to a few 
hundred dispersion lengths \cite{PNC2010,PC2012}. 
Further analysis showed that a major mechanism for transmission destabilization in these systems 
is associated with resonant emission of radiation during intersequence soliton collisions, 
where the emitted radiation undergoes unstable growth and develops 
into radiative sidebands \cite{PNT2016,PNH2017,CPN2016}. 
Significant increase in the stable propagation distances was achieved by 
the introduction of frequency dependent linear gain-loss in 
$N$-waveguide couplers \cite{PNT2016,PNH2017,PC2016,CPN2016}. 
It was shown in these works that the presence of frequency dependent linear gain-loss leads to 
efficient suppression of the instability due to resonant radiation emission. 
The limiting cause for transmission instability in $N$-waveguide couplers 
with frequency dependent linear gain-loss was associated with non-resonant radiation emission 
due to the effects of the dissipation on single-soliton propagation \cite{PNT2016,PC2016}. 
Therefore, the latter process is a serious obstacle for further enhancement of 
transmission stability in nonlinear optical waveguide systems, 
where conventional optical solitons are used.

In two of the recent works, where stable long-distance multisequence transmission with conventional 
solitons was demonstrated, the effects of delayed Raman response were taken into account 
in addition to the effects of frequency dependent linear gain-loss \cite{PNT2016,PNH2017}. 
The stable transmission distances achieved in these studies were larger by two orders of 
magnitude compared with the distances obtained in earlier studies, where the effects of frequency dependent 
linear gain-loss and delayed Raman response were not taken into account \cite{PNC2010,PC2012}. 
It is known that the most important effect of delayed Raman response on single-soliton propagation 
in nonlinear optical waveguides is a continuous downshift of the soliton's frequency, 
which is called the Raman self-frequency shift \cite{Mitschke86,Gordon86,Kodama87}. 
In view of the findings in Refs. \cite{PNT2016,PNH2017}, it is important to investigate 
whether the combination of frequency dependent linear gain-loss and one of the 
effects associated with delayed Raman response, such as the Raman self-frequency shift, 
can indeed lead to significant enhancement of transmission stability in soliton-based optical 
waveguide systems. If such transmission stabilization is possible, it is important to characterize 
the mechanism leading to the stabilization.

In the current paper, we take on these important tasks by studying propagation 
of a single soliton in nonlinear optical waveguides with weak linear gain-loss, cubic loss,   
and delayed Raman response. We characterize transmission quality and stability 
by calculating the transmission quality integral, 
which measures the deviation of the pulse shape obtained in numerical simulations 
with perturbed NLS equations from the shape expected by the perturbation theory 
for the NLS soliton. In addition, we compare the dynamics of the soliton's amplitude 
and frequency obtained in the simulations with the dynamics expected by the perturbation theory.      
We first study soliton propagation in the absence of delayed Raman response. 
Our numerical simulations with the perturbed NLS equations show that 
transmission quality in waveguides with frequency independent linear gain and cubic loss is 
comparable to transmission quality in waveguides with frequency dependent linear gain-loss and cubic loss. 
We then include the effects of delayed Raman response in the perturbed NLS model. 
Our numerical simulations show that in waveguides with frequency independent linear gain, 
cubic loss, and delayed Raman response, the soliton's spectrum becomes separated 
from the radiation's spectrum due to the Raman-induced self-frequency shift experienced by the soliton. 
However, in this case transmission quality is not improved compared with transmission quality 
in the absence of delayed Raman response due to the lack of an efficient mechanism 
for suppression of radiation emission. Furthermore, for waveguides with 
frequency dependent linear gain-loss, cubic loss, and delayed Raman response, 
we observe significant enhancement of transmission quality compared with 
transmission quality in the absence of delayed Raman response. 
The enhancement of transmission quality in the latter waveguides is enabled by the 
separation of the soliton's spectrum from the radiation's spectrum due to the Raman self-frequency shift 
and by the efficient suppression of radiation emission due to the frequency dependent linear 
gain-loss. Additionally, we show by further numerical simulations that enhancement of transmission quality  
in waveguides with frequency dependent linear gain-loss, cubic loss, and delayed Raman response 
is similar to transmission quality enhancement in waveguides with weak linear gain, cubic loss, 
and guiding filters with a varying central frequency. 
More specifically, we demonstrate that the variation of the central filtering frequency leads 
to separation of the soliton's spectrum from the radiation's spectrum, 
while the presence of the guiding filters leads to efficient suppression of radiation emission.

We choose to study pulse propagation in optical waveguides with linear 
gain or loss and cubic loss as a major example for waveguides, in which linear and nonlinear dissipation 
plays an important role in pulse dynamics. The waveguide's cubic loss can arise due to 
two-photon absorption (2PA) or gain/loss saturation \cite{Boyd2008,Agrawal2007a,Dekker2007,Prasad2008}. 
Pulse propagation in the presence of 2PA or cubic loss has been studied in 
many previous works \cite{Malomed89,Stegeman89,Silberberg90,Aceves92,
Prasad95,Kivshar95,Tsoy2001,Silberberg2008,PCDN2009,PNC2010,Gaeta2012}. 
The subject received further attention in recent years due to the importance of 2PA in silicon nanowaveguides,  
which are expected to play a key role in many applications in optoelectronic devices
\cite{Agrawal2007a,Dekker2007,Gaeta2008,Soref2006}. These applications include 
modulators \cite{Cohen2005a,Lipson2005}, switches \cite{Jalali2004,Liang2005}, 
regeneration \cite{Gaeta2007}, pulse compression \cite{Boyraz2007}, 
logical gates \cite{Liang2006,Lei2013}, and supercontinuum generation \cite{Agrawal2007b}. 
In many of the applications it is desired to achieve a steady state, in which the pulse 
propagates along the waveguide with a constant amplitude. 
This can be realized by providing linear gain via Raman amplification 
\cite{Islam2004,Agrawal2005,Jalali2003,Lipson2004,Cohen2005b}.  
We also point out that it was recently demonstrated that waveguide spans 
with linear gain and cubic loss can be used for robust transmission switching 
and transmission recovery in hybrid soliton-based nonlinear waveguide 
systems \cite{NPT2015,PNH2017}.

The rest of the paper is organized as follows. In Section \ref{no_shifting}, 
we study transmission stabilization in waveguides with linear gain or loss and cubic loss, 
considering frequency independent linear gain in Section \ref{no_shifting1}  
and frequency dependent linear gain-loss in Section \ref{no_shifting2}.  
In Section \ref{Raman_sfs}, we investigate transmission stabilization 
in waveguides with linear gain or loss, cubic loss, and delyaed Raman response. 
We consider frequency independent linear gain in Section \ref{Raman_sfs1} 
and frequency dependent linear gain-loss in Section \ref{Raman_sfs2}.  
In Section \ref{filters}, we study transmission stabilization in waveguides with linear gain, 
cubic loss, and guiding optical filters,  
considering a constant central filtering frequency in Section \ref{filters1} 
and a varying central filtering frequency in Section \ref{filters2}.  
Our conclusions are summarized in Section \ref{conclusions}. 
In \ref{appendA}, we present a brief summary of the adiabatic perturbation 
theory for the NLS soliton. In \ref{appendB}, we describe the calculation 
of the transmission quality integral, while in \ref{appendC}, we derive the equation 
for dynamics of the soliton's amplitude in the presence of frequency dependent 
linear gain-loss.

\section{Pulse dynamics in waveguides with linear gain-loss and cubic loss}
\label{no_shifting}
\subsection{Waveguides with frequency independent linear gain and cubic loss}
\label{no_shifting1}
We consider propagation of an optical pulse in a nonlinear optical waveguide 
in the presence of second-order dispersion, Kerr nonlinearity, weak frequency 
independent linear gain, and weak cubic loss. The frequency independent linear gain 
can be realized by distributed Raman amplification \cite{Islam2004,Agrawal2005}. 
The propagation is described by the following perturbed NLS equation \cite{Agrawal2007a,Dekker2007,Silberberg90}: 
\begin{eqnarray}
i\partial_z\psi+\partial_t^2\psi+2|\psi|^2\psi=
ig_{0}\psi/2 - i\epsilon_{3}|\psi|^2\psi,
\label{sfs1}
\end{eqnarray}
where $\psi$ is proportional to the envelope of the electric field, 
$z$ is propagation distance, $t$ is time, and $g_{0}$ and 
$\epsilon_{3}$ are the linear gain and cubic loss coefficients \cite{dimensions1}. 
These coefficients satisfy $0<g_{0}\ll 1$ and $0< \epsilon_{3} \ll 1$. 
The second and third terms on the left-hand side of Eq. (\ref{sfs1}) 
are due to second-order dispersion and Kerr nonlinearity,
while the first and second terms on the right-hand side of Eq. (\ref{sfs1}) 
are due to linear gain and cubic loss.   
In the current paper we study transmission stabilization for fundamental solitons 
of the unperturbed NLS equation. The envelope of the electric field 
for these solitons is given by:   
\begin{eqnarray} 
\psi_{s}(t,z)\!=\!
\eta\exp(i\chi)/\cosh(x),
\label{sfs2}
\end{eqnarray}
where $x=\eta\left(t-y+2\beta z\right)$, 
$\chi=\alpha-\beta(t-y)+\left(\eta^2-\beta^{2}\right)z$, 
and $\eta$, $\beta$, $y$, and $\alpha$ are 
the soliton amplitude, frequency, position, and phase.

The equations for the dynamics of the soliton amplitude and frequency 
are obtained by using the adiabatic perturbation theory for the NLS soliton, 
see, e.g., Refs. \cite{Hasegawa95,Kaup90,Kaup91,CCDG2003} and \ref{appendA}. 
In the case of soliton propagation in the presence of linear gain and cubic loss, we obtain: 
\begin{eqnarray}&&
\frac{d\eta}{dz}=g_{0}\eta-\frac{4}{3}\epsilon_{3}\eta^{3},  
\label{sfs3}
\end{eqnarray} 
and $d\beta/dz=0$. Since we are interested in realizing stable transmission of the soliton 
with a constant amplitude, we require that $\eta=\eta_{0}>0$ is an equilibrium point of Eq. (\ref{sfs3}).   
This requirement yields $g_{0}=4\eta_{0}^{2}/3$. 
Thus, the equation for amplitude dynamics is:
\begin{equation}
\frac{d\eta}{dz}=\frac{4}{3}\epsilon_{3}\eta\left(\eta_{0}^{2} - \eta^{2}\right).
\label{sfs4}
\end{equation}
The solution of this equation for a soliton with an initial amplitude $\eta(0)$ is
\begin{equation}
\eta(z) = \eta_{0}\left[1 + \left(\frac{\eta_{0}^{2}}{\eta^{2}(0)}-1\right)
\exp\left(-8\eta_{0}^{2}\epsilon_{3}z/3\right)\right]^{-\frac{1}{2}} . 
\label{sfs5}
\end{equation}
It is clear from both Eqs. (\ref{sfs4}) and (\ref{sfs5}) that the equilibrium point at   
$\eta=\eta_{0}$ is stable, while the one at $\eta=0$ is unstable.

{\it Numerical simulations.} The prediction for stable dynamics of the soliton amplitude that was 
obtained in the previous paragraph was based on an adiabatic perturbation description, 
which neglects the effects of radiation emission. However, radiation emission effects can become 
significant at large propagation distances and this can lead to pulse shape distortion 
and to the breakdown of the adiabatic perturbation description of Eqs.  (\ref{sfs4}) and (\ref{sfs5}). 
This is especially true in waveguides with linear gain, since the presence of linear gain leads 
to unstable growth of small amplitude waves that are associated with radiation.   
It is therefore important to check the predictions obtained with the adiabatic perturbation theory 
by numerical simulations with the perturbed NLS model (\ref{sfs1}).

Equation  (\ref{sfs1}) is numerically integrated on a domain $[t_{\mbox{min}},t_{\mbox{max}}]=[-1600,1600]$ 
using the split-step method with periodic boundary conditions \cite{Agrawal2001,Yang2010}.              
The initial condition is in the form of a single NLS soliton $\psi_{s}$ with 
amplitude $\eta(0)$, frequency $\beta(0)=0$, position $y(0)=0$, and phase $\alpha(0)=0$. 
For concreteness, we present here the results of numerical simulations with $\epsilon_{3}=0.01$ and $\eta(0)=0.8$. 
We emphasize, however, that similar results are obtained for other values of the physical parameters. 
To avoid dealing with effects due to radiation leaving the computational domain at one boundary 
and re-entering it at the other boundary, we employ damping near the domain boundaries. 
The same method for suppressing re-entry of radiation into the computational domain was 
successfully used in many earlier studies of pulse propagation in nonlinear optical waveguides,  
see, e.g., Refs. \cite{Kuznetsov95,SP2004,PCDN2009}. Physically, the damping at the 
boundaries can be realized by employing filters at the waveguide ends \cite{Agrawal2001,Mollenauer2006}. 
Thus, the numerical simulations in the current section correspond to transmission in an open optical waveguide.

Transmission quality at a distance $z$ is measured from the results of the numerical simulations 
by calculating the transmission quality integral $I(z)$ in Eq. (\ref{Iz4}) in \ref{appendB}. 
This integral measures the deviation of the numerically obtained pulse shape $|\psi^{(num)}(t,z)|$ 
from the soliton's shape expected by the adiabatic perturbation theory $|\psi^{(th)}(t,z)|$, 
which is given by Eq. (\ref{Iz1}). Thus, $I(z)$ measures both distortion in the pulse shape due 
to radiation emission and deviations in the numerically obtained values of the soliton's parameters 
from the values predicted by the adiabatic perturbation theory.
Transmission quality is further quantified by measuring the transmission quality distance $z_{q}$, 
which is the distance at which the value of $I(z)$ first exceeds 0.075. In the numerical simulation 
with $\epsilon_{3}=0.01$ and $\eta(0)=0.8$ we find $z_{q}=432$. 
To characterize pulse shape degradation at larger distances, we run the simulation up to 
a final propagation distance $z_{f}$ at which the value of $I(z)$ first exceeds 0.655.  
In the simulation with $\epsilon_{3}=0.01$ and $\eta(0)=0.8$ we find $z_{f}=750$.

\begin{figure}[ptb]
\begin{tabular}{cc}
\epsfxsize=5.8cm  \epsffile{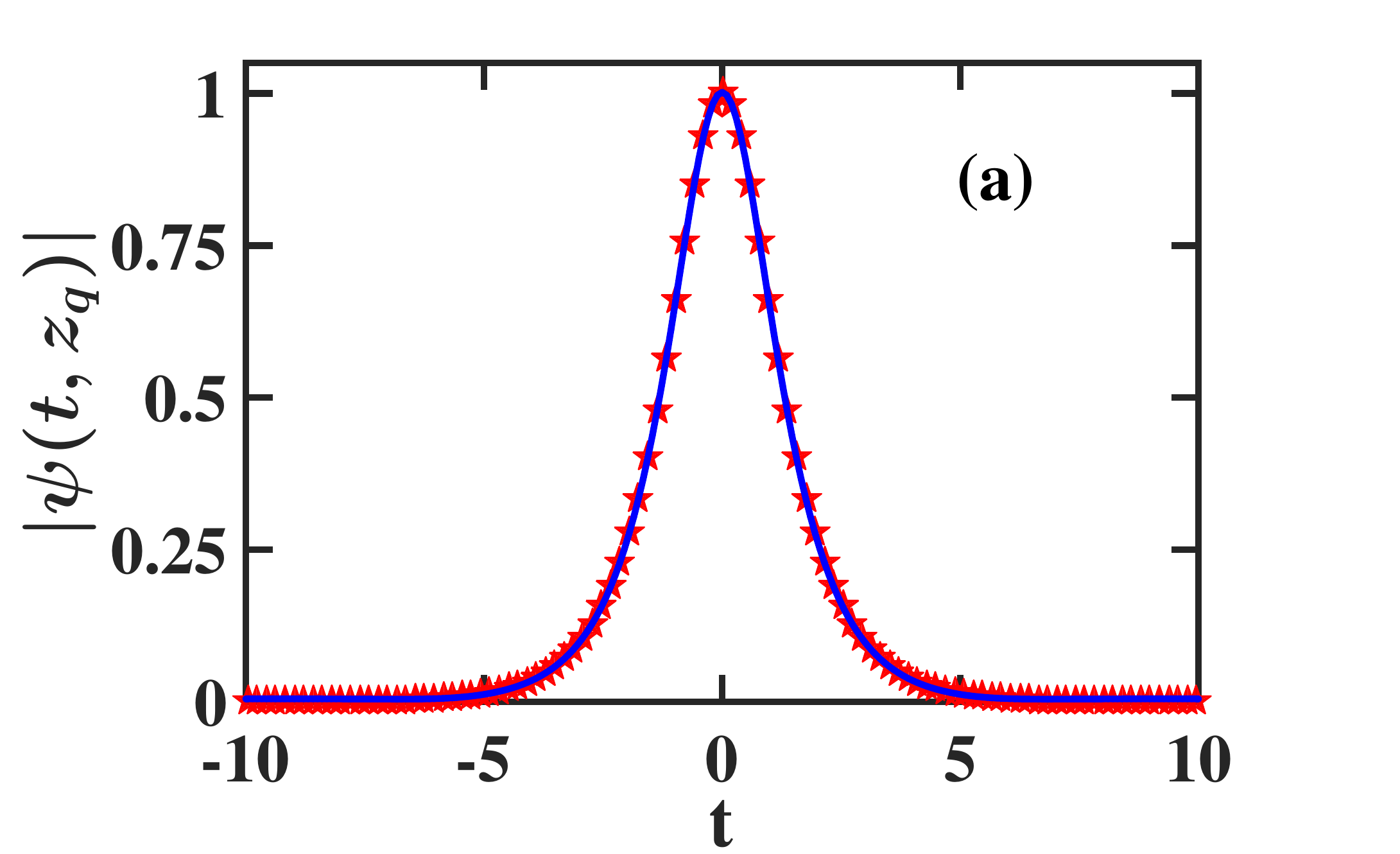} &
\epsfxsize=5.8cm  \epsffile{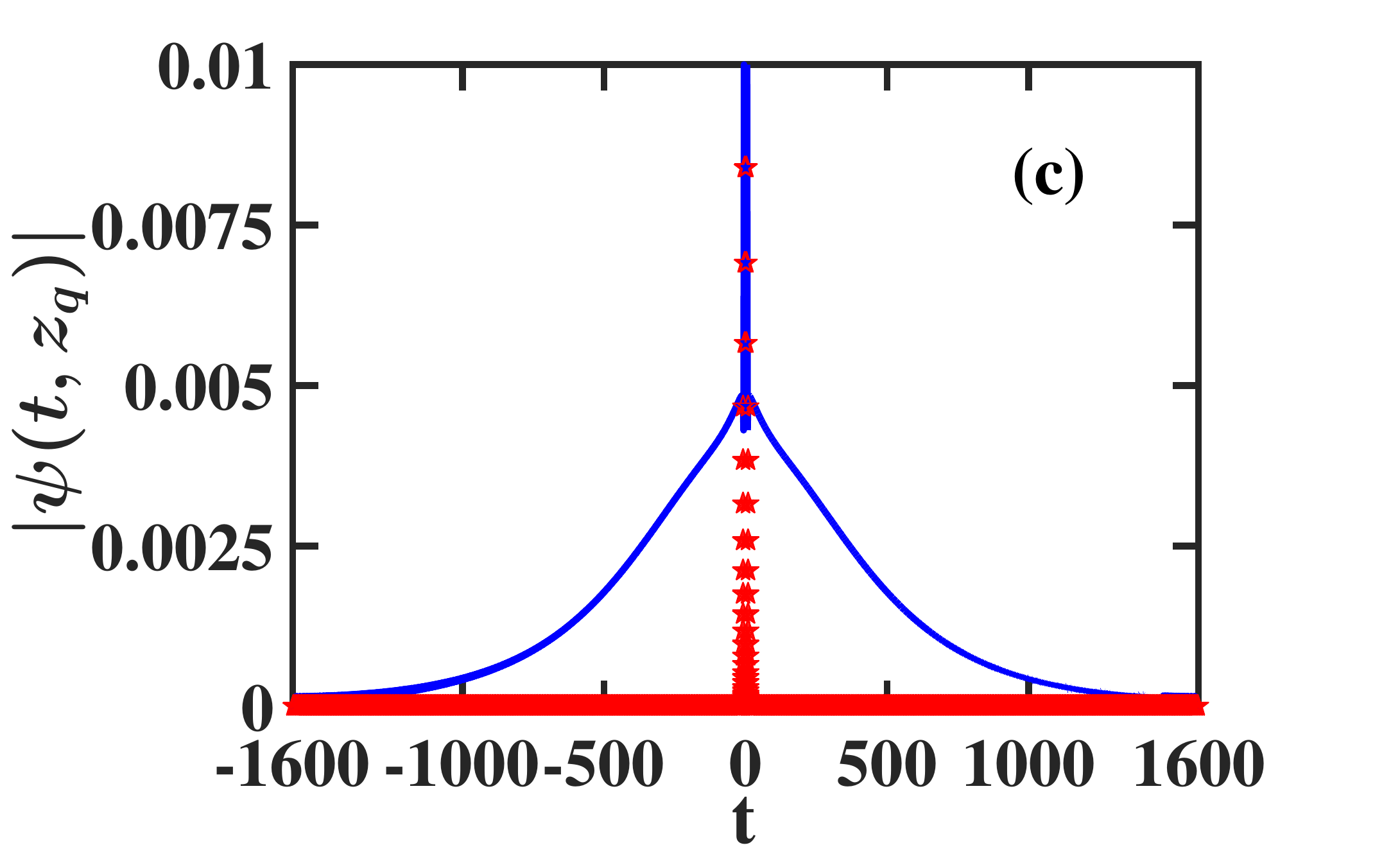} \\
\epsfxsize=5.8cm  \epsffile{fig1c.eps} &
\epsfxsize=5.8cm  \epsffile{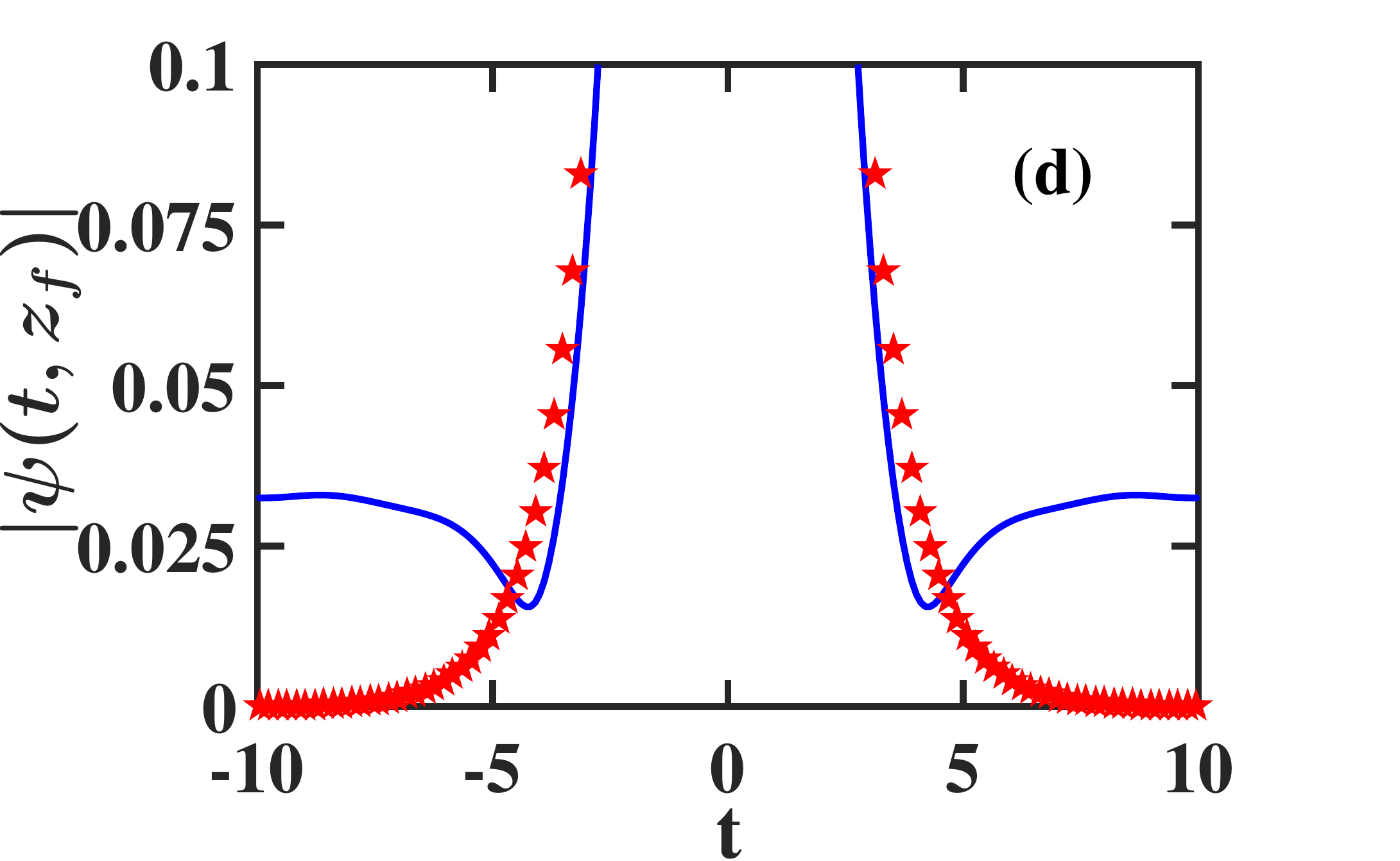}\\
\epsfxsize=5.8cm  \epsffile{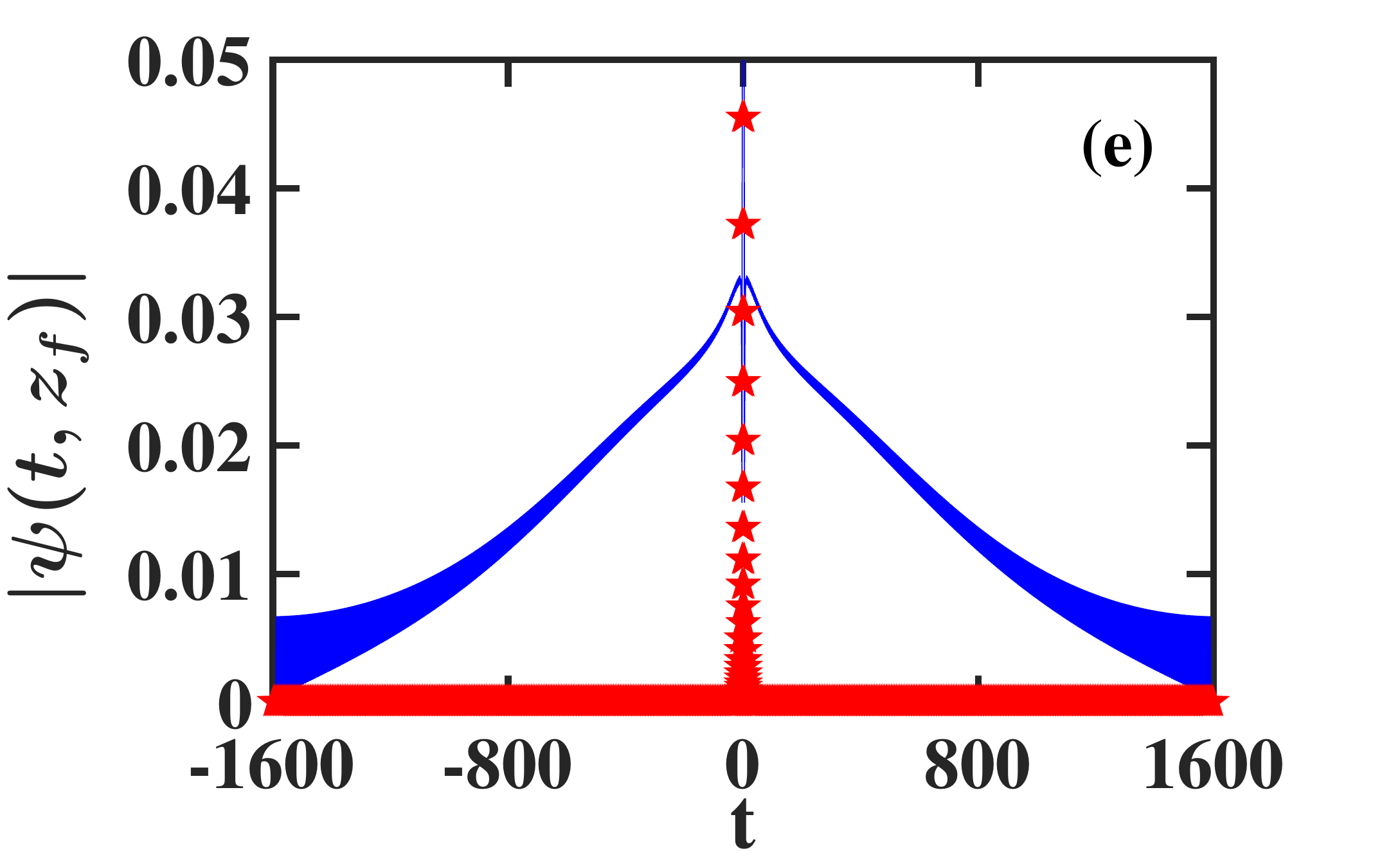}
\end{tabular}
\caption{The pulse shape $|\psi(t,z)|$ at $z_{q}=432$ [(a), (b), and (c)]  
and at $z_{f}=750$ [(d) and (e)] for soliton propagation in an open optical waveguide 
with weak frequency independent linear gain and cubic loss. 
The cubic loss coefficient is $\epsilon_{3}=0.01$ and the 
initial soliton amplitude is $\eta(0)=0.8$. 
The solid blue curve corresponds to the result obtained by numerical 
simulations with Eq.  (\ref{sfs1}), while the red stars correspond to 
the perturbation theory prediction of Eqs. (\ref{Iz1}) and (\ref{sfs5}).}
 \label{fig1}
\end{figure}

\begin{figure}[ptb]
\begin{tabular}{cc}
\epsfxsize=10cm  \epsffile{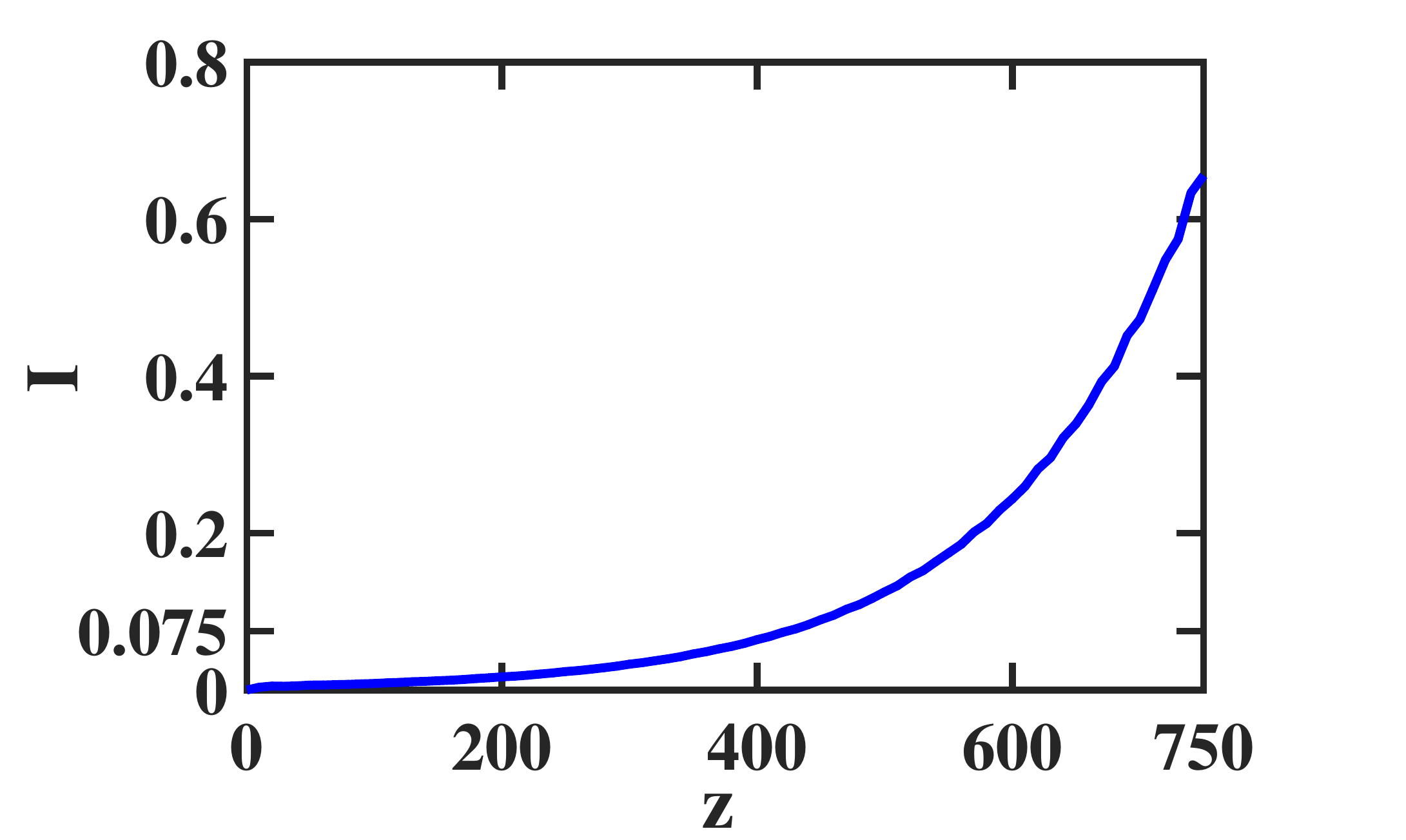} 
\end{tabular}
\caption{The $z$ dependence of the transmission quality integral $I(z)$ obtained 
by numerical simulations with Eq. (\ref{sfs1}) for the same optical waveguide setup 
considered in Fig. \ref{fig1}.}
\label{fig2}
\end{figure}

Figure \ref{fig1} shows the pulse shape $|\psi(t,z)|$ at $z=z_{q}$ and at $z=z_{f}$, 
as obtained by the numerical simulations. Also shown is the analytic prediction of Eqs. (\ref{Iz1}) 
and (\ref{sfs5}), which is obtained by the adiabatic perturbation theory. 
As seen in Figs. \ref{fig1}(a) and  \ref{fig1}(b), the pulse shape obtained by the simulations 
at $z=z_{q}$ is close to the analytic prediction. However, the comparison of the analytic 
prediction with the numerical result for small $|\psi(t,z_{q})|$ values in Fig. \ref{fig1}(c) reveals 
that an appreciable radiative tail exists already at $z=z_{q}$. As the soliton continues to propagate 
along the waveguide the radiative tail continues to grow [see Figs. \ref{fig1}(d) and \ref{fig1}(e)].             
The growth of the radiative tail is also manifested in Fig. \ref{fig2}, which shows the values 
of the integral $I(z)$ obtained in the simulations. As seen in this figure, the value of $I(z)$ 
increases from 0.075 at $z_{q}=432$ to 0.6556 at $z_{f}=750$. 

\begin{figure}[ptb]
\begin{center}
\begin{tabular}{cc}
\epsfxsize=5.8cm  \epsffile{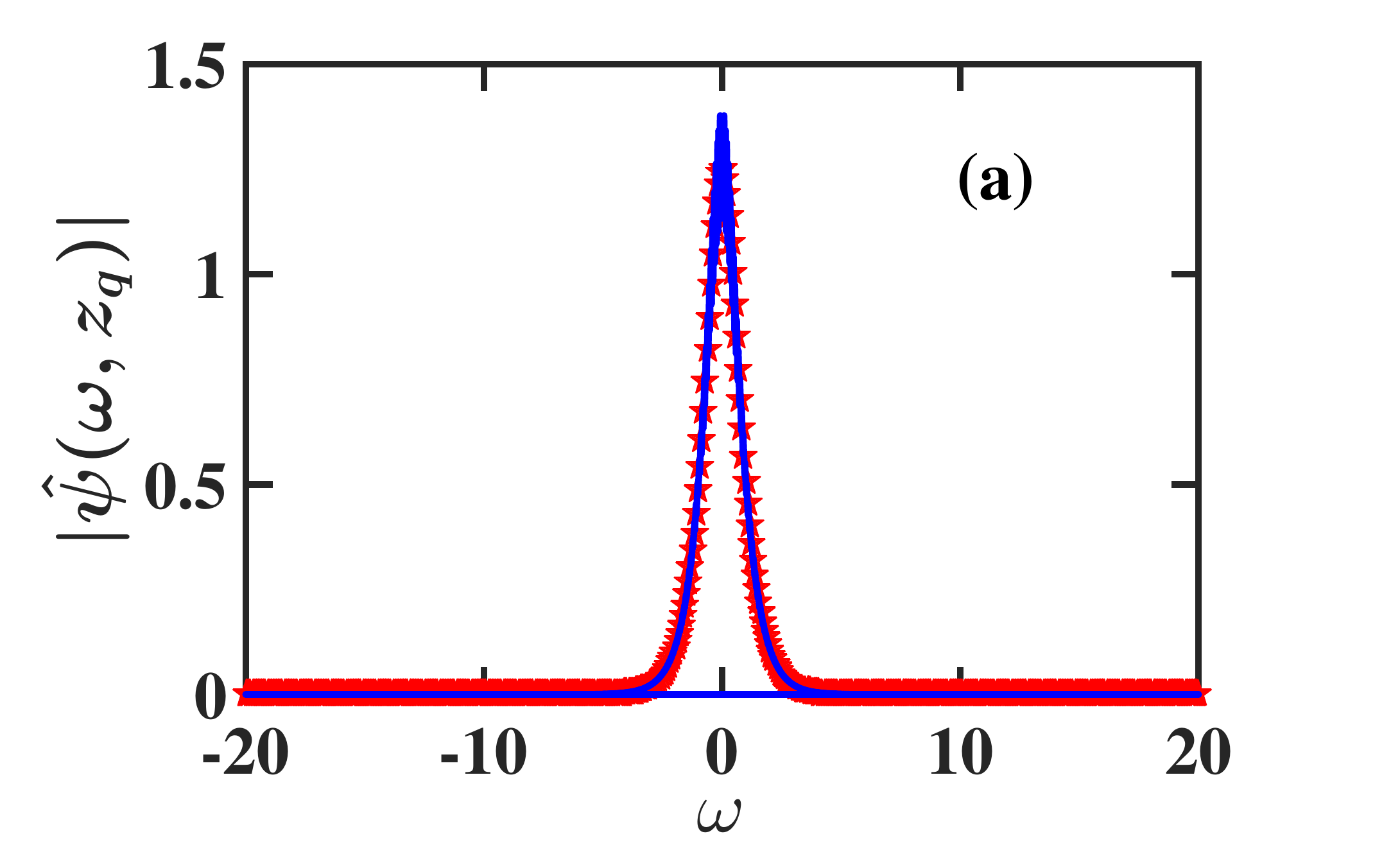} &
\epsfxsize=5.8cm  \epsffile{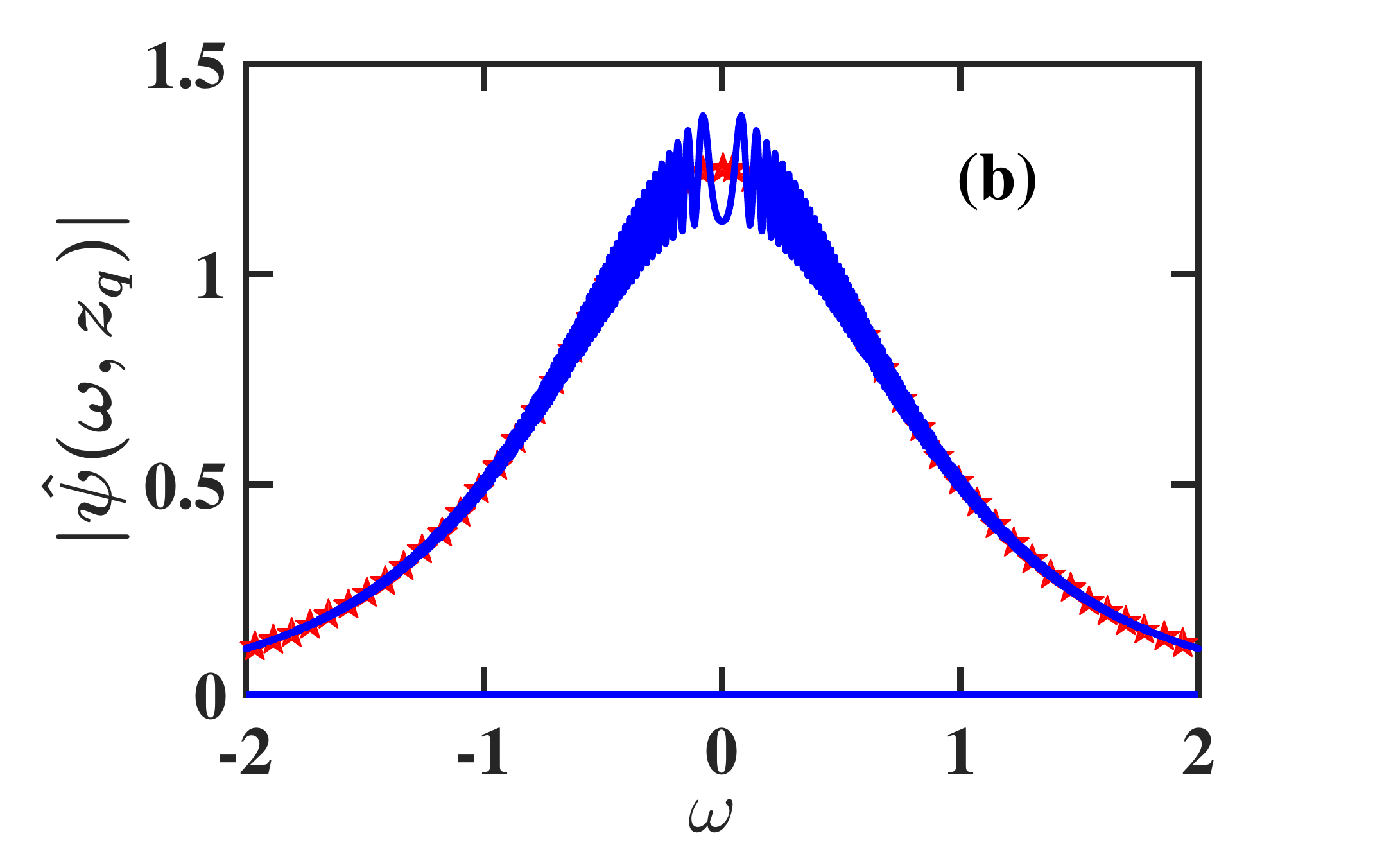} \\
\epsfxsize=5.8cm  \epsffile{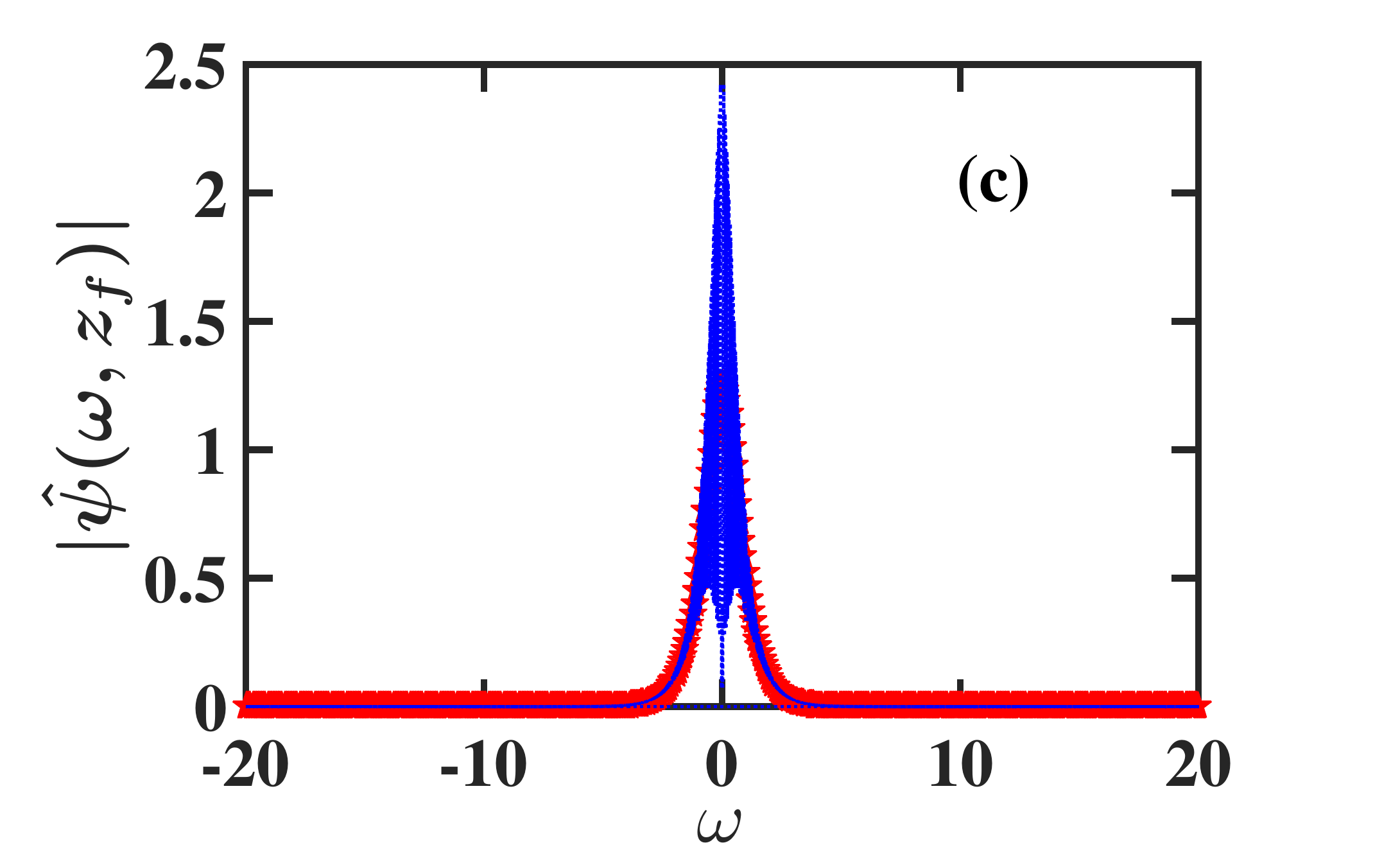} 
\end{tabular}
\end{center}
\caption{The Fourier transform of the pulse shape $|\hat\psi(\omega,z)|$ 
at $z_{q}=432$ [(a) and (b)] and at $z_{f}=750$ (c) 
for soliton propagation in an open optical waveguide 
with weak frequency independent linear gain and cubic loss. 
The physical parameter values are the same as in Fig. \ref{fig1}. 
The solid blue curve represents the result obtained by numerical 
simulations with Eq.  (\ref{sfs1}), while the red stars correspond to 
the prediction of the adiabatic perturbation theory, 
obtained with Eqs. (\ref{Iz3}) and (\ref{sfs5}).}                        
 \label{fig3}
\end{figure}

The growth of the radiative tail can be further characterized by the shape of the 
Fourier transform of the pulse $|\hat\psi(\omega,z)|$.  
Figure \ref{fig3} shows the Fourier transform $|\hat\psi(\omega,z)|$ 
at $z=z_{q}$ and at $z=z_{f}$, as obtained by the numerical simulations.  
Also shown is the prediction of the adiabatic perturbation theory, 
obtained with Eqs. (\ref{Iz3}) and (\ref{sfs5}). 
As seen in Figs. \ref{fig3}(a) and \ref{fig3}(b), 
the deviation of the numerical result from the analytic prediction is noticeable 
already at $z=z_{q}$. This deviation appears as fast oscillations in the numerical 
curve of $|\hat\psi(\omega,z)|$, which are most pronounced near $\omega=0$, 
i.e., at relatively small frequencies. Furthermore, the difference between the 
numerical result and the analytic prediction continues to grow as the pulse 
continues to propagate along the waveguide. Indeed, as seen in Fig. \ref{fig3}(c), 
the difference between the analytic prediction and the numerical result at $z=z_{f}$ 
is already of order 1.

\begin{figure}[ptb]
\begin{tabular}{cc}
\epsfxsize=10cm  \epsffile{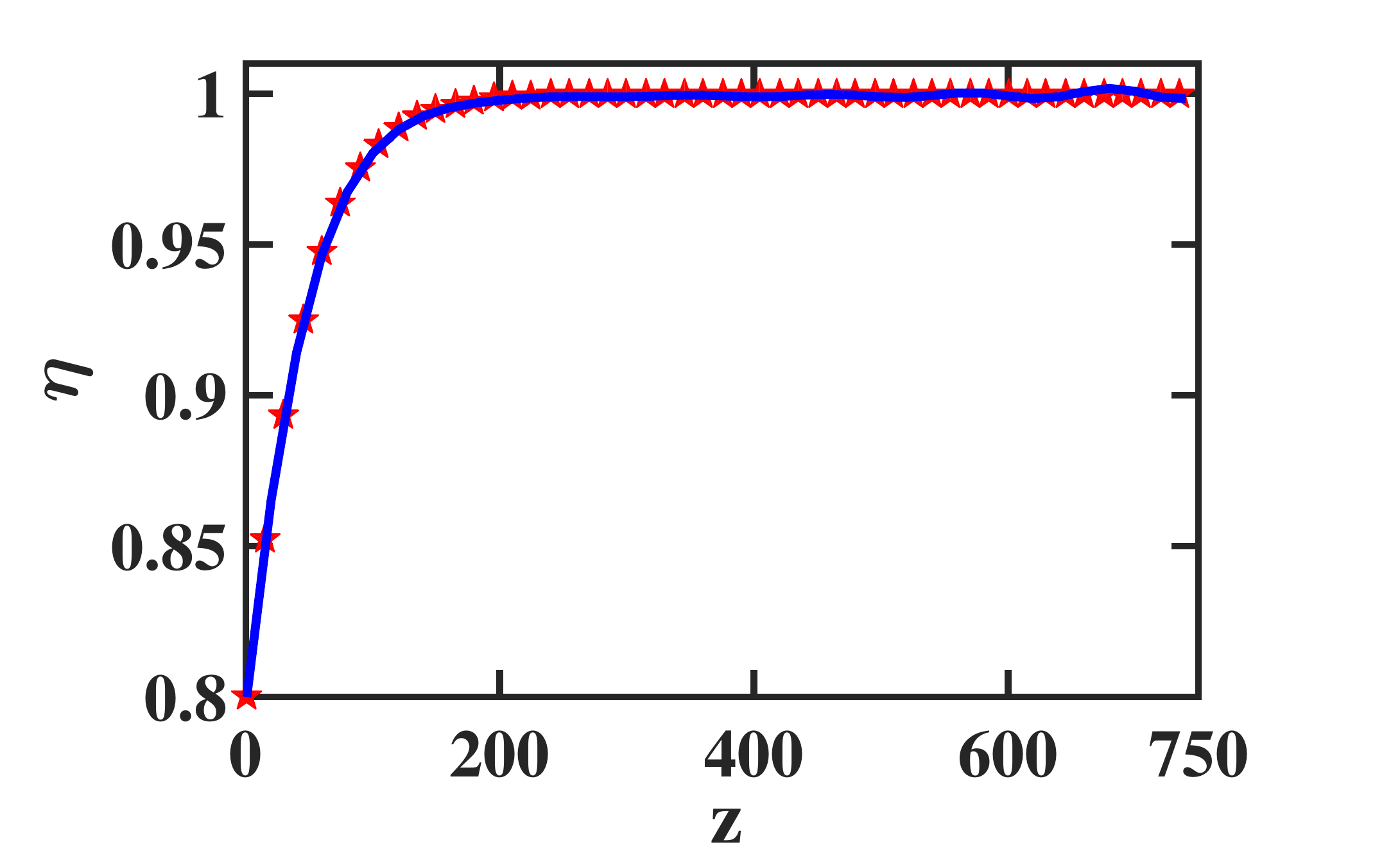} 
\end{tabular}
\caption{The $z$ dependence of the soliton amplitude $\eta(z)$ for the open 
waveguide setup considered in Figs. 1-3. The solid blue curve represents the result 
obtained by numerical simulations with Eq.  (\ref{sfs1}). 
The red stars represent the perturbation theory prediction of Eq. (\ref{sfs5}).}
\label{fig4}
\end{figure}

The $z$ dependence of the soliton amplitude obtained in the simulations 
is shown in Fig. \ref{fig4} along with the analytic prediction of Eq. (\ref{sfs5}). 
We observe good agreement between the numerical and analytic results 
for $0 \le z \le 600$. For $600 < z \le 750$, the difference between 
the numerical result and the analytic prediction becomes noticeable. 
The good agreement between the analytic prediction and the numerical result for $\eta(z)$ 
can be attributed to the fact that radiation emission affects the dynamics of $\eta$ only 
in second order of the small perturbation parameter $\epsilon_{3}$ (see, e.g., Refs. \cite{Kaup91,CCDG2003}).

We emphasize that the effects of radiation emission due to weak perturbations can have 
much stronger impact on soliton dynamics and stability compared with the impact observed 
here for single-soliton propagation in an open optical waveguide. 
More specifically, in the case of transmission of a soliton sequence through an optical waveguide, 
the emitted radiation leads to long-range interaction between the solitons, 
which in turn leads to the breakup of the soliton pattern \cite{CCDG2003}. 
Furthermore, in the case of transmission of multiple soliton sequences through an optical waveguide, 
the radiation emitted by the solitons in a given sequence can resonantly interact with solitons 
from other sequences \cite{CPN2016,PNT2016,PNH2017}. This resonant interaction 
leads to severe pulse pattern distortion and eventually to the destruction of 
the soliton sequences \cite{CPN2016,PNT2016,PNH2017,PNC2010,PC2012}.   
Finally, in the case of a soliton propagating in a closed waveguide loop,  
the accumulation of the emitted radiation and its interaction with the soliton 
will also lead to pulse shape distortion and to the destruction of the soliton. 
This latter scenario will be discussed and demonstrated in sections 3 and 4.

\subsection{Waveguides with frequency dependent linear gain-loss and cubic loss}
\label{no_shifting2}
As seen in section \ref{no_shifting1}, transmission quality in a waveguide with frequency 
independent linear gain and cubic loss is degraded at relatively short distances due to radiation emission. 
It is therefore important to look for waveguide setups, in which radiation emission might be suppressed. 
A possible way for achieving this goal is by employing frequency dependent linear gain-loss, 
such that the weak effects of cubic loss are balanced by weak linear gain in a frequency interval 
centered around the soliton frequency, while radiation emission effects are mitigated by relatively 
strong linear loss outside this frequency interval \cite{CPN2016,PNT2016,PNH2017,PC2016}.
Indeed, it was shown in several recent works that the implementation of such frequency dependent 
linear gain-loss can lead to significant enhancement of transmission stability in {\it multisequence} 
soliton-based optical waveguide systems \cite{CPN2016,PNT2016,PNH2017,PC2016}. 
We therefore turn to investigate soliton propagation in the presence of 
frequency dependent linear gain-loss and weak cubic loss. The propagation 
is described by the perturbed NLS equation \cite{PNH2017,PC2016}  
\begin{eqnarray}
i\partial_z\psi+\partial_t^2\psi+2|\psi|^2\psi=
i{\cal F}^{-1}(\hat g(\omega) \hat\psi)/2 - i\epsilon_{3}|\psi|^2\psi,
\label{sfs11}
\end{eqnarray}          
where $\omega$ is frequency, $\hat\psi$ is the Fourier transform of $\psi$ 
with respect to time, $\hat g(\omega)$ is the frequency dependent linear gain-loss,   
and ${\cal F}^{-1}$ is the inverse Fourier transform with respect to time.

The form of $\hat g(\omega)$ is chosen such that radiation emission effects are mitigated,
while the soliton amplitude still approaches $\eta_{0}$ with increasing propagation distance. 
In particular, we choose the form \cite{CPN2016}: 
\begin{eqnarray} &&
\hat g(\omega) = -g_{L} + \frac{1}{2}\left(g_{0} + g_{L}\right)
\left[\tanh \left\lbrace \rho \left[\omega + \beta(0)+W/2\right] 
\right\rbrace 
\right.
\nonumber \\&&
\left.
- \tanh \left\lbrace \rho \left[\omega + \beta(0)- W/2\right] 
\right\rbrace\right], 
\!\!\!\!\!\!\!
\label{sfs12}
\end{eqnarray}         
where $\beta(0)$ is the initial soliton frequency, $g_{L}$ is an $O(1)$ positive constant, 
and the constants $W$ and $\rho$ satisfy $W \gg 1$ and $\rho\gg 1$. 
In the limit $\rho\gg 1$, the linear gain-loss $\hat g(\omega)$ 
can be approximated by a step function,  
which is equal to $g_{0}$ inside a frequency interval of width $W$ centered about 
$-\beta(0)$, and to $-g_{L}$ elsewhere: 
\begin{eqnarray} &&
\hat g(\omega) \simeq 
\left\{ \begin{array}{l l}
g_{0} &  \mbox{ if $-\beta(0)-W/2 < \omega \le -\beta(0)+W/2$,}\\
- g_{L} &  \mbox{elsewhere.}\\
\end{array} \right. 
\label{sfs13}
\end{eqnarray}     
The potential advantages of using the frequency dependent linear gain-loss 
function (\ref{sfs12}) can be explained with the help of the approximate expression (\ref{sfs13}).  
The weak linear gain $g_{0}$ in the frequency interval $(-\beta(0)-W/2, -\beta(0)+W/2]$ 
balances the effects of cubic loss, such that the soliton amplitude tends to $\eta_{0}$ with increasing $z$. 
The relatively strong linear loss $g_{L}$ leads to suppression of emission of radiation with frequencies 
outside of the interval $(-\beta(0)-W/2, -\beta(0)+W/2]$. 
The flat gain in the interval $(-\beta(0)-W/2, -\beta(0)+W/2]$ 
can be realized by flat-gain amplifiers \cite{Becker99}, 
and the strong loss outside of this interval can be achieved by filters \cite{Becker99} 
or by waveguide impurities \cite{Agrawal2001}.

In \ref{appendC}, we show that within the framework of the adiabatic perturbation theory,  
the dynamics of the soliton amplitude is described by  
\begin{eqnarray} &&
\frac{d\eta}{dz} =
\left[ -g_{L}  + \left( g_{0} + g_{L} \right) \tanh(V) - 4\epsilon_{3}\eta^{2}/3\right]\eta ,
\label{sfs14}
\end{eqnarray}
where $V=\pi W/(4\eta)$. To realize stable transmission with a constant amplitude $\eta_{0}>0$, 
we require that $\eta=\eta_{0}>0$ is a stable equilibrium point of Eq. (\ref{sfs14}). 
This requirement yields 
\begin{equation}
g_{0} = g_{L}\left[\frac{1}{\tanh(V_{0})}-1\right] + \frac{4\epsilon_{3}\eta_{0}^{2}}{3\tanh(V_{0})}, 
\label{sfs15}
\end{equation} 
where $V_{0}=\pi W/(4\eta_{0})$. Substituting  Eq. (\ref{sfs15}) into  Eq. (\ref{sfs14}), we obtain: 
\begin{equation}
\frac{d\eta}{dz} = \eta\left\lbrace g_{L}\left[\frac{\tanh(V)}{\tanh\left(V_{0}\right)} - 1 \right]
+\frac{4}{3}\epsilon_{3}\left[\eta_{0}^{2}\frac{\tanh(V)}{\tanh\left(V_{0}\right)} - \eta^{2} \right]
\right\rbrace . 
\label{sfs16}
\end{equation}  
In  \ref{appendC}, we show that the only equilibrium points of Eq. (\ref{sfs16}) with $\eta \ge 0$ 
are $\eta=\eta_{0}$ and  $\eta=0$. In addition, we show that $\eta=\eta_{0}$ is a stable equilibrium point, 
while $\eta=0$ is an unstable equilibrium point. Thus, the number, locations, and stability properties 
of the equilibrium points of Eq. (\ref{sfs16}) and Eq. (\ref{sfs4}) are the same. 
In other words, the introduction of the frequency dependent linear gain-loss 
does not change the equilibrium points properties.    
We also note that in the typical transmission setup that we consider in the current work, 
$\eta_{0}$ is of order 1, $\eta$ is of order 1 or smaller, and $W \gg 1$ \cite{V_setup}. 
Therefore, in this case both $V_{0}$ and $V$ satisfy $V_{0} \gg 1$ and $V \gg 1$, 
and one can obtain an approximate form of Eq. (\ref{sfs16}) by    
expanding its right hand side in a Taylor series with respect to $e^{-2V_{0}}$ 
and $e^{-2V}$. Keeping terms up to first order in the expansion, we obtain: 
\begin{eqnarray} &&
\frac{d\eta}{dz} =
\left[2g_{L}\left(e^{-2V_{0}} - e^{-2V}\right)
+\frac{4}{3}\epsilon_{3}\left(\eta_{0}^{2}-\eta^{2}\right)
\right.
\nonumber \\&&
\left.
+\frac{8}{3}\epsilon_{3}\eta_{0}^{2}\left(e^{-2V_{0}} - e^{-2V}\right)
\right]\eta .
\label{sfs17}
\end{eqnarray}   
Comparing Eq. (\ref{sfs17}) with Eq. (\ref{sfs4}) we see that in the typical transmission setup, 
the correction terms that appear in the equation for amplitude dynamics due to the introduction 
of frequency dependent linear gain-loss are exponentially small in both $V_{0}$ and $V$.

{\it Numerical simulations}.                  
To check whether the introduction of frequency dependent linear gain-loss leads 
to enhanced transmission stability, we carry out numerical simulations with Eq. (\ref{sfs11}) 
and the linear gain-loss (\ref{sfs12}). The equation is solved on a domain 
$[t_{\mbox{min}},t_{\mbox{max}}]=[-1600,1600]$ with periodic boundary conditions
and with the same damping at the boundaries as in subsection \ref{no_shifting1}.  
Therefore, the numerical simulations correspond to open waveguide transmission. 
The initial condition is in the form of a single NLS soliton with amplitude $\eta(0)$, 
frequency $\beta(0)=0$, position $y(0)=0$, and phase $\alpha(0)=0$. 
To enable comparison with the results of numerical simulations for transmission 
in the presence of frequency independent linear gain, we use the same parameter 
values that were used in subsection \ref{no_shifting1}: 
$\epsilon_{3}=0.01$ and $\eta(0)=0.8$.  
In addition, the values of the parameters $W$, $\rho$, and $g_{L}$ of the frequency 
dependent linear gain-loss $\hat g(\omega)$ are similar to the values used in 
Refs. \cite{CPN2016,PNT2016,PNH2017,PC2016} in studies of multisequence 
soliton-based transmission: $W=10$, $\rho=10$, and $g_{L}=0.5$. 
These values were found to lead to enhanced stability of soliton propagation in 
multisequence transmission systems \cite{CPN2016,PNT2016,PNH2017,PC2016}.
The values of the transmission quality distance and the final propagation distance  
obtained in the simulations were $z_{q}=432$ and $z_{f}=750$, which are 
the same as the values found for waveguides with frequency independent linear gain.

\begin{figure}[ptb]
\begin{tabular}{cc}
\epsfxsize=5.8cm  \epsffile{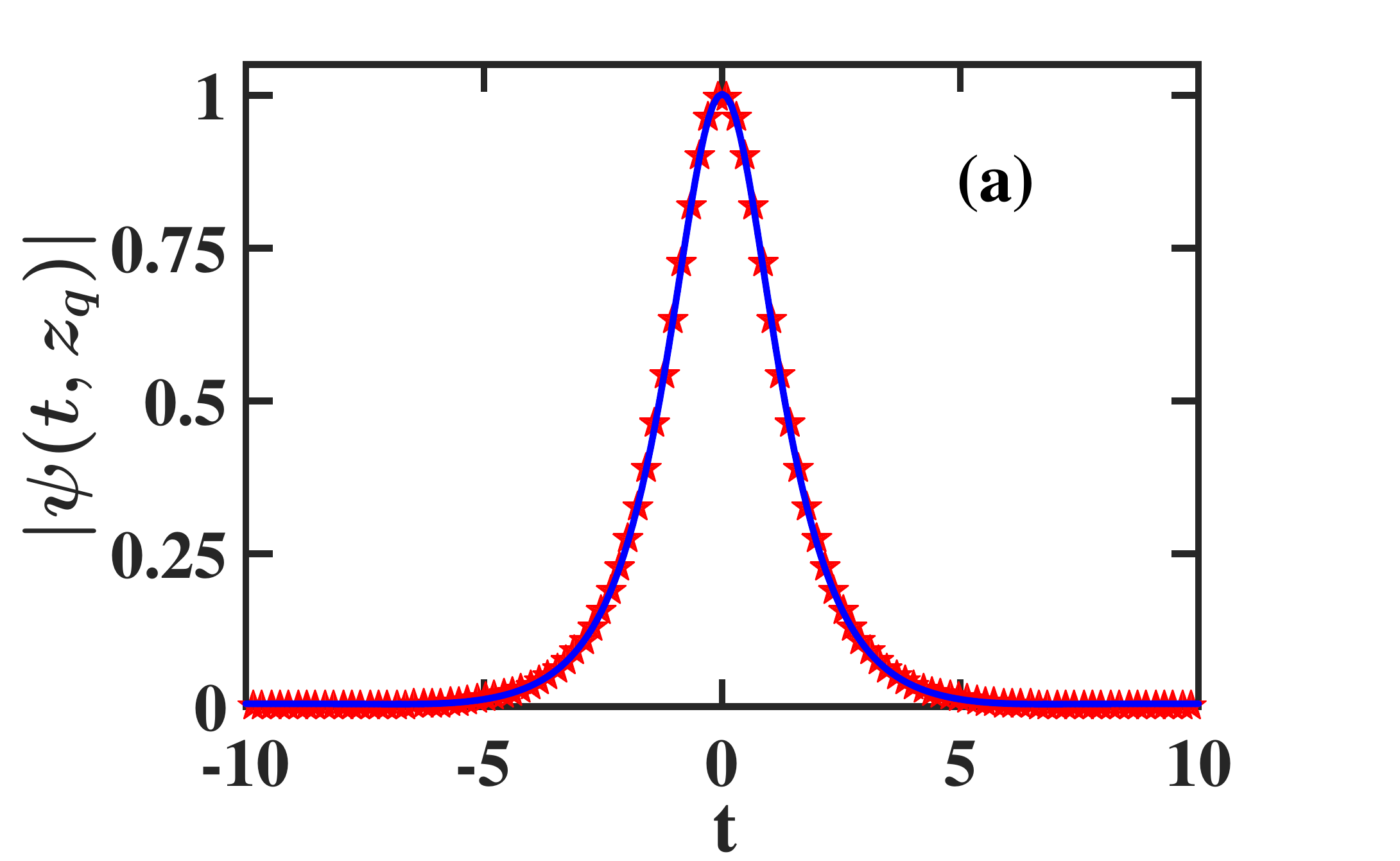} &
\epsfxsize=5.8cm  \epsffile{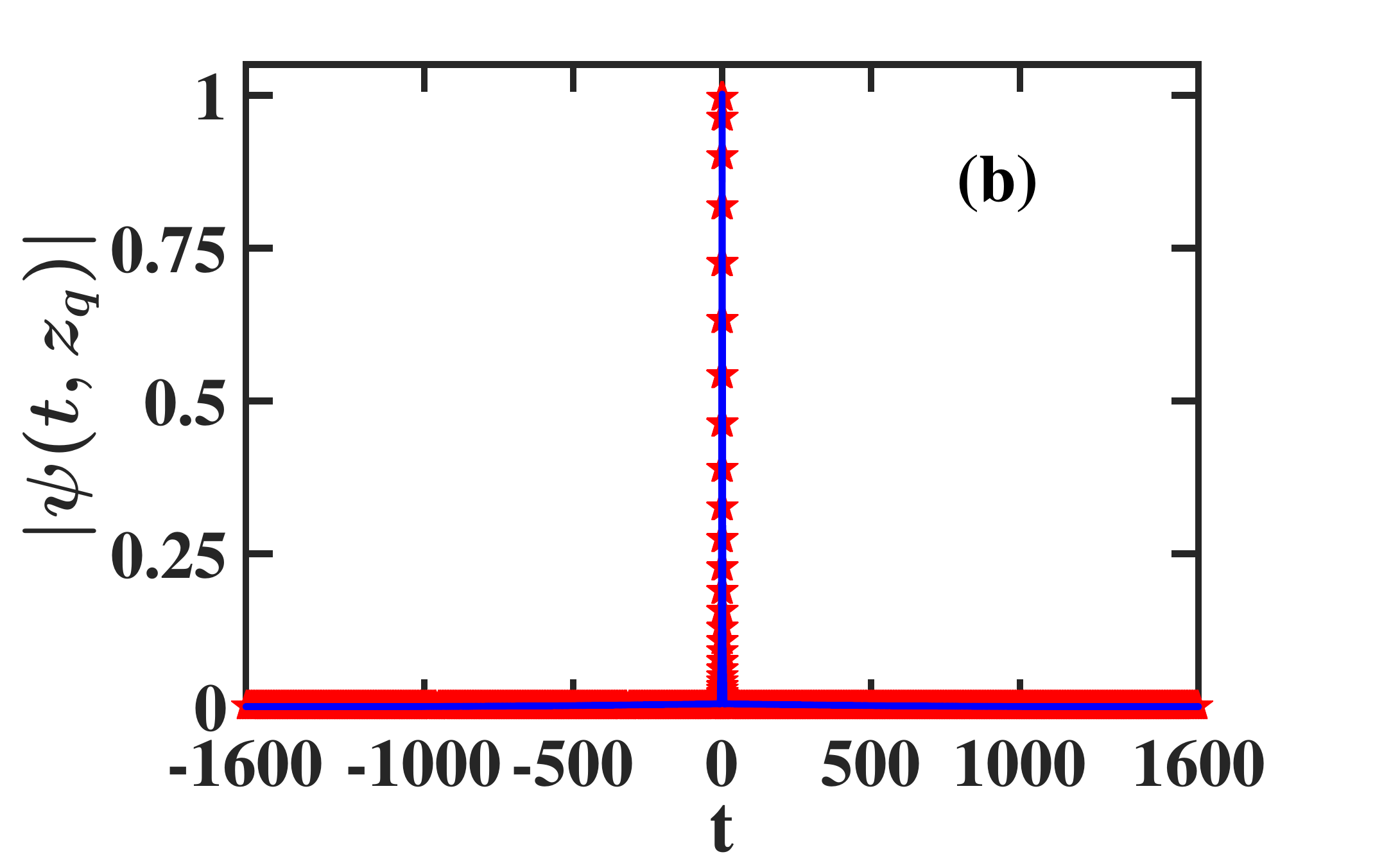} \\
\epsfxsize=5.8cm  \epsffile{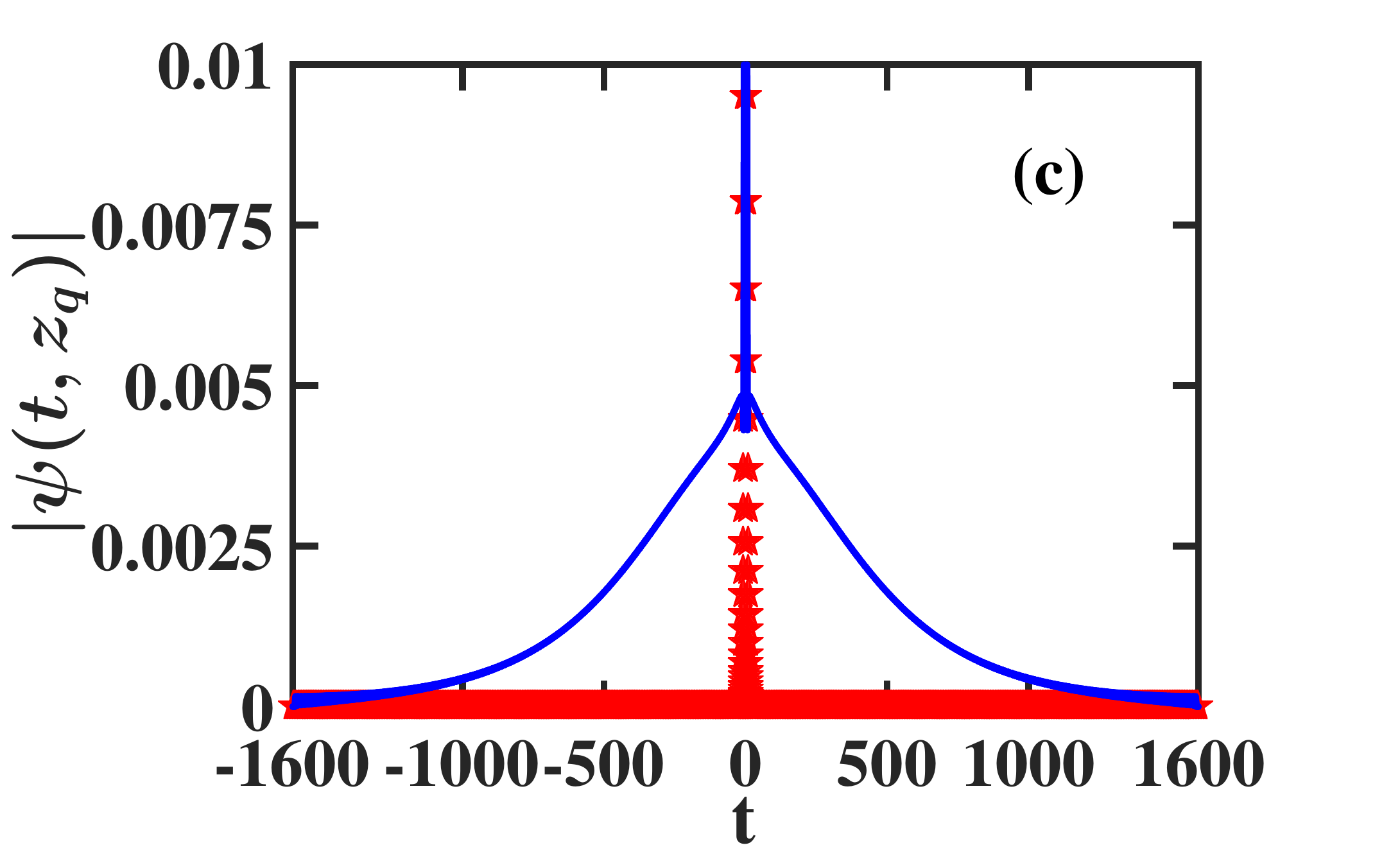} &
\epsfxsize=5.8cm  \epsffile{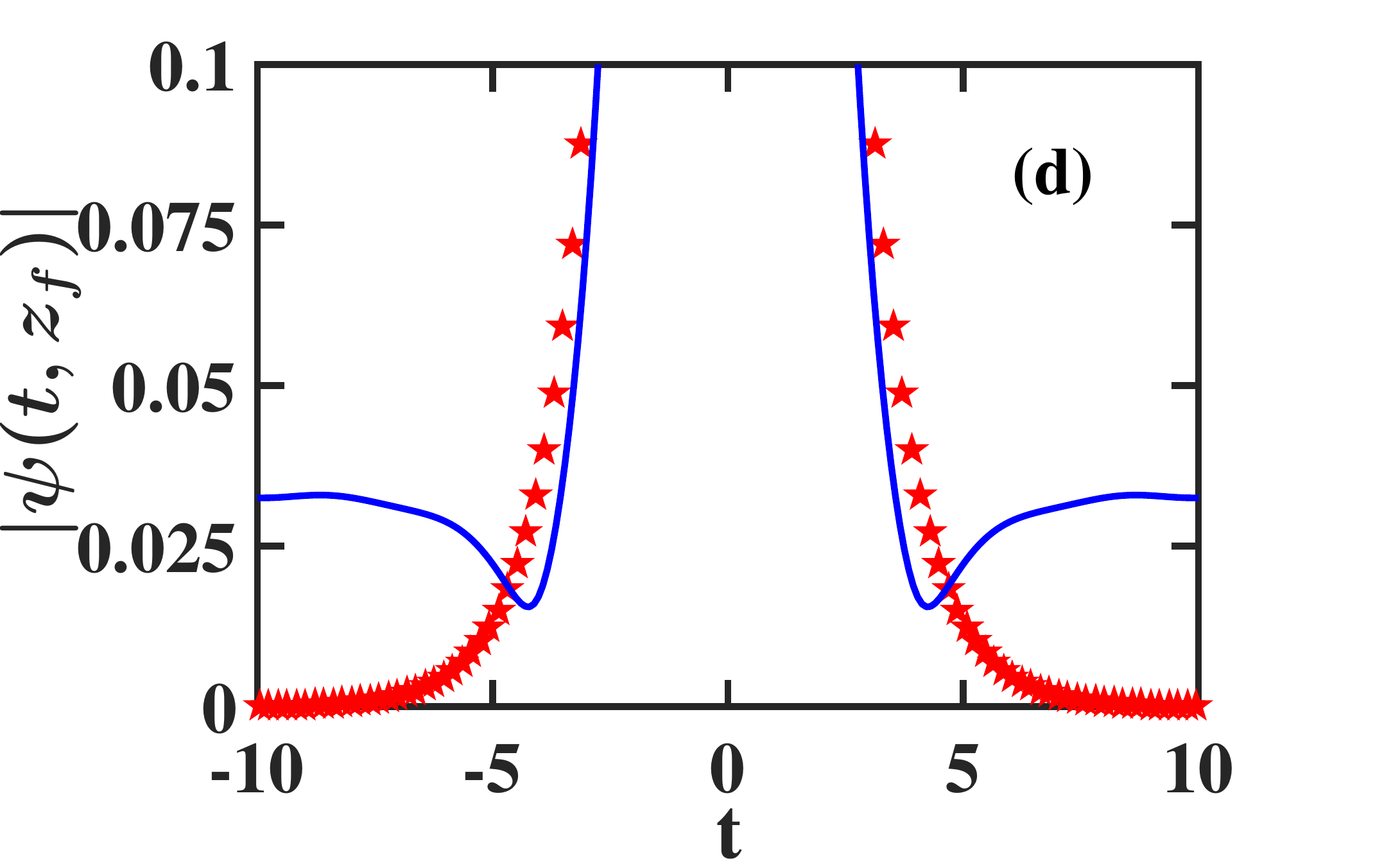} \\
\epsfxsize=5.8cm  \epsffile{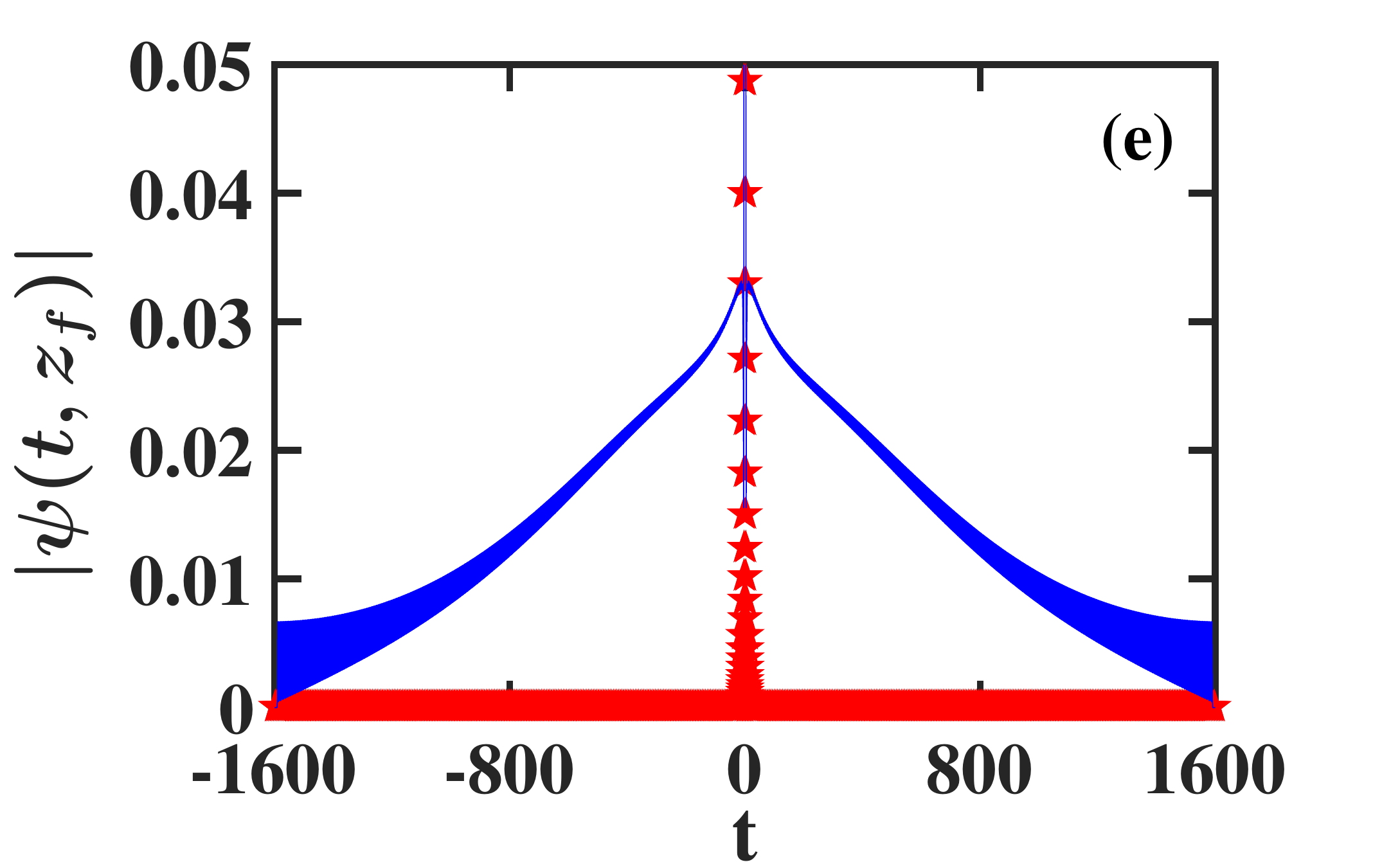}
\end{tabular}
\caption{The pulse shape $|\psi(t,z)|$ at $z_{q}=432$ [(a), (b), and (c)]  
and at $z_{f}=750$ [(d) and (e)] for soliton propagation in an open optical waveguide 
with weak frequency dependent linear gain-loss and cubic loss. 
The cubic loss coefficient is $\epsilon_{3}=0.01$, 
the initial soliton amplitude is $\eta(0)=0.8$, 
and the parameters of the linear gain-loss  $\hat g(\omega)$ 
in Eq. (\ref{sfs12}) are $W=10$, $\rho=10$, and $g_{L}=0.5$.  
The solid blue curve corresponds to the result obtained by numerical 
simulations with Eqs.  (\ref{sfs11}) and (\ref{sfs12}), while the red stars correspond to 
the prediction of the perturbation theory, obtained with Eqs. (\ref{Iz1}) and (\ref{sfs16}).}
 \label{fig5}
\end{figure}

\begin{figure}[ptb]
\begin{tabular}{cc}
\epsfxsize=10cm  \epsffile{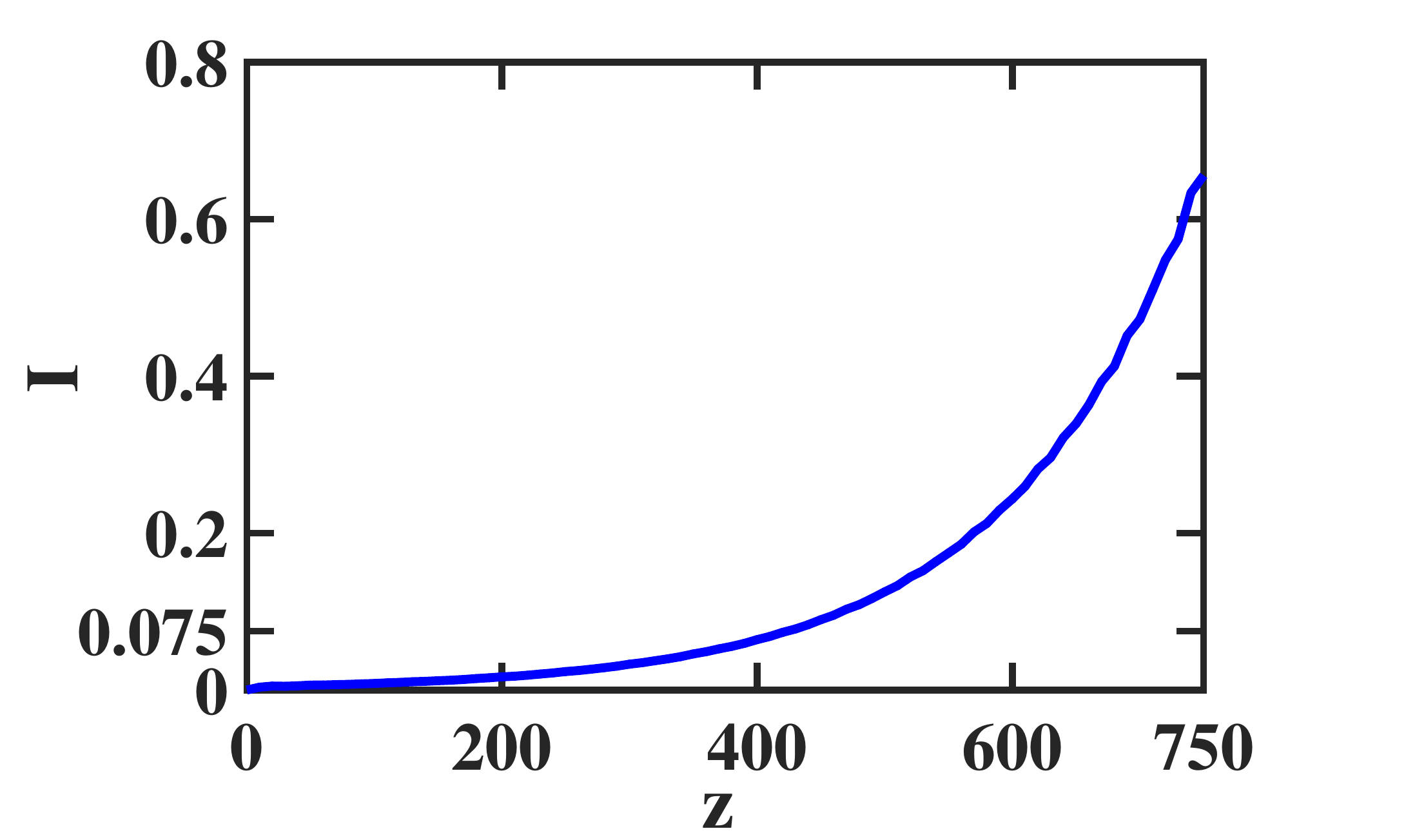} 
\end{tabular}
\caption{The $z$ dependence of the transmission quality integral $I(z)$ obtained 
by numerical simulations with Eqs. (\ref{sfs11}) and (\ref{sfs12}) 
for the same optical waveguide setup considered in Fig. \ref{fig5}.}
\label{fig6}
\end{figure}

The pulse shape $|\psi(t,z)|$ obtained in the simulations at $z=z_{q}$ and at $z=z_{f}$
is shown in Fig. \ref{fig5}. Also shown is the prediction of the adiabatic perturbation theory, 
obtained with Eqs. (\ref{Iz1}) and (\ref{sfs16}). 
We observe that the evolution of the pulse shape is very similar to the one obtained for 
waveguides with frequency independent linear gain. More specifically, the numerical result 
for the pulse shape at $z=z_{q}$ is very close to the prediction of the adiabatic perturbation theory 
[see Figs. \ref{fig5}(a) and \ref{fig5}(b)]. 
However, as seen in Fig. \ref{fig5}(c), the soliton develops an appreciable radiative tail at 
$z=z_{q}$ with a shape that is very similar to the one observed for waveguides with frequency 
independent linear gain. Additionally, as seen in Figs. \ref{fig5}(d) and \ref{fig5}(e), 
the radiative tail keeps growing as the soliton continues to propagate along the waveguide.     
The growth of the radiative tail with increasing $z$ leads to an increase of the integral $I(z)$ 
with values that are very close to the values obtained for waveguides with frequency independent 
linear gain. Indeed, as seen in Fig. \ref{fig6}, the value of $I(z)$ 
for a waveguide with frequency dependent linear gain-loss increases from 0.075 at $z_{q}=432$ 
to 0.6557 at $z_{f}=750$, compared with an increase from  0.075 at $z_{q}=432$ 
to 0.6556 at $z_{f}=750$ for a waveguide with frequency independent linear gain.

\begin{figure}[ptb]
\begin{center}
\begin{tabular}{cc}
\epsfxsize=5.8cm  \epsffile{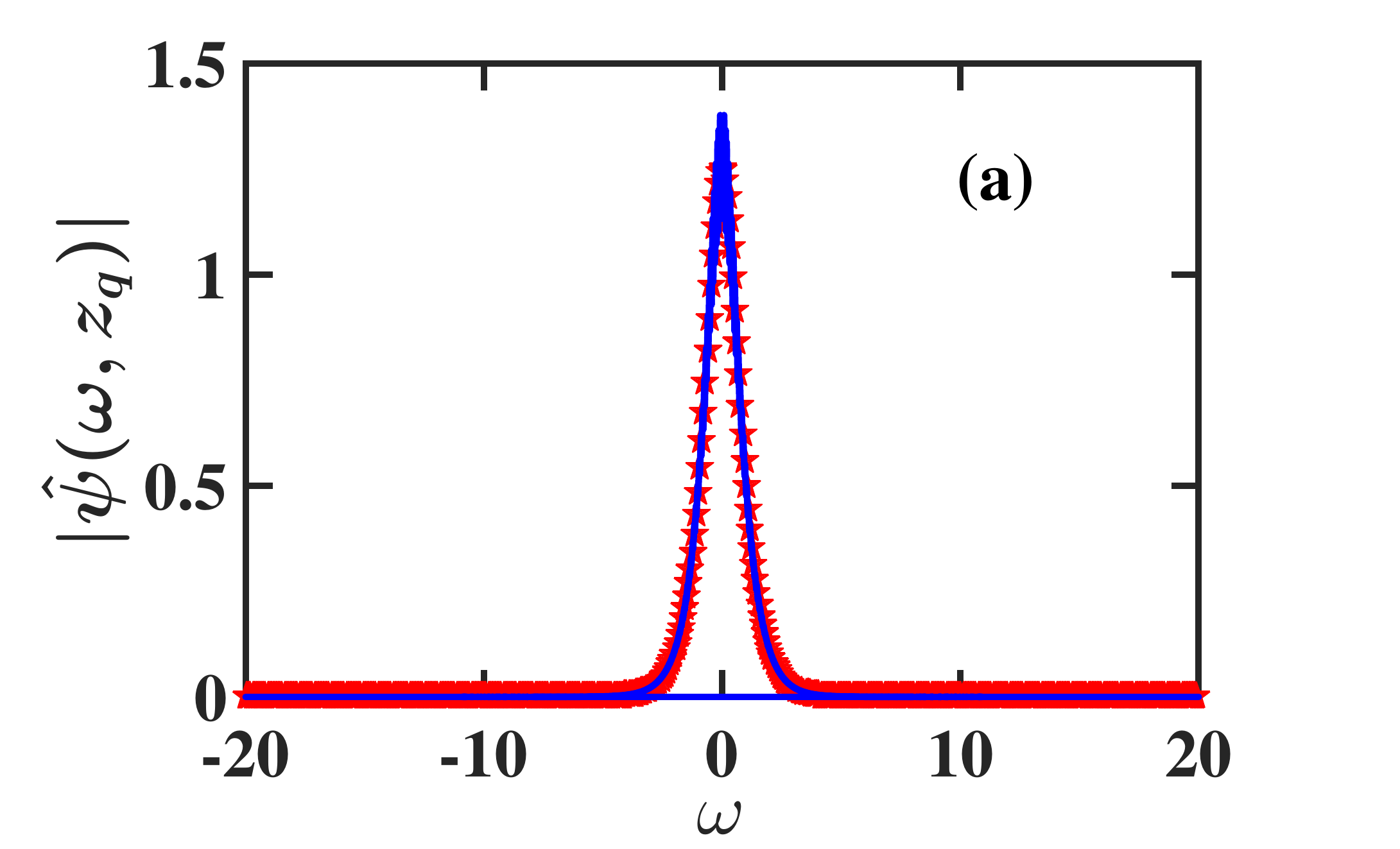} &
\epsfxsize=5.8cm  \epsffile{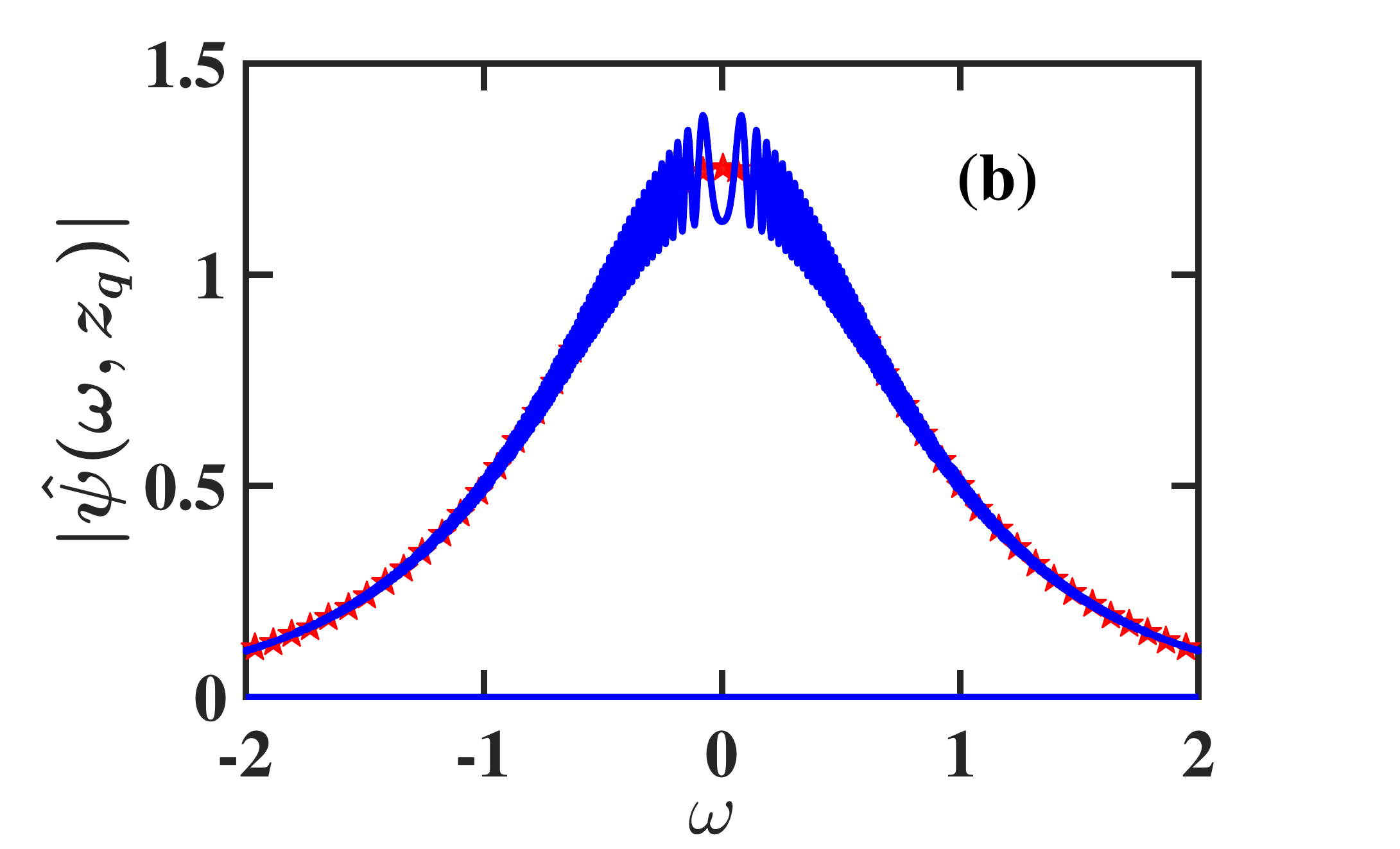} \\
\epsfxsize=5.8cm  \epsffile{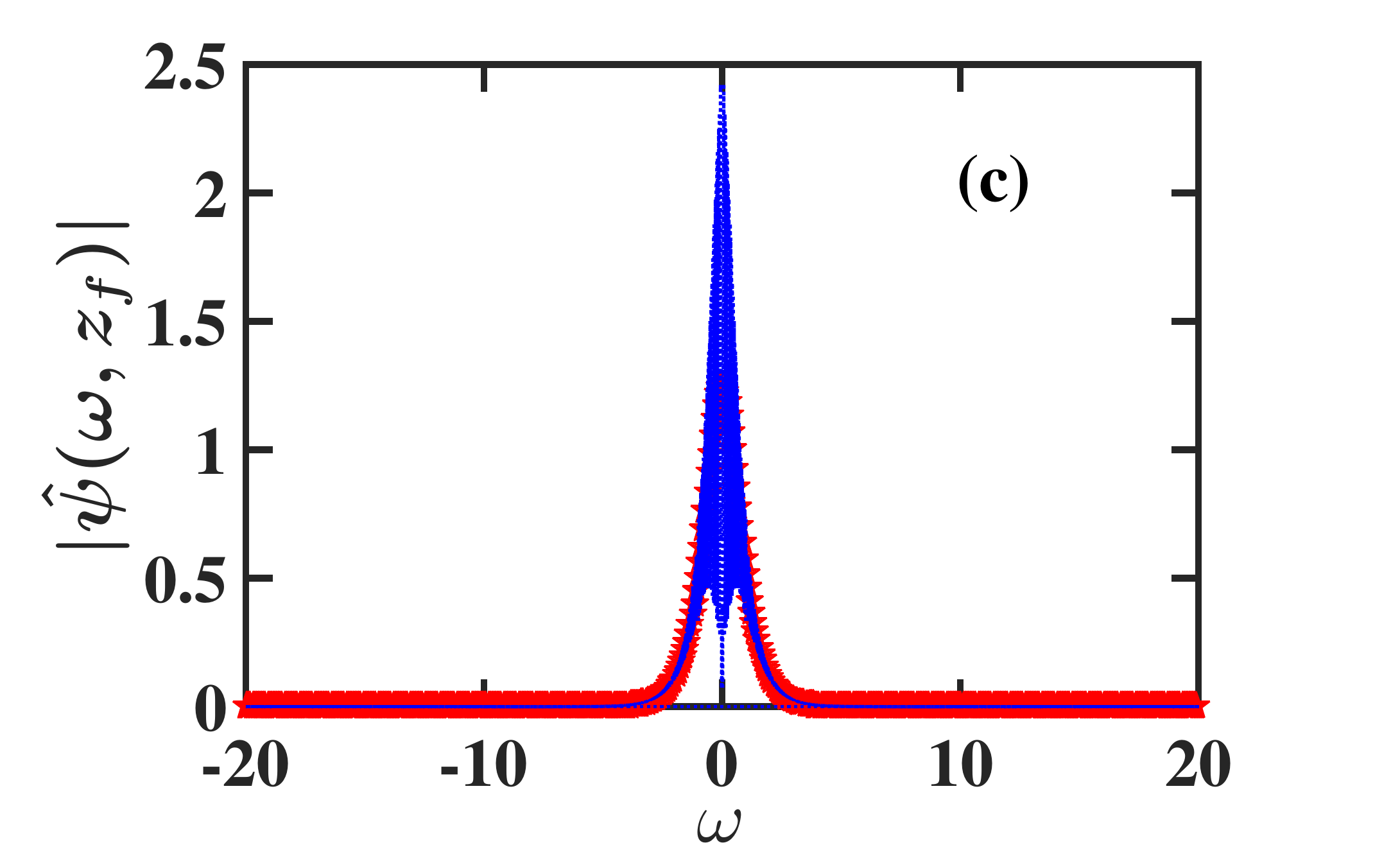} 
\end{tabular}
\end{center}
\caption{The Fourier transform of the pulse shape $|\hat\psi(\omega,z)|$ 
at $z_{q}=432$ [(a) and (b)] and at $z_{f}=750$ (c) 
for soliton propagation in an open optical waveguide 
with weak frequency dependent linear gain-loss and cubic loss. 
The physical parameter values are the same as in Fig. \ref{fig5}. 
The solid blue curve represents the result obtained by numerical 
simulations with Eqs. (\ref{sfs11}) and (\ref{sfs12}), 
while the red stars correspond to the prediction of the adiabatic perturbation theory, 
obtained with Eqs. (\ref{Iz3}) and (\ref{sfs16}).}                        
 \label{fig7}
\end{figure}

The similarity between pulse dynamics in the presence of frequency dependent 
linear gain-loss and frequency independent linear gain can be understood with the 
help of the Fourier transform of the pulse $|\hat\psi(\omega,z)|$.  
Figure \ref{fig7} shows the numerically obtained $|\hat\psi(\omega,z)|$ 
at $z=z_{q}$ and at $z=z_{f}$ along with the prediction of 
the adiabatic perturbation theory, obtained with Eqs. (\ref{Iz3}) and (\ref{sfs16}).
We observe that the graphs of $|\hat\psi(\omega,z_{q})|$ and $|\hat\psi(\omega,z_{f})|$ 
vs $\omega$ are very similar to the graphs obtained for waveguides with frequency independent 
linear gain. More specifically, the deviation of the numerical result from the analytic prediction 
is noticeable already at $z=z_{q}$ and is of order 1 at $z=z_{f}$. 
Additionally, the deviation appears as fast oscillations in the graph of the 
numerically obtained $|\hat\psi(\omega,z)|$ vs $\omega$, 
which are most pronounced at small $\omega$ values.  
Moreover, there is no observable separation between the Fourier spectrum of 
the soliton and the Fourier spectrum of the radiation. As a result, the introduction 
of the frequency dependent linear gain-loss with $W$ values satisfying $W \gg 1$ 
does not lead to efficient mitigation of radiation emission in the current waveguide setup. 
We will demonstrate in sections \ref{Raman_sfs2} and \ref{filters2} that the situation 
changes dramatically due to the effects of delayed Raman response or due to the effects 
of guiding optical filters with a varying central frequency.

\begin{figure}[ptb]
\begin{tabular}{cc}
\epsfxsize=10cm  \epsffile{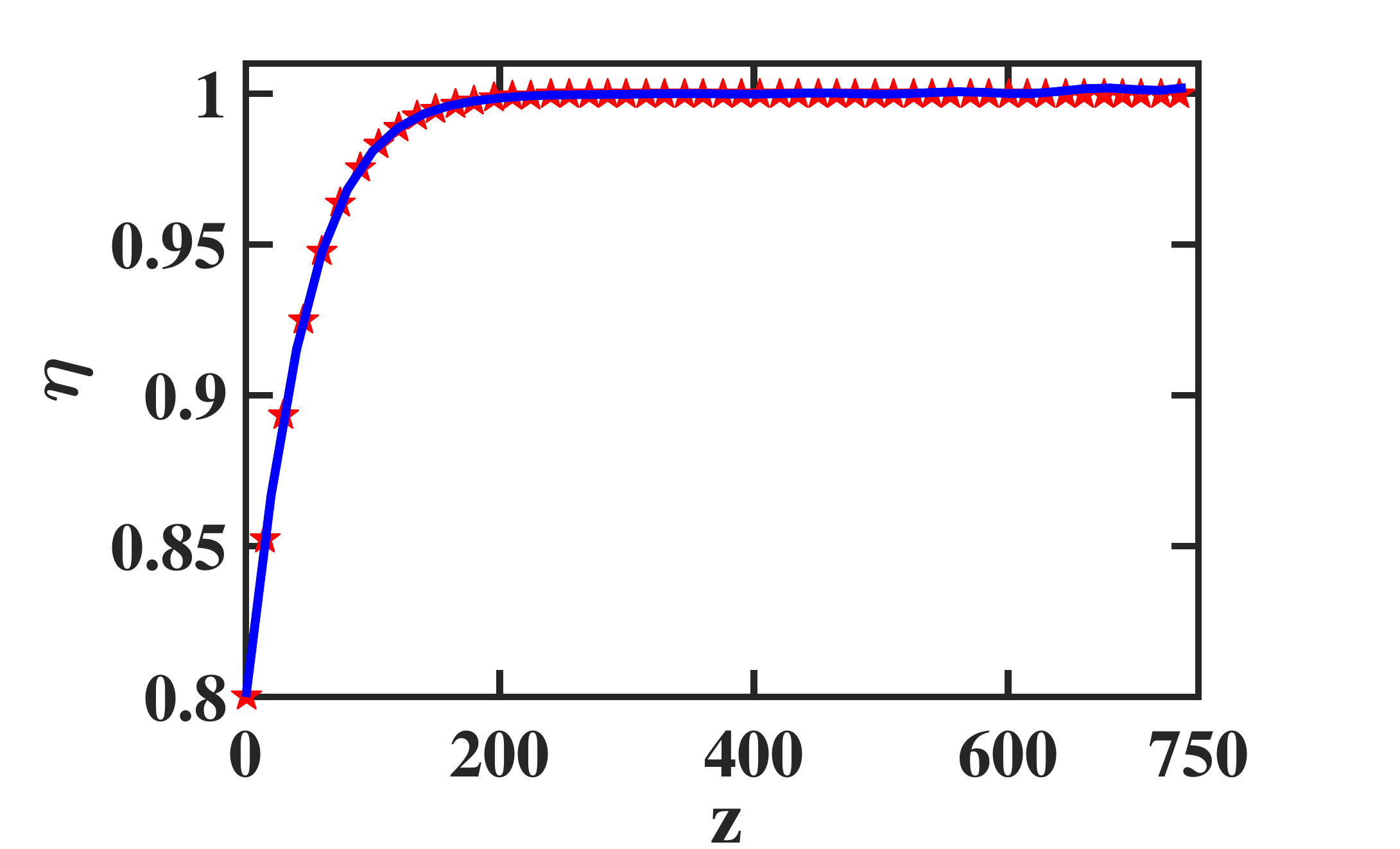} 
\end{tabular}
\caption{The $z$ dependence of the soliton amplitude $\eta(z)$ for the open 
waveguide setup considered in Figs. \ref{fig5}-\ref{fig7}. The solid blue curve represents the result 
obtained by numerical simulations with Eqs.  (\ref{sfs11}) and (\ref{sfs12}). 
The red stars represent the perturbation theory prediction of Eq. (\ref{sfs16}).}
\label{fig8}
\end{figure}

Figure \ref{fig8} shows the $z$ dependence of the soliton amplitude obtained 
in the simulations together with the analytic prediction of Eq. (\ref{sfs16}). 
We observe good agreement between the numerical and analytic results 
for $0 \le z \le 600$, while for $600 < z \le 750$, the difference between 
the numerical result and the analytic prediction becomes noticeable.  
Thus, similar to the situation in open optical waveguides with frequency independent 
linear gain, the dynamics of the soliton amplitude is still stable in the interval $0 \le z \le 750$.

\section{Pulse dynamics in waveguides with linear gain-loss, cubic loss, and delayed Raman response}
\label{Raman_sfs}
{\it Introduction.} As seen in section \ref{no_shifting2}, the replacement 
of frequency independent linear gain by frequency dependent linear 
gain-loss does not lead to significant enhancement of transmission quality. On the other hand, 
numerical simulations of multisequence soliton-based transmission show that transmission stability is 
significantly enhanced when the effects of delayed Raman response 
and frequency dependent linear gain-loss are both taken into account \cite{PNT2016,PNH2017}.  
It is therefore important to investigate whether the presence 
of delayed Raman response can improve transmission quality in the single-soliton propagation 
problem considered in the current paper. We now turn to address this question.

\subsection{Waveguides with frequency independent linear gain, cubic loss, and delayed Raman response}
\label{Raman_sfs1}

We start by considering the impact of delayed Raman response on the propagation of 
a single soliton in nonlinear optical waveguides with weak frequency independent linear gain and cubic loss. 
The propagation is described by the following perturbed NLS 
equation \cite{Agrawal2007a,Dekker2007}: 
\begin{eqnarray}
i\partial_z\psi+\partial_t^2\psi+2|\psi|^2\psi=
ig_{0}\psi/2 - i\epsilon_{3}|\psi|^2\psi
+\epsilon_{R}\psi\partial_{t}|\psi|^2,
\label{sfs21}
\end{eqnarray}  
where the Raman coefficient $\epsilon_{R}$ satisfies $0 < \epsilon_{R} \ll 1$ \cite{dimensions2,Chi89}. 
The third term on the right-hand side of Eq. (\ref{sfs21}) describes the effects of delayed Raman response.

A calculation based on the adiabatic perturbation theory shows that the main effect 
of delayed Raman response on single-soliton propagation is a frequency shift, 
whose rate is given by \cite{Mitschke86,Gordon86,Kodama87}:  
\begin{equation}
\frac{d\beta}{dz} = -\frac{8}{15}\epsilon_{R}\eta^{4}.
\label{sfs22}
\end{equation} 
The soliton amplitude is not affected by delayed Raman response in $O(\epsilon_{R})$ \cite{Mitschke86,Gordon86,Kodama87}. 
Therefore, the dynamics of the soliton amplitude is still given by Eqs. (\ref{sfs4}) and (\ref{sfs5}). 
Substituting $\eta(z)$ from Eq. (\ref{sfs5}) into Eq. (\ref{sfs22}) and integrating with respect to $z$, 
we obtain: 
\begin{eqnarray}
\beta(z) = \beta(0) 
-\frac{\epsilon_{R}\eta_{0}^{2}}{5\epsilon_{3}}
\left\{\ln\left[
\frac{\eta_{0}^{2} - \eta^{2}(0) + \eta^{2}(0)\exp\left(8\epsilon_{3}\eta_{0}^{2}z/3\right)}{\eta_{0}^{2}}\right]
\right.
\nonumber \\
\left.
+\frac{\eta^{2}(0)}{\eta_{0}^{2}}
-\frac{\eta^{2}(0)}{\eta^{2}(0)+\left[\eta_{0}^{2} - \eta^{2}(0)\right]
\exp\left(-8\epsilon_{3}\eta_{0}^{2}z/3\right)}
\right\} .
\label{sfs23}
\end{eqnarray}      
The soliton position and phase are affected by the perturbations only via 
the dependence of $\eta$ and $\beta$ on $z$.

{\it Nemerical simulations.} Equation  (\ref{sfs21}) is numerically solved on a domain $[t_{\mbox{min}},t_{\mbox{max}}]=[-400,400]$ 
with periodic boundary conditions. The initial condition is in the form of a single NLS soliton with 
amplitude $\eta(0)$, frequency $\beta(0)=0$, position $y(0)=0$, and phase $\alpha(0)=0$. 
As a typical example, we present here the results of the simulations with 
$\epsilon_{3}=0.01$, $\epsilon_{R}=0.04$, and $\eta(0)=0.8$. 
We point out that similar results are obtained for other physical parameters values. 
Due to the presence of delayed Raman response and the relatively large propagation 
distance, the soliton experiences a very large position shift. 
For example, for $\epsilon_{R}=0.04$, $\eta(0)=1$, and $\tilde z=750$, 
we find using the adiabatic perturbation theory that the soliton position shift at $\tilde z=750$ 
is $y(\tilde z)=8\epsilon_{R}\eta^{4}(0)\tilde z^{2}/15=12000$. 
Carrying out numerical simulations for transmission in an open optical waveguide setup, i.e., 
in a setup in which the soliton does not reach the computational domain's boundaries, 
is prohibitively time consuming, since one has to employ a computational domain with a 
size exceeding 12000. We therefore choose to work with 
a numerical simulations setup, in which the soliton passes 
through the computational domain's boundaries multiple times during the simulation. 
In such setup, we do {\it not} use damping at the boundaries, since such damping 
leads to the soliton's destruction. Note that the numerical simulations setup used in 
the current section corresponds to soliton propagation in a closed optical waveguide loop. 
This setup is very relevant for applications, since many long-distance 
transmission experiments are carried out in closed waveguide loops, see. e.g., 
Refs. \cite{Mollenauer2006,Mollenauer97,MM98,Nakazawa2000,Nakazawa91,MMN96,Mollenauer2003} 
and references therein. The values of the transmission quality distance and the final propagation distance   
obtained in the simulations were $z_{q}=378$ and $z_{f}=785$.

\begin{figure}[ptb]
\begin{tabular}{cc}
\epsfxsize=5.8cm  \epsffile{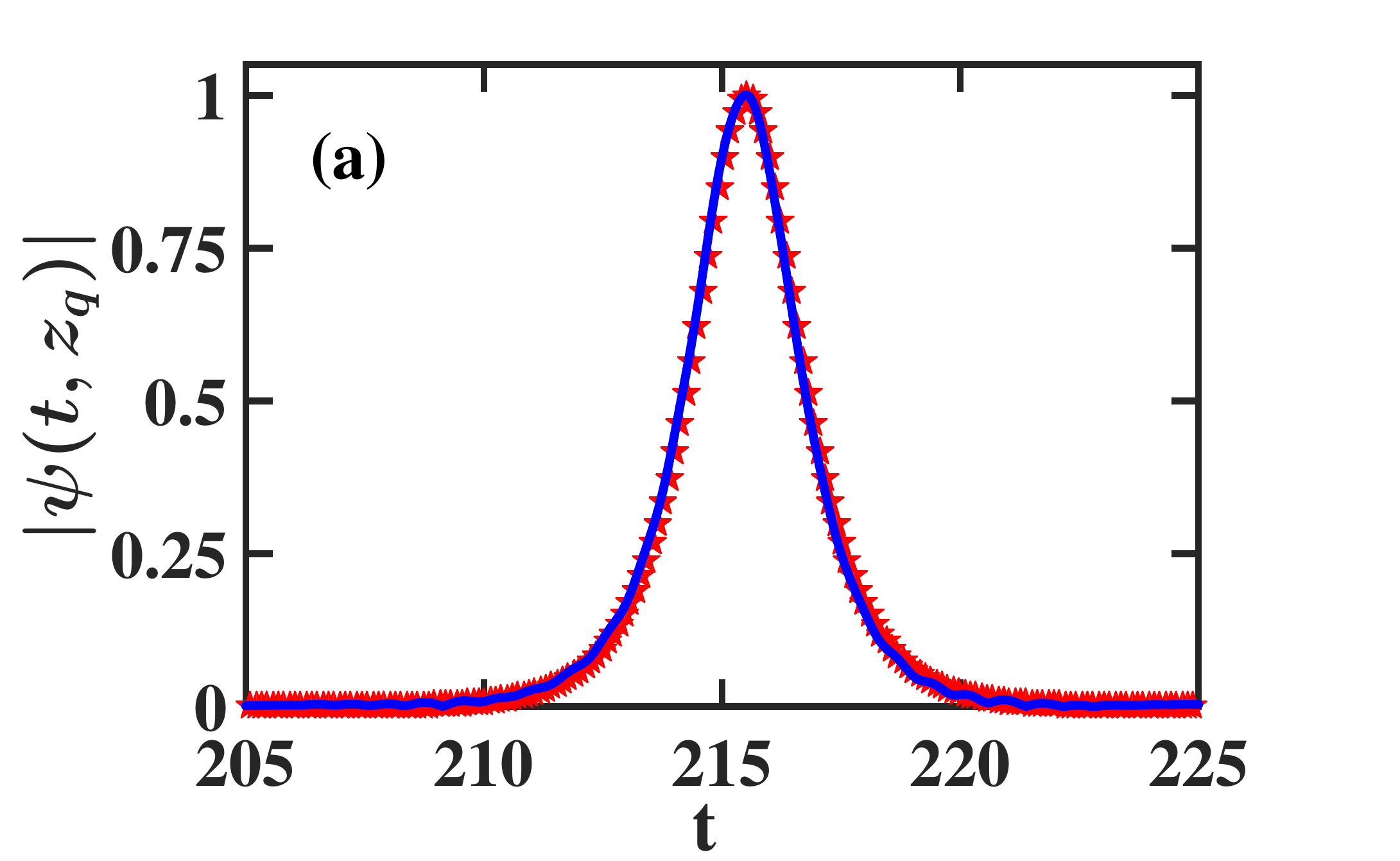} &
\epsfxsize=5.8cm  \epsffile{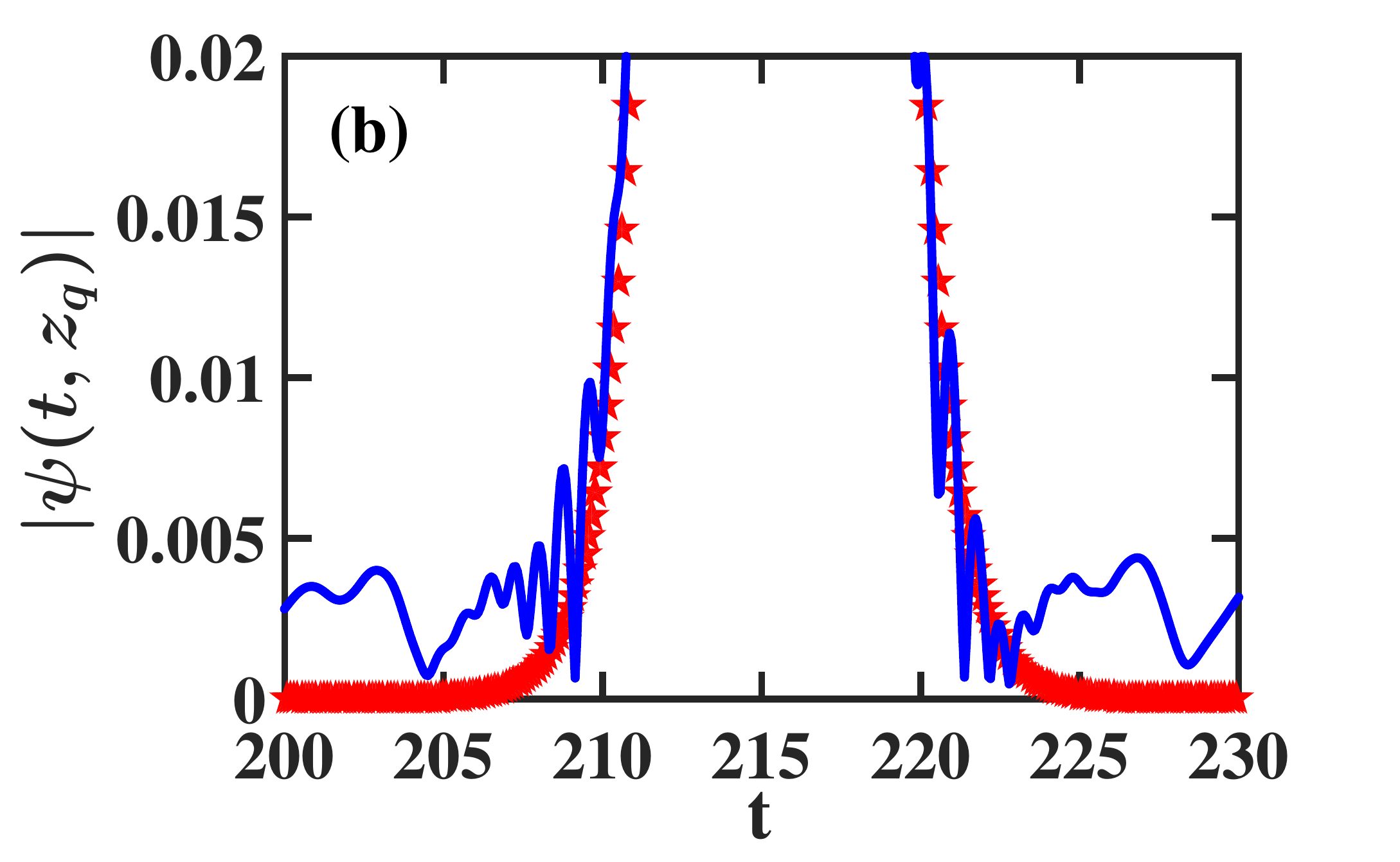} \\
\epsfxsize=5.8cm  \epsffile{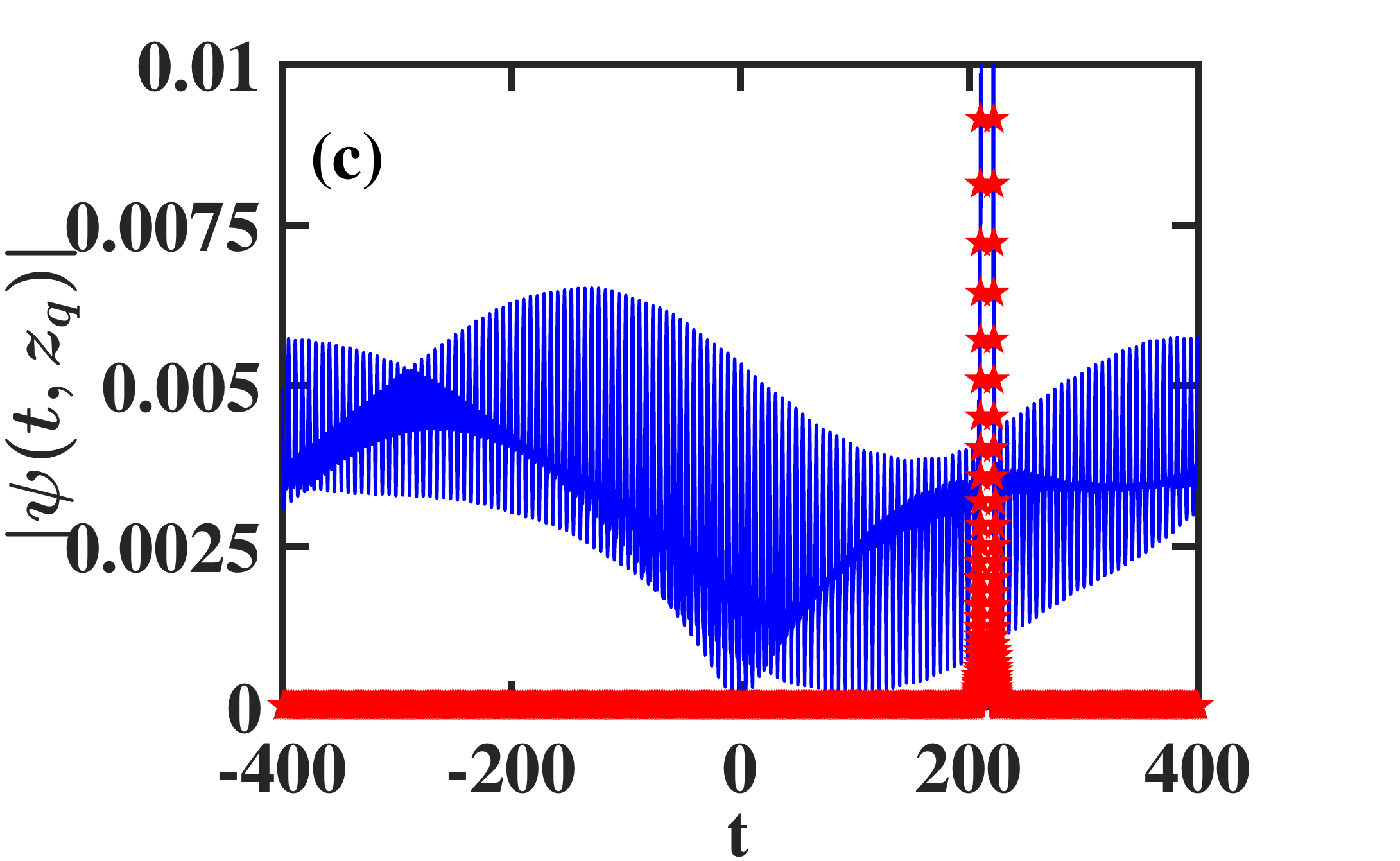} &
\epsfxsize=5.8cm  \epsffile{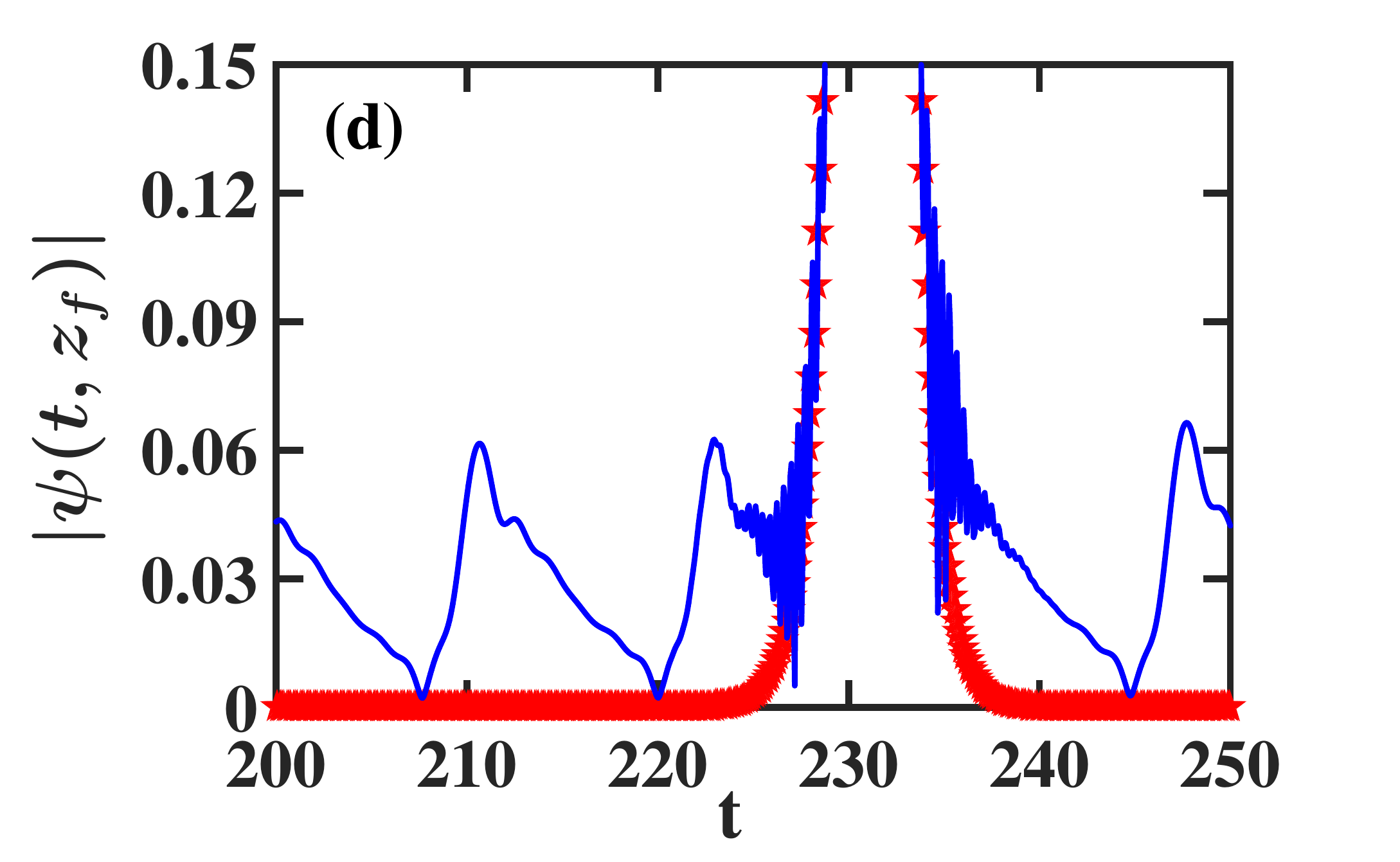} \\
\epsfxsize=5.8cm  \epsffile{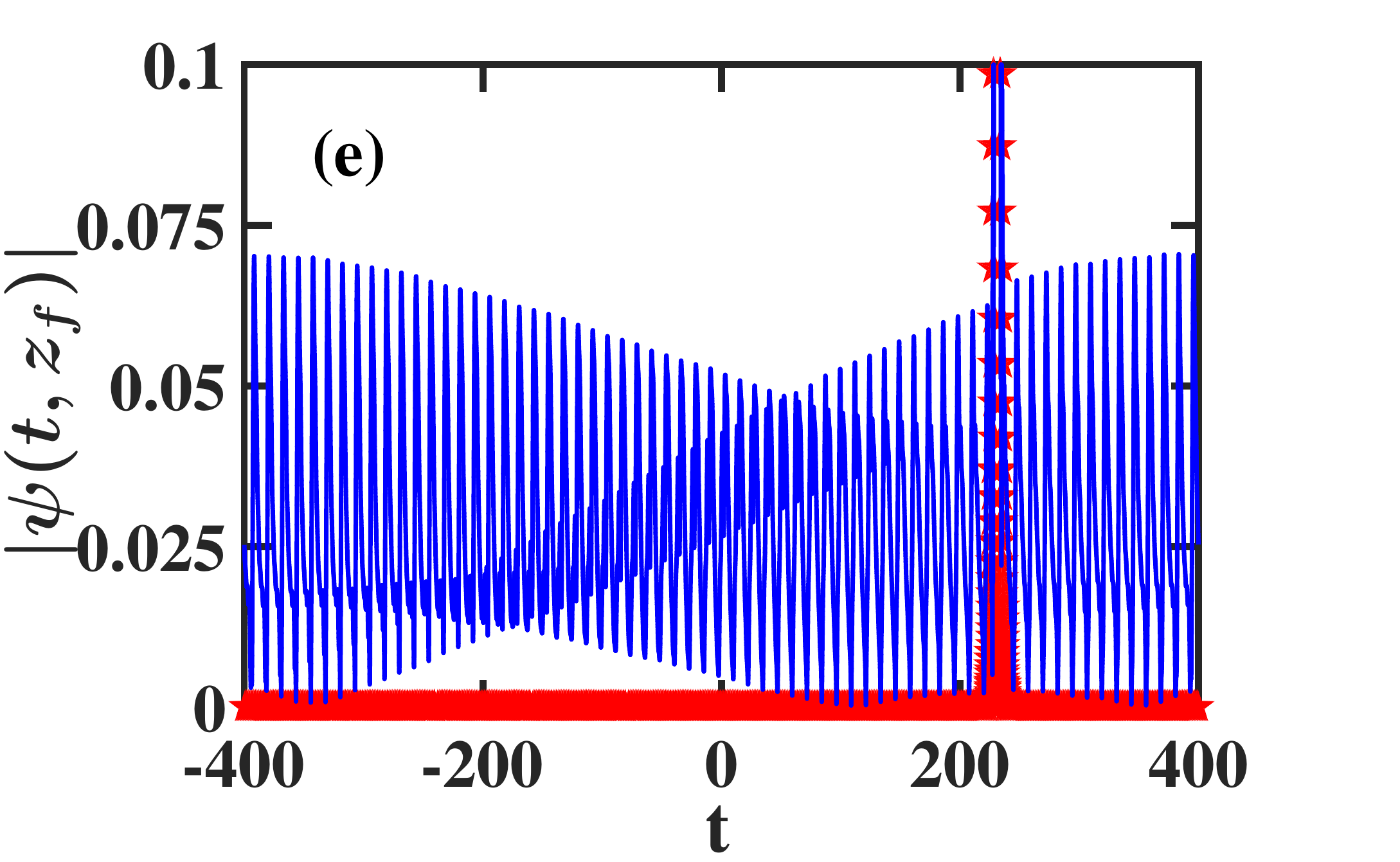}
\end{tabular}
\caption{The pulse shape $|\psi(t,z)|$ at $z_{q}=378$ [(a), (b), and (c)]  
and at $z_{f}=785$ [(d) and (e)] for soliton propagation in a closed optical waveguide loop  
with weak frequency independent linear gain, cubic loss, and delayed Raman response. 
The cubic loss coefficient is $\epsilon_{3}=0.01$, the Raman coefficient is $\epsilon_{R}=0.04$,  
and the initial soliton amplitude is $\eta(0)=0.8$. 
The solid blue curve corresponds to the result obtained by numerical 
simulations with Eq.  (\ref{sfs21}), while the red stars correspond to 
the perturbation theory prediction of Eqs. (\ref{Iz1}) and (\ref{sfs5}).}
\label{fig9}
\end{figure}

\begin{figure}[ptb]
\begin{tabular}{cc}
\epsfxsize=10cm  \epsffile{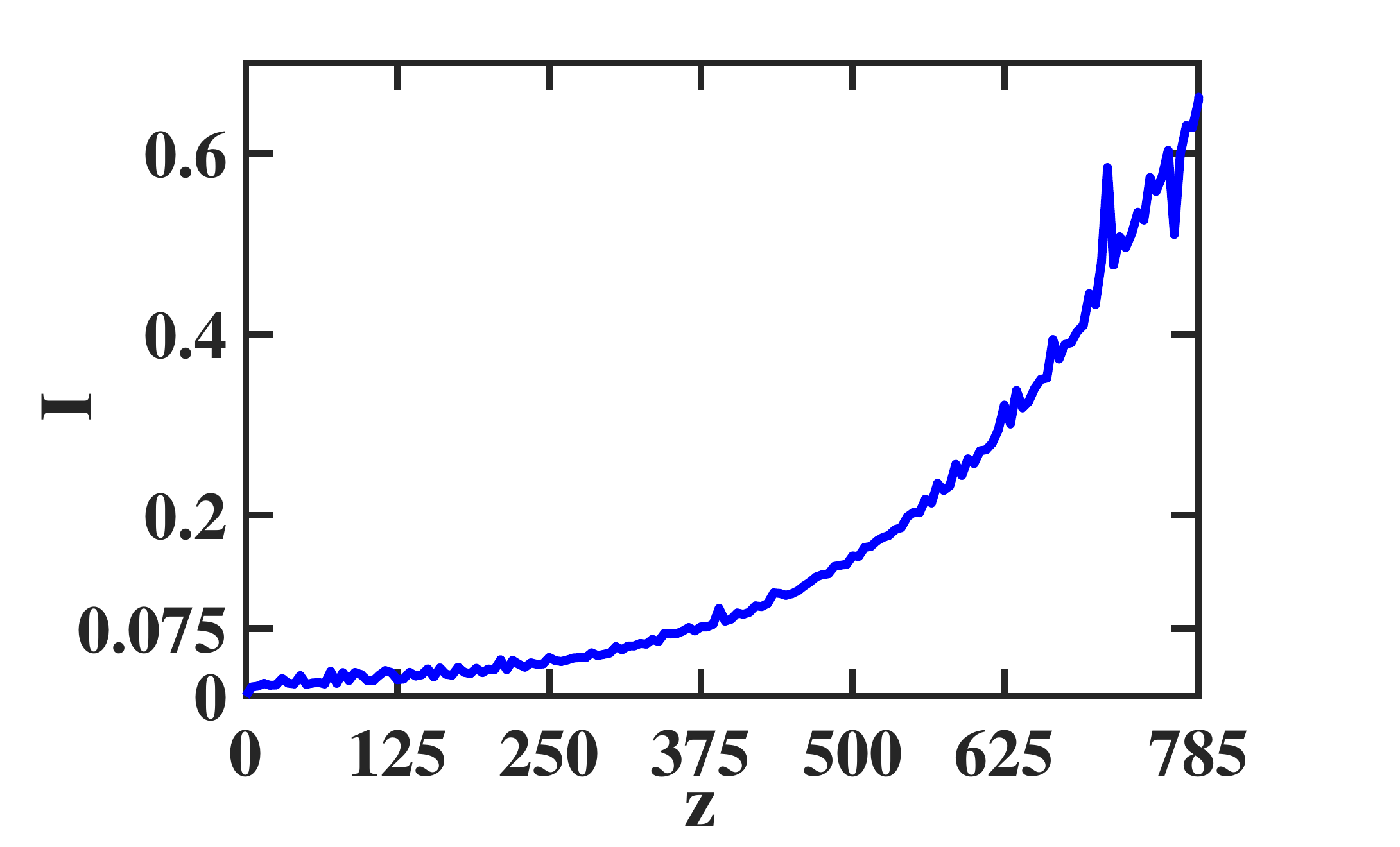} 
\end{tabular}
\caption{The $z$ dependence of the transmission quality integral $I(z)$ obtained 
by numerical simulations with Eq. (\ref{sfs21}) for the same optical waveguide 
setup considered in Fig. \ref{fig9}.}
\label{fig10}
\end{figure}

Figure \ref{fig9} shows the pulse shape $|\psi(t,z)|$ at $z=z_{q}$ and at $z=z_{f}$, 
obtained in the simulations together with the prediction of the adiabatic perturbation theory, 
obtained with Eqs. (\ref{Iz1}) and (\ref{sfs5}). 
As seen in Figs. \ref{fig9}(a),  \ref{fig9}(b), and \ref{fig9}(c),  
the numerically obtained pulse shape at $z=z_{q}$ is close to the analytic prediction, 
although, a noticeable radiative tail exists at this distance. 
We observe that the radiative tail is highly oscillatory and is  
spread over the entire computational domain at $z=z_{q}$. 
The oscillatory nature of the radiative tail is attributed to the presence 
of delayed Raman response. The spread of radiation over the 
entire computational domain is due to additional emission of radiation induced 
by the presence of delayed Raman response, the closed waveguide loop setup, 
which leads to accumulation of radiation, and the smaller size of 
the computational time domain compared with the one used 
in the simulations in section \ref{no_shifting}. 
We also observe that the radiative tail continues to grow 
as the soliton continues to propagate along the waveguide [see Figs. \ref{fig9}(d) and \ref{fig9}(e)].           
As a result, the value of the transmission quality integral $I(z)$
increases from 0.075 at $z_{q}=378$ to 0.6565 at $z_{f}=785$ [see Fig. \ref{fig10}].

\begin{figure}[ptb]
\begin{center}
\begin{tabular}{cc}
\epsfxsize=5.8cm  \epsffile{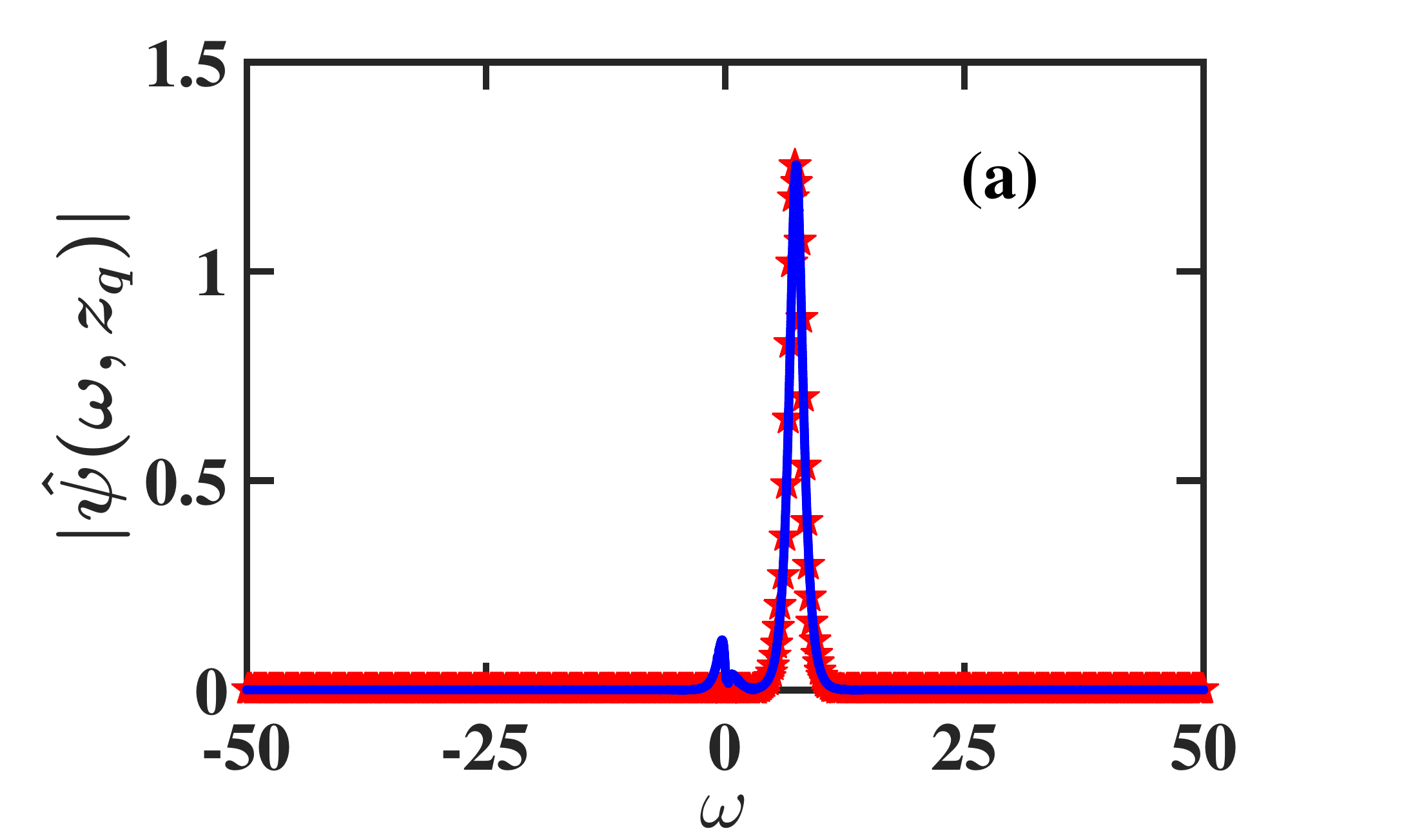} &
\epsfxsize=5.8cm  \epsffile{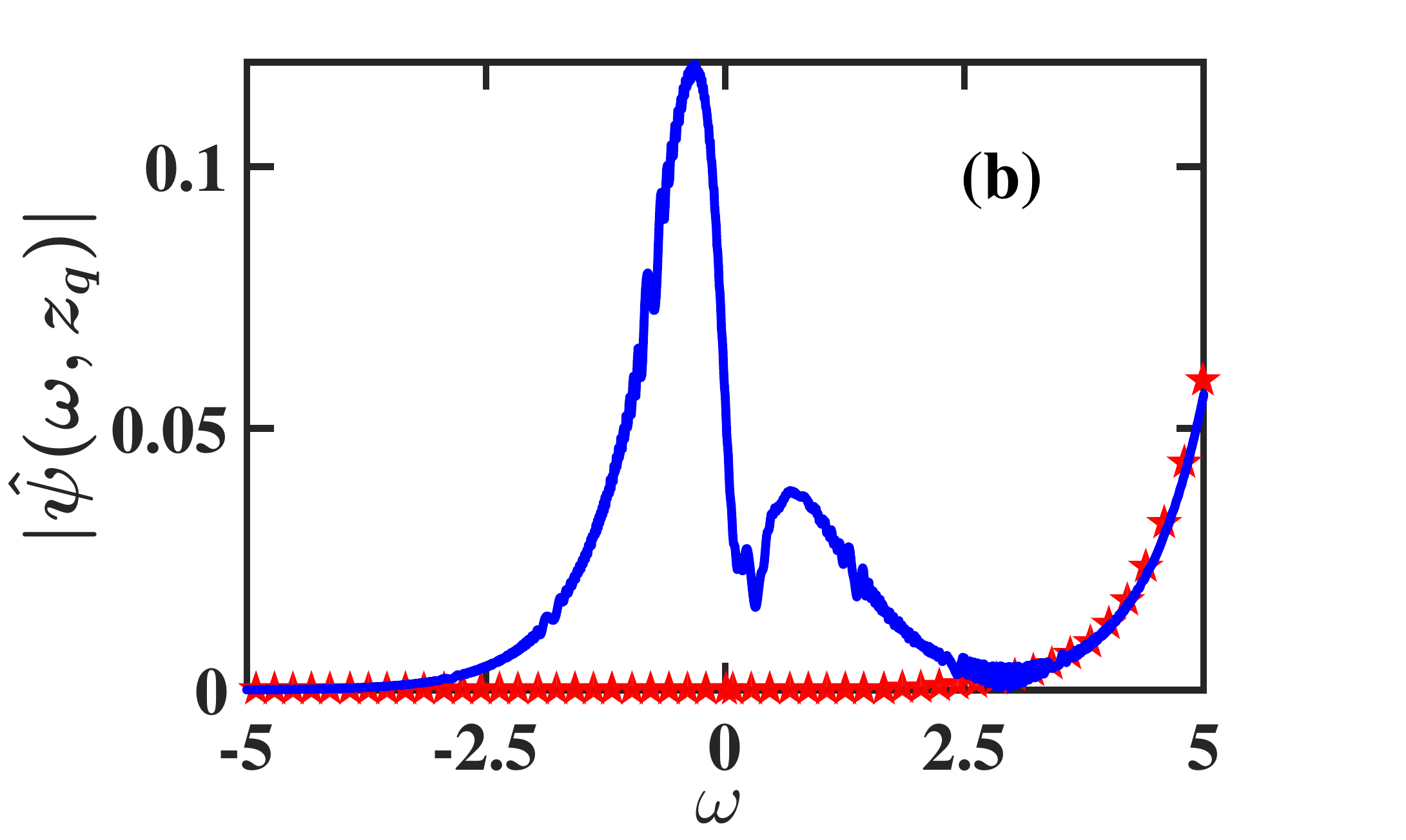} \\
\epsfxsize=5.8cm  \epsffile{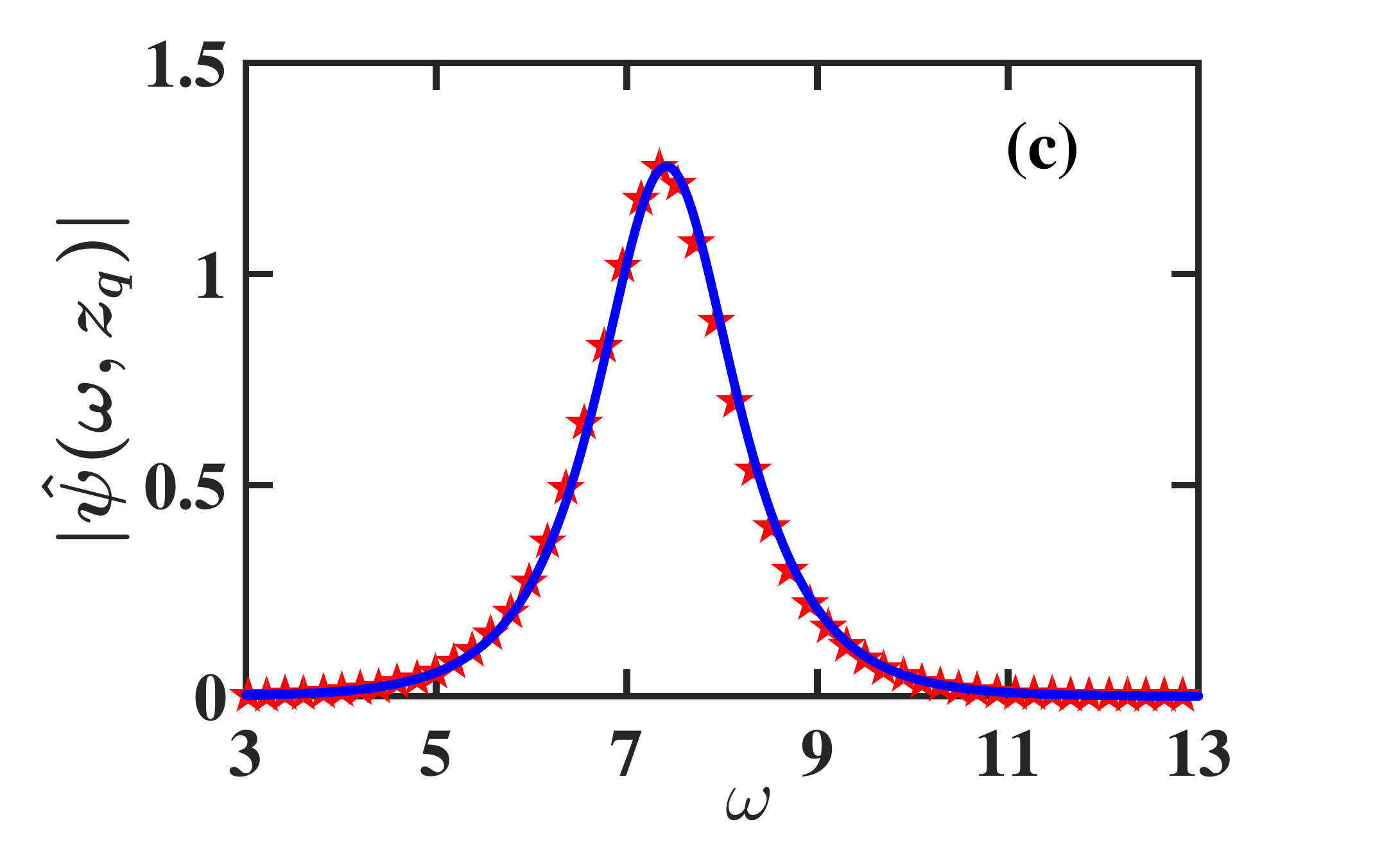} &
\epsfxsize=5.8cm  \epsffile{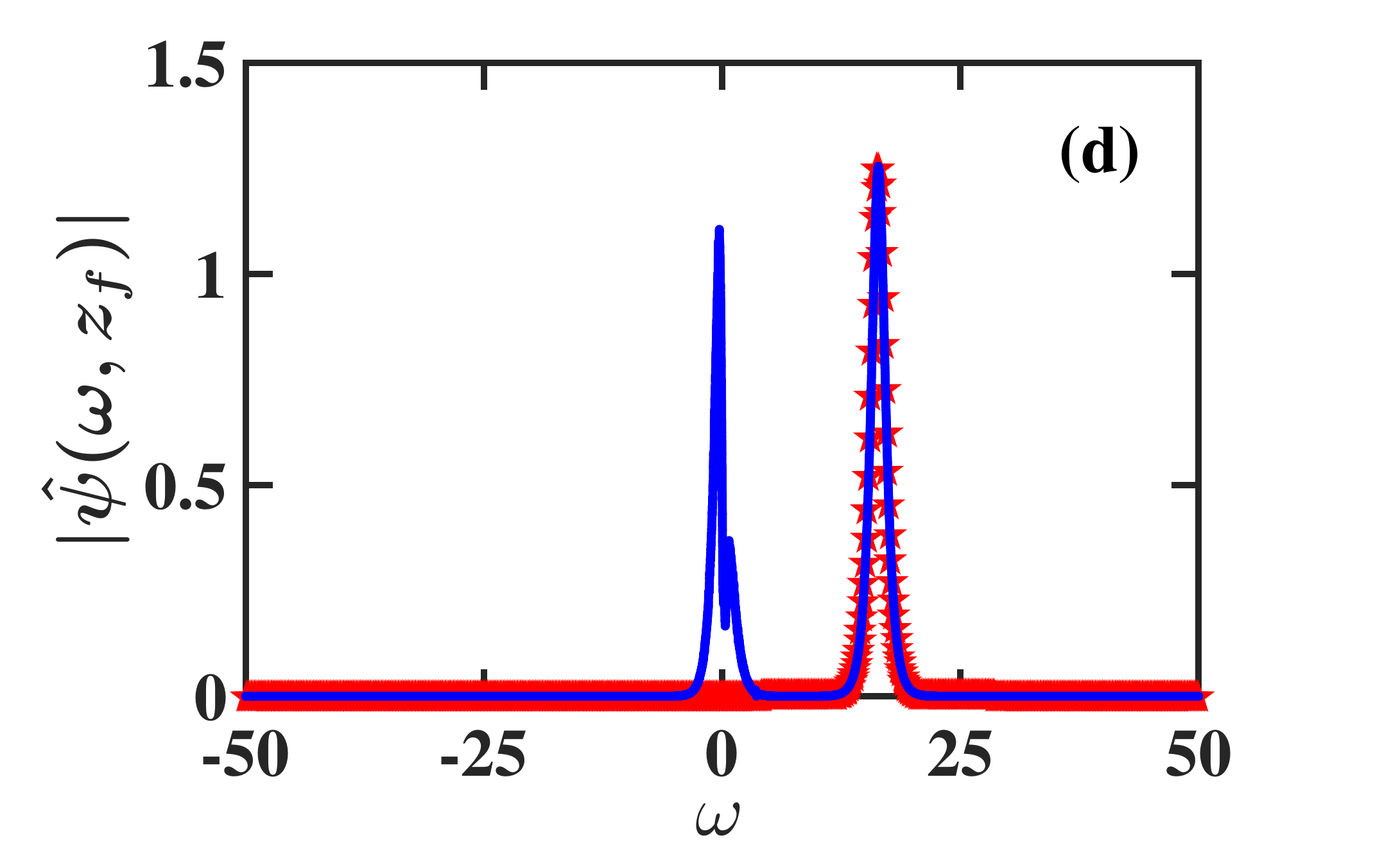}  
\end{tabular}
\end{center}
\caption{The Fourier transform of the pulse shape $|\hat\psi(\omega,z)|$ 
at $z_{q}=378$ [(a), (b), and (c)] and at $z_{f}=785$ (d) 
for soliton propagation in a closed optical waveguide loop  
with weak frequency independent linear gain, cubic loss, and delayed Raman response. 
The physical parameter values are the same as in Fig. \ref{fig9}. 
The solid blue curve represents the result obtained by numerical 
simulations with Eq. (\ref{sfs21}) and the red stars correspond 
to the prediction of the adiabatic perturbation theory, 
obtained with Eqs. (\ref{Iz3}), (\ref{sfs5}), and  (\ref{sfs23}).}                        
 \label{fig11}
\end{figure}

Further insight into transmission quality degradation and pulse dynamics is 
gained from the shape of the Fourier spectrum $|\hat\psi(\omega,z)|$. 
Figure \ref{fig11} shows the numerically obtained $|\hat\psi(\omega,z)|$ 
at $z=z_{q}$ and at $z=z_{f}$ together with the prediction of 
the adiabatic perturbation theory, obtained with Eqs. (\ref{Iz3}), (\ref{sfs5}), and  (\ref{sfs23}).
It is clear that the Fourier spectrum of the optical field for waveguides with 
frequency independent linear gain, cubic loss, and delayed Raman response is very different 
from the Fourier spectrum observed in section \ref{no_shifting} for soliton 
propagation in the absence of delayed Raman response. 
More specifically, the soliton's Fourier spectrum in the current waveguide setup is centered 
about the nonzero $z$ dependent soliton's frequency $\beta(z)$ 
and is shifted relative to the radiation's spectrum, which is centered near $\omega=0$. 
The separation between the soliton's spectrum and the radiation's spectrum, 
which is a result of the Raman self-frequency shift experienced by the soliton, 
is already very clear at $z=z_{q}$ [see Figs. \ref{fig11}(a) and \ref{fig11}(b)].  
It continues to grow with increasing $z$ due to the increase in $|\beta(z)|$ [see Fig. \ref{fig11}(d)].   
As a result of the separation between the two spectra, the soliton part of the numerically 
obtained graph of $|\hat\psi(\omega,z)|$ does not contain fast oscillations and is very 
close to the prediction of the adiabatic perturbation theory [see Fig. \ref{fig11}(c)].       
In contrast, in the waveguides considered in section \ref{no_shifting}, 
the Fourier spectrum of the entire optical field (soliton $+$ radiation) is 
centered about $\omega=0$. That is, there is no significant separation between 
the soliton and the radiation spectra. Therefore, for the waveguides considered in section \ref{no_shifting}, 
the deviation of the numerically obtained Fourier spectrum from the spectrum expected for 
an NLS soliton is significant already at $z=z_{q}$.

\begin{figure}[ptb]
\begin{tabular}{cc}
\epsfxsize=8cm  \epsffile{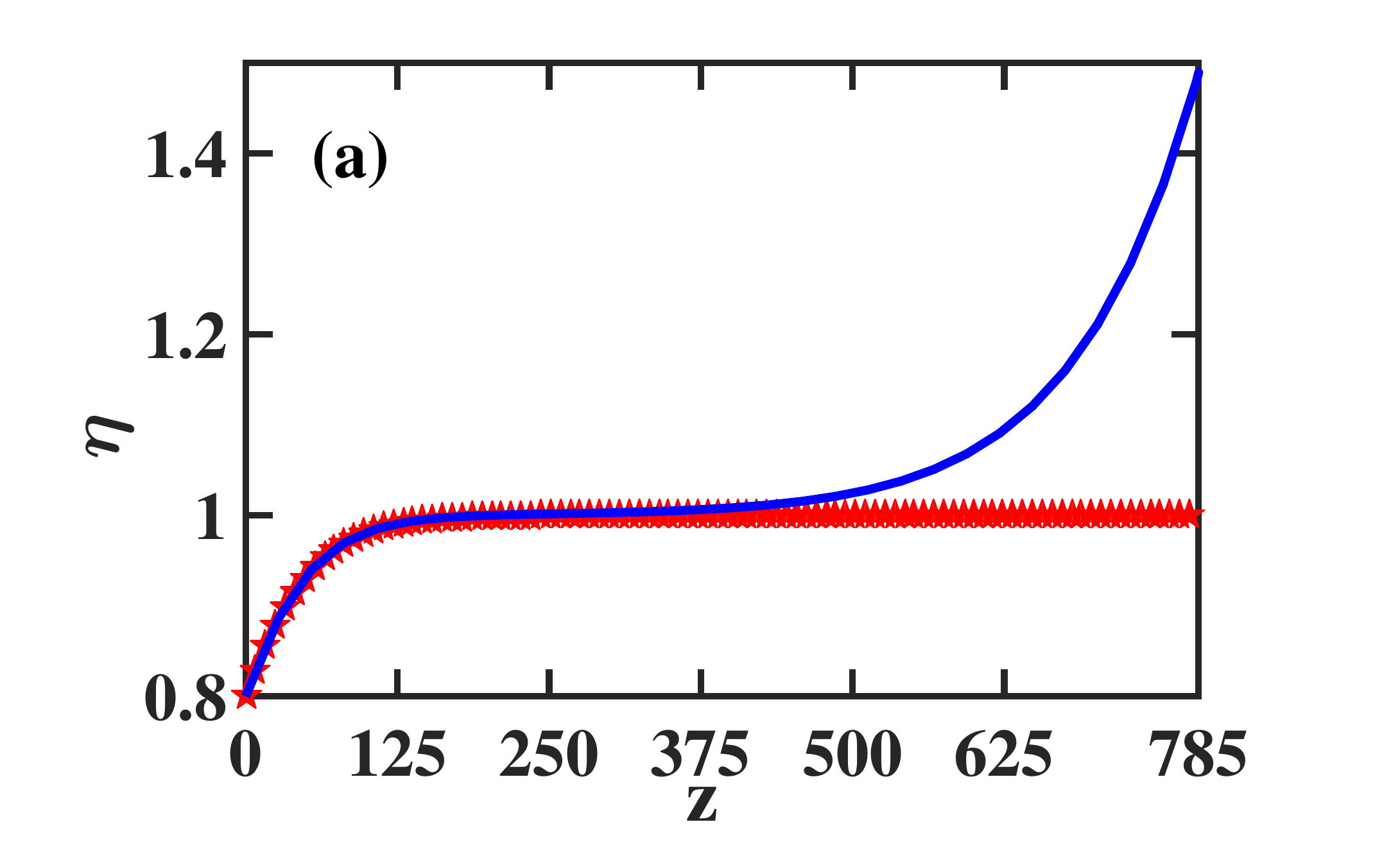} \\
\epsfxsize=8cm  \epsffile{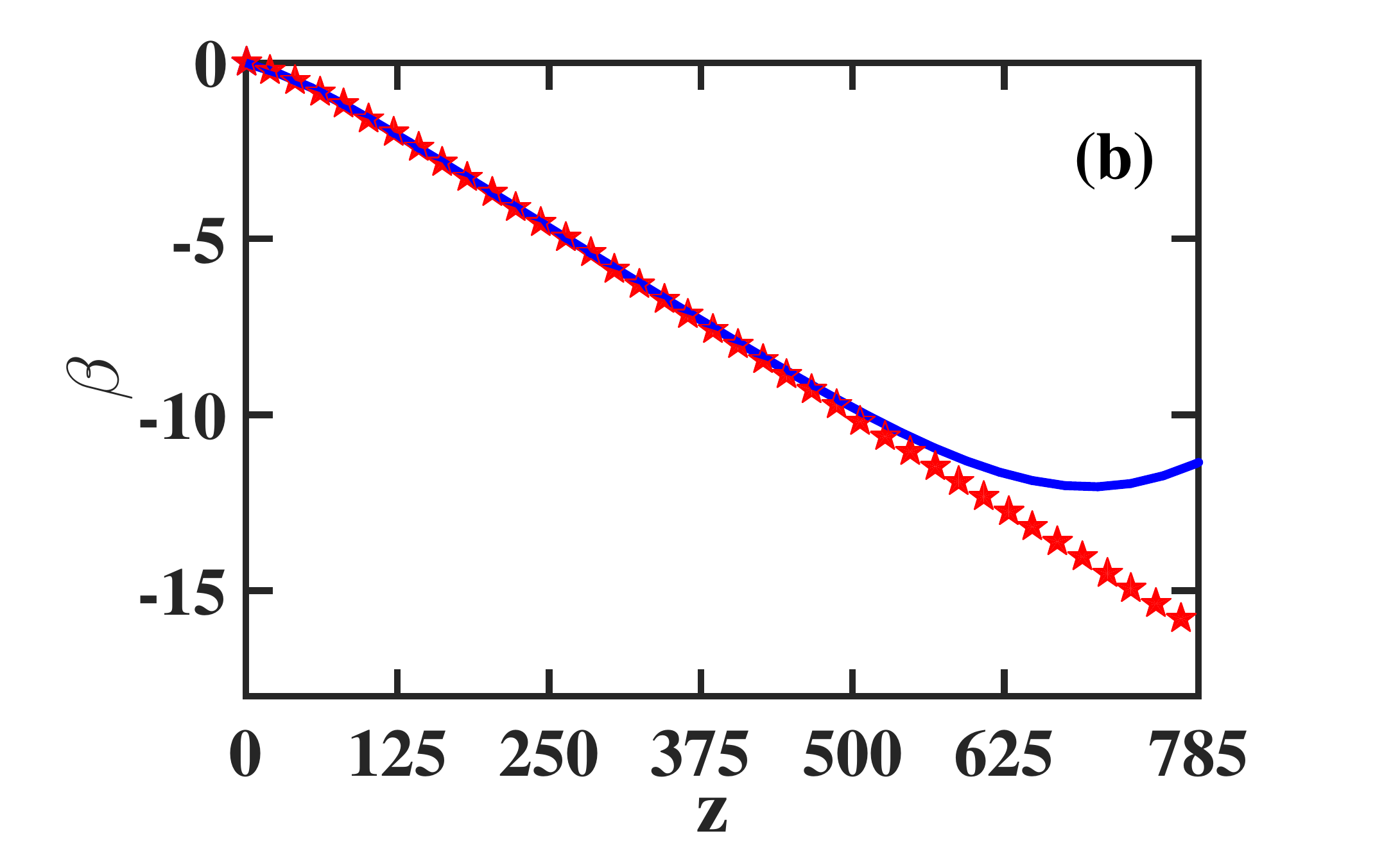}
\end{tabular}
\caption{The $z$ dependence of the soliton amplitude $\eta(z)$ (a) and frequency $\beta(z)$ (b)  
for the closed optical waveguide loop setup considered in Figs. \ref{fig9}-\ref{fig11}. 
The solid blue curves represent the results obtained by numerical simulations with Eq.  (\ref{sfs21}). 
The red stars correspond to the perturbation theory predictions of Eq. (\ref{sfs5}) 
in (a) and of Eq. (\ref{sfs23}) in (b).}
\label{fig12}
\end{figure}

The $z$ dependence of the soliton's amplitude and frequency obtained in the simulations 
is shown in Figs. \ref{fig12}(a) and \ref{fig12}(b). Also shown are the adiabatic perturbation 
theory predictions for $\eta(z)$ and $\beta(z)$, which are given by Eqs. (\ref{sfs5}) 
and (\ref{sfs23}), respectively. In both graphs we observe good agreement 
between the numerical and analytic results for $0 \le z \le 500$, 
whereas for $500 < z \le 785$, the difference between the two results becomes significant.   
Based on this comparison, we conclude that the dynamics of soliton amplitude and frequency 
becomes unstable for distances larger than 500.              
We notice that the deviation of the numerical result from the analytic result for $\eta(z)$ 
in the current waveguide setup is larger compared with the deviation found for soliton propagation 
in the absence of delayed Raman response in section \ref{no_shifting}. 
We attribute this larger deviation to the presence of a larger radiative tail 
[compare Fig. \ref{fig9}(e) with Figs. \ref{fig1}(e) and \ref{fig5}(e)]. 
The radiative tail in the current waveguide setup is larger compared with 
the radiative tail in the waveguide setups of section \ref{no_shifting} due to 
additional emission of radiation induced by the presence of delayed Raman response, 
the closed waveguide loop setup, which leads to accumulation of radiation, 
and the smaller size of the computational time domain used in the simulations.

\subsection{Waveguides with frequency dependent linear gain-loss, cubic loss, and delayed Raman response}
\label{Raman_sfs2}
We saw in section \ref{Raman_sfs1} that the presence of delayed Raman response 
in optical waveguides with frequency independent linear gain and cubic loss leads to strong separation of the 
soliton's Fourier spectrum from the radiation's Fourier spectrum. Thus, we expect that 
the replacement of the frequency independent linear gain by frequency dependent linear gain-loss 
of a from similar to the one in Eq. (\ref{sfs12}) 
will lead to efficient suppression of radiation emission and to significant enhancement of 
transmission quality. We therefore turn to investigate soliton propagation in nonlinear optical waveguides 
in the presence of weak frequency dependent linear gain-loss, cubic loss, and delayed Raman response. 
The propagation is described by the following perturbed NLS equation \cite{PNT2016,PNH2017}:    
\begin{eqnarray}
i\partial_z\psi+\partial_t^2\psi+2|\psi|^2\psi=
i{\cal F}^{-1}(\hat g(\omega,z) \hat\psi)/2 - i\epsilon_{3}|\psi|^2\psi
+\epsilon_{R}\psi\partial_{t}|\psi|^2. 
\label{sfs31}
\end{eqnarray}  
The form of the frequency and distance dependent linear gain-loss $\hat g(\omega,z)$ 
is similar to the one in Eq. (\ref{sfs12}), apart from a replacement of 
the initial soliton frequency $\beta(0)$ by the $z$ dependent soliton frequency $\beta(z)$. 
Thus, $\hat g(\omega,z)$ is given by: 
\begin{eqnarray} &&
\hat g(\omega,z) = -g_{L} + \frac{1}{2}\left(g_{0} + g_{L}\right)
\left[\tanh \left\lbrace \rho \left[\omega + \beta(z)+W/2\right] 
\right\rbrace 
\right.
\nonumber \\&&
\left.
- \tanh \left\lbrace \rho \left[\omega + \beta(z)- W/2\right] 
\right\rbrace\right]. 
\!\!\!\!\!\!\!
\label{sfs32}
\end{eqnarray}       
A similar form was used in Refs. \cite{PNT2016,PNH2017} in studies of multisequence 
soliton-based transmission in the presence of delayed Raman response and 
different transmission stabilizing mechanisms based on frequency dependent gain-loss.         
In the limit $\rho\gg 1$, $\hat g(\omega,z)$ can be approximated 
by the following step function: 
\begin{eqnarray} &&
\hat g(\omega,z) \simeq 
\left\{ \begin{array}{l l}
g_{0} &  \mbox{ if $-\beta(z)-W/2 < \omega \le -\beta(z)+W/2$,}\\
-g_{L} &  \mbox{elsewhere.}\\
\end{array} \right. 
\label{sfs33}
\end{eqnarray}       
We observe that the weak linear gain $g_{0}$ in the frequency interval $(-\beta(z)-W/2, -\beta(z)+W/2]$ 
balances the effects of cubic loss, such that the soliton amplitude approaches the equilibrium value $\eta_{0}$ with increasing $z$. 
Additionally, the relatively strong linear loss $g_{L}$ leads to suppression of emission of radiation with frequencies 
outside of the interval $(-\beta(z)-W/2, -\beta(z)+W/2]$. 
Thus, due to the relatively large separation between the soliton's spectrum and the radiation's spectrum 
expected for the current waveguide setup, the introduction of the frequency dependent linear gain-loss 
$\hat g(\omega,z)$ of Eq. (\ref{sfs32}) is expected to lead to efficient suppression of radiation emission 
and to significant enhancement of transmission quality.

Since the soliton amplitude is not affected by delayed Raman response in $O(\epsilon_{R})$, 
the dynamics of the amplitude is still described by Eq. (\ref{sfs16}). In addition, the dynamics 
of the soliton frequency is given by Eq. (\ref{sfs22}). The soliton position and phase are affected 
by the perturbations only via the dependence of $\eta$ and $\beta$ on $z$.

{\it Numerical simulations}.                  
To check whether the interplay between frequency dependent linear gain-loss 
and delayed Raman response leads to enhanced transmission quality, 
we perform numerical simulations with Eqs. (\ref{sfs31}) and (\ref{sfs32}). 
The equations are numerically integrated on a domain 
$[t_{\mbox{min}},t_{\mbox{max}}]=[-400,400]$ with periodic boundary conditions. 
The initial condition is in the form of a single NLS soliton with amplitude $\eta(0)$, 
frequency $\beta(0)=0$, position $y(0)=0$, and phase $\alpha(0)=0$. 
To enable comparison with the results of the numerical simulations 
in sections \ref{no_shifting} and \ref{Raman_sfs1}, we use 
the same parameter values that were used in those sections. 
That is, we carry out the simulations with $\epsilon_{3}=0.01$, 
$\epsilon_{R}=0.04$, $\eta(0)=0.8$, $W=10$, $\rho=10$, and $g_{L}=0.5$. 
We emphasize, however, that similar results are obtained for other physical parameters values.  
Similar to the simulations in section \ref{Raman_sfs1}, the soliton passes 
multiple times through the computational domain's boundaries during the simulation,  
i.e., the simulation describes soliton propagation in a closed waveguide loop.  
To avoid soliton destruction,  we do not employ damping at the boundaries. 
The simulation is run up to a pre-determined final propagation distance $z_{f}=2000$, 
at which the value of the transmission quality integral is still smaller than 0.075.

\begin{figure}[ptb]
\begin{tabular}{cc}
\epsfxsize=5.8cm  \epsffile{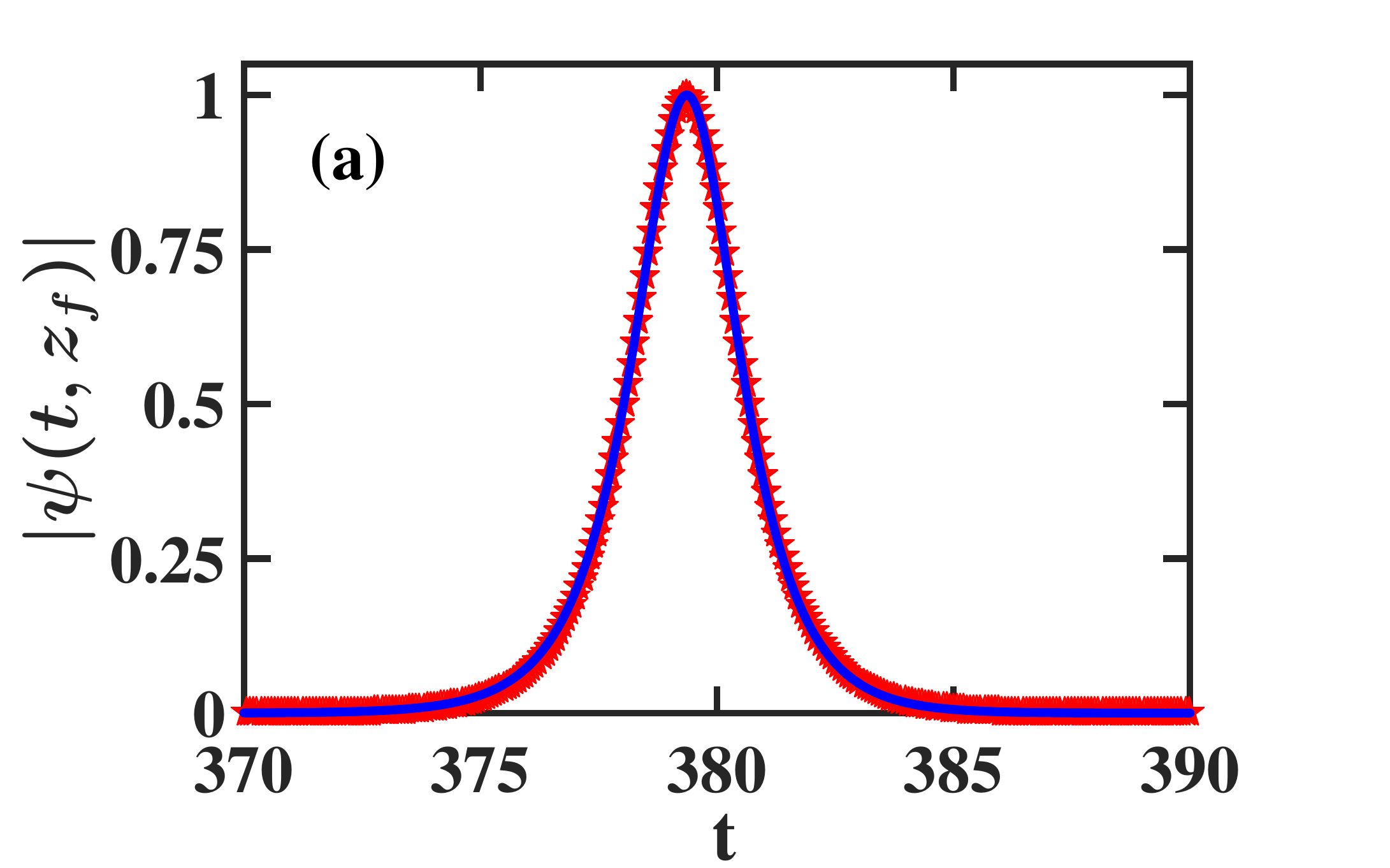} &
\epsfxsize=5.8cm  \epsffile{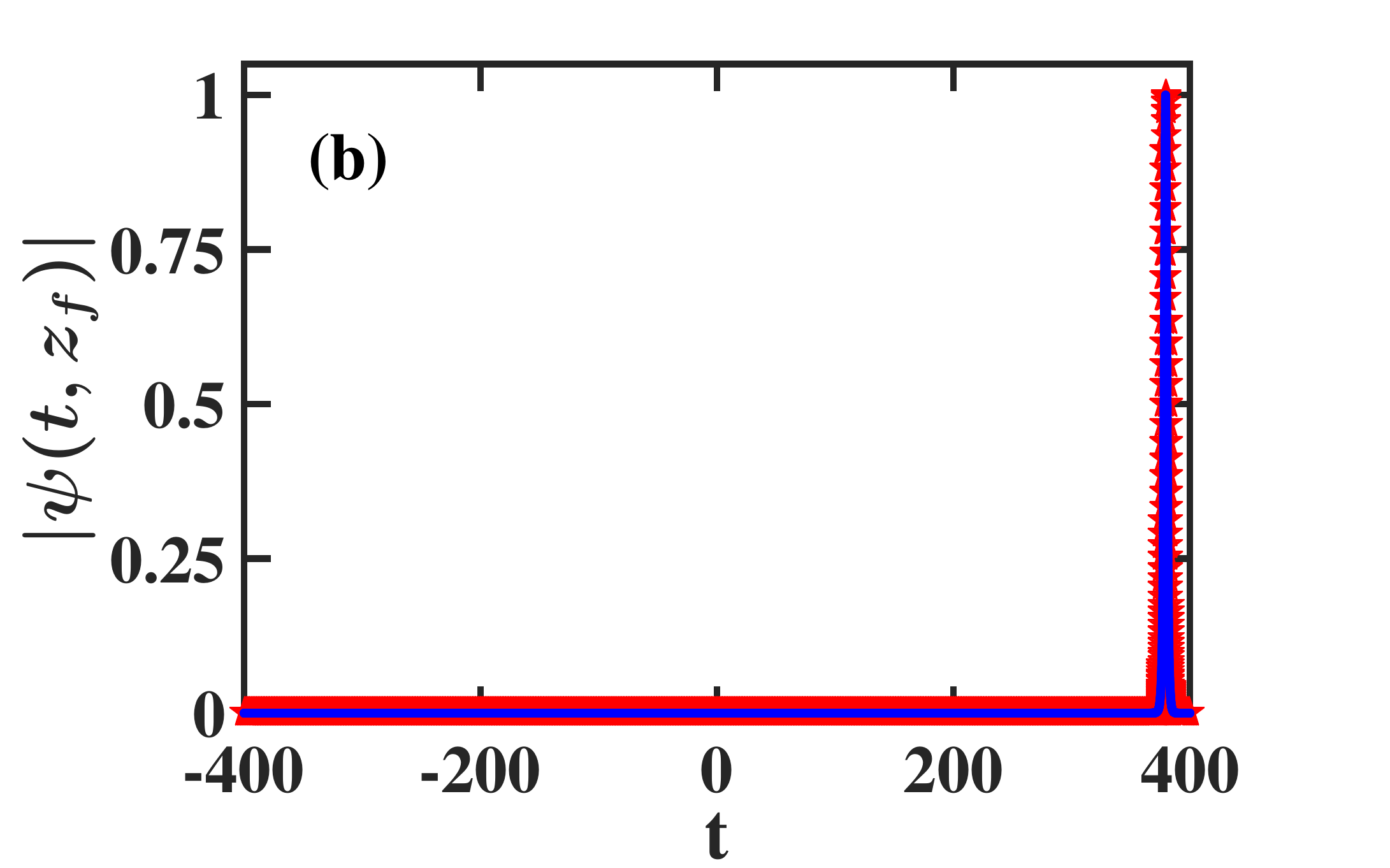} \\
\epsfxsize=5.8cm  \epsffile{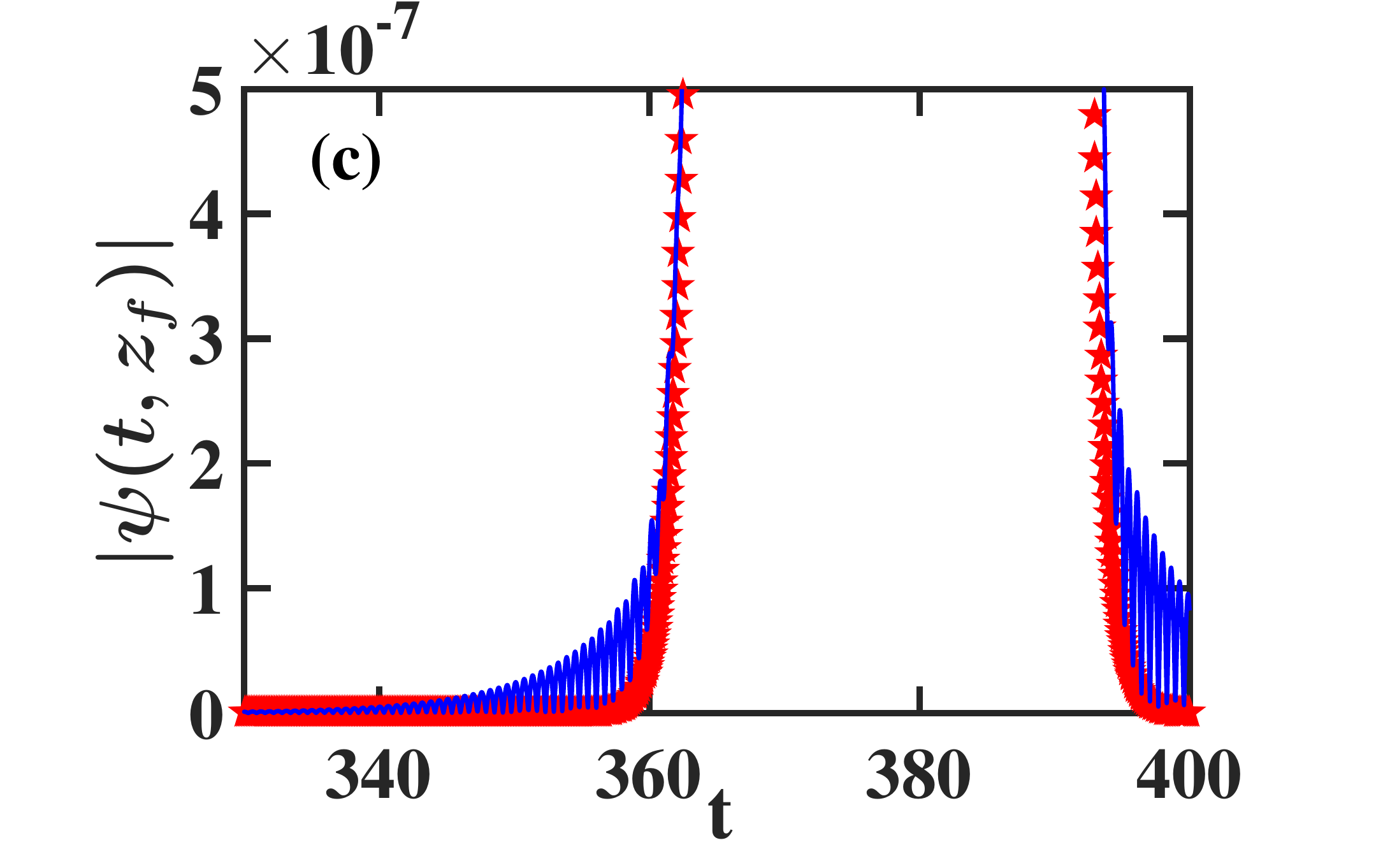} 
\end{tabular}
\caption{The pulse shape $|\psi(t,z_{f})|$,  where $z_{f}=2000$, for soliton propagation 
in a closed optical waveguide loop with weak frequency dependent linear gain-loss, 
cubic loss, and delayed Raman response. 
The cubic loss coefficient is $\epsilon_{3}=0.01$, the Raman coefficient is $\epsilon_{R}=0.04$,  
the initial soliton amplitude is $\eta(0)=0.8$, and the parameters of 
the linear gain-loss  $\hat g(\omega,z)$ 
in Eq. (\ref{sfs32}) are $W=10$, $\rho=10$, and $g_{L}=0.5$.  
The solid blue curve corresponds to the result obtained by numerical 
simulations with Eqs. (\ref{sfs31}) and (\ref{sfs32}).  
The red stars correspond to the prediction of the adiabatic perturbation theory, 
obtained with Eqs. (\ref{Iz1}) and (\ref{sfs16}).}
 \label{fig13}
\end{figure}

\begin{figure}[ptb]
\begin{tabular}{cc}
\epsfxsize=10cm  \epsffile{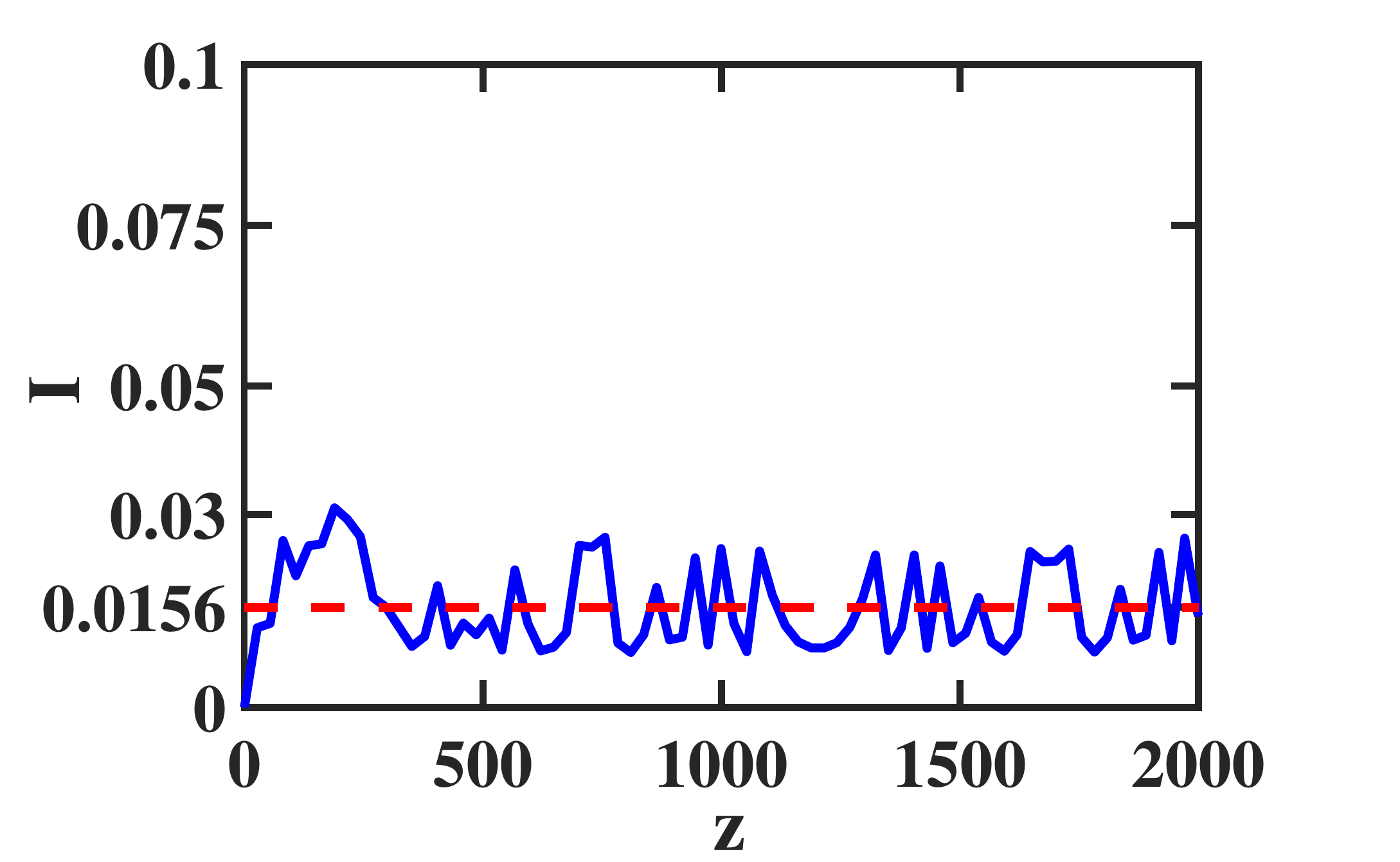} 
\end{tabular}
\caption{The $z$ dependence of the transmission quality integral $I(z)$ obtained 
by numerical simulations with Eqs. (\ref{sfs31}) and (\ref{sfs32}) for 
the same optical waveguide setup considered in Fig. \ref{fig13}.
The solid blue curve represents $I(z)$ and the dashed red horizontal line 
corresponds to $\langle I(z) \rangle$.}
\label{fig14}
\end{figure}

Figure \ref{fig13} shows the pulse shape $|\psi(t,z)|$ at $z=z_{f}$, as obtained in the simulations.  
The prediction of the adiabatic perturbation theory, obtained with Eqs. (\ref{Iz1}) and (\ref{sfs16}), 
is also shown. As seen in Figs. \ref{fig13}(a) and  \ref{fig13}(b),   
the numerically obtained pulse shape at $z=z_{f}$ is very close to the analytic prediction 
and no significant radiative tail is observed. Moreover, as seen in Fig. \ref{fig13}(c), 
the deviation of the numerical result for $|\psi(t,z_{f})|$ from the theoretical one is smaller than 
$10^{-6}$ for all $t$ values. Thus, the interplay between frequency dependent linear gain-loss 
and delayed Raman response does lead to significant enhancement of transmission quality 
compared with the waveguide setups considered in sections \ref{no_shifting} and \ref{Raman_sfs1}. 
The enhancement of transmission quality is also demonstrated in Fig. \ref{fig14}, which shows the 
numerically obtained $I(z)$ curve and the average $\langle I(z) \rangle$, 
which is defined by $\langle I(z) \rangle \equiv \int_{0}^{z_{f}} dz' I(z') /z_{f}$. 
As seen in this figure, the value of $I(z)$ remains smaller than 0.032 
throughout the propagation and $\langle I(z) \rangle=0.0156$.

\begin{figure}[ptb]
\begin{center}
\begin{tabular}{cc}
\epsfxsize=8.0cm  \epsffile{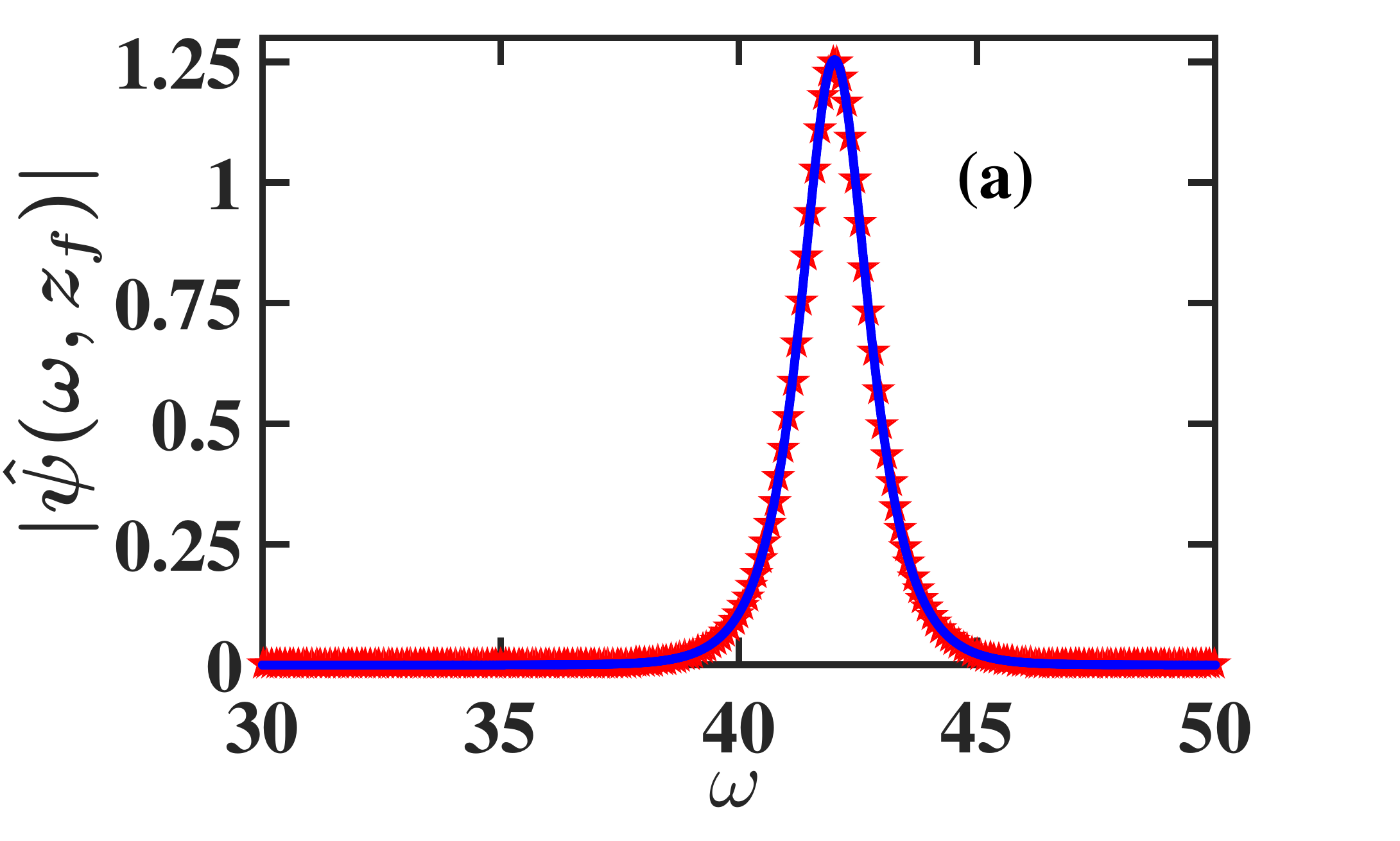} \\
\epsfxsize=8.0cm  \epsffile{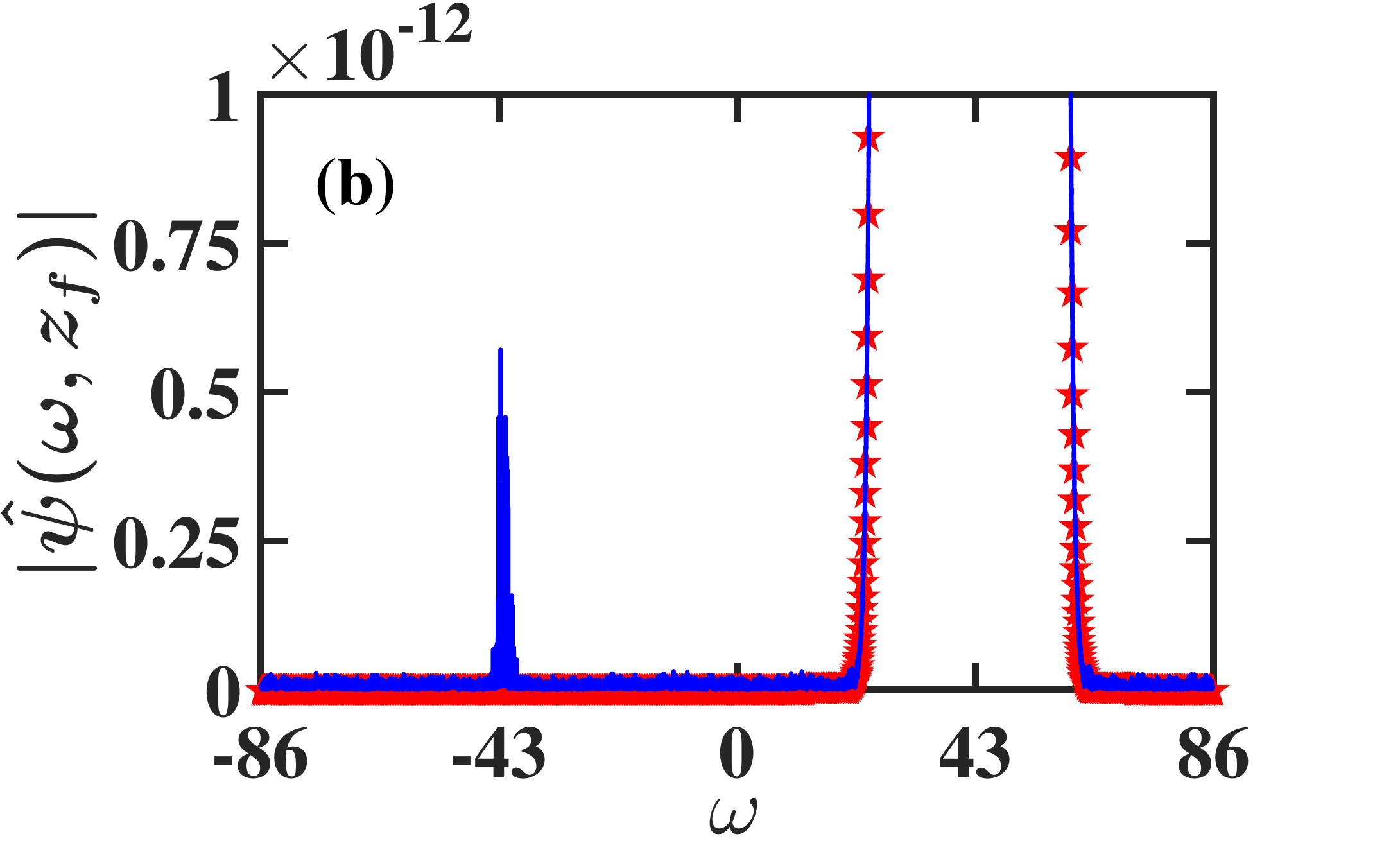} 
\end{tabular}
\end{center}
\caption{The Fourier transform of the pulse shape $|\hat\psi(\omega,z)|$ at $z_{f}=2000$ 
for soliton propagation in a closed optical waveguide loop  
with weak frequency dependent linear gain-loss, cubic loss, and delayed Raman response. 
The physical parameter values are the same as in Fig. \ref{fig13}. 
The solid blue curve represents the result obtained by numerical 
simulations with Eqs. (\ref{sfs31}) and (\ref{sfs32}). The red stars 
correspond to the prediction of the adiabatic perturbation theory, 
obtained with Eqs. (\ref{Iz3}), (\ref{sfs16}), and (\ref{sfs22}).}                        
 \label{fig15}
\end{figure}

The enhanced transmission quality can be explained  with the 
help of the Fourier transform of the pulse $|\hat\psi(\omega,z)|$.  
Figure \ref{fig15} shows the numerically obtained Fourier transform $|\hat\psi(\omega,z)|$ 
at $z=z_{f}$ together with the prediction of the adiabatic perturbation theory, 
obtained with Eqs. (\ref{Iz3}), (\ref{sfs16}), and (\ref{sfs22}). 
We observe very good agreement between the two results. 
More specifically, in both curves, the Fourier spectrum of the soliton 
is strongly downshifted and is centered about the frequency $\omega_{m}=-\beta(z_{f})=42.0$. 
Additionally, the numerically obtained curve of $|\hat\psi(\omega,z_{f})|$ 
does not contain any fast oscillations in the main peak such as the oscillations seen 
in Figs. \ref{fig3} and \ref{fig7} (in section \ref{no_shifting}) 
for soliton propagation in the absence of delayed Raman response. 
Furthermore, the numerically obtained curve of $|\hat\psi(\omega,z_{f})|$
does not contain any significant ``radiation peaks'' such as the one seen in Fig. \ref{fig11} 
(in section \ref{Raman_sfs1}) for waveguides with frequency independent linear gain, 
cubic loss, and delayed Raman response.     
Based on these observations we conclude that the presence of delayed Raman response 
leads to separation of the soliton's spectrum from the radiation's spectrum via the 
soliton self-frequency shift, while the frequency dependent linear gain-loss leads to 
efficient suppression of radiation emission. As a result, transmission quality is significantly enhanced 
in waveguides with frequency dependent linear gain-loss, cubic loss, and delayed Raman response 
compared with the waveguide setups considered in sections \ref{no_shifting} and \ref{Raman_sfs1}.

\begin{figure}[ptb]
\begin{tabular}{cc}
\epsfxsize=8cm  \epsffile{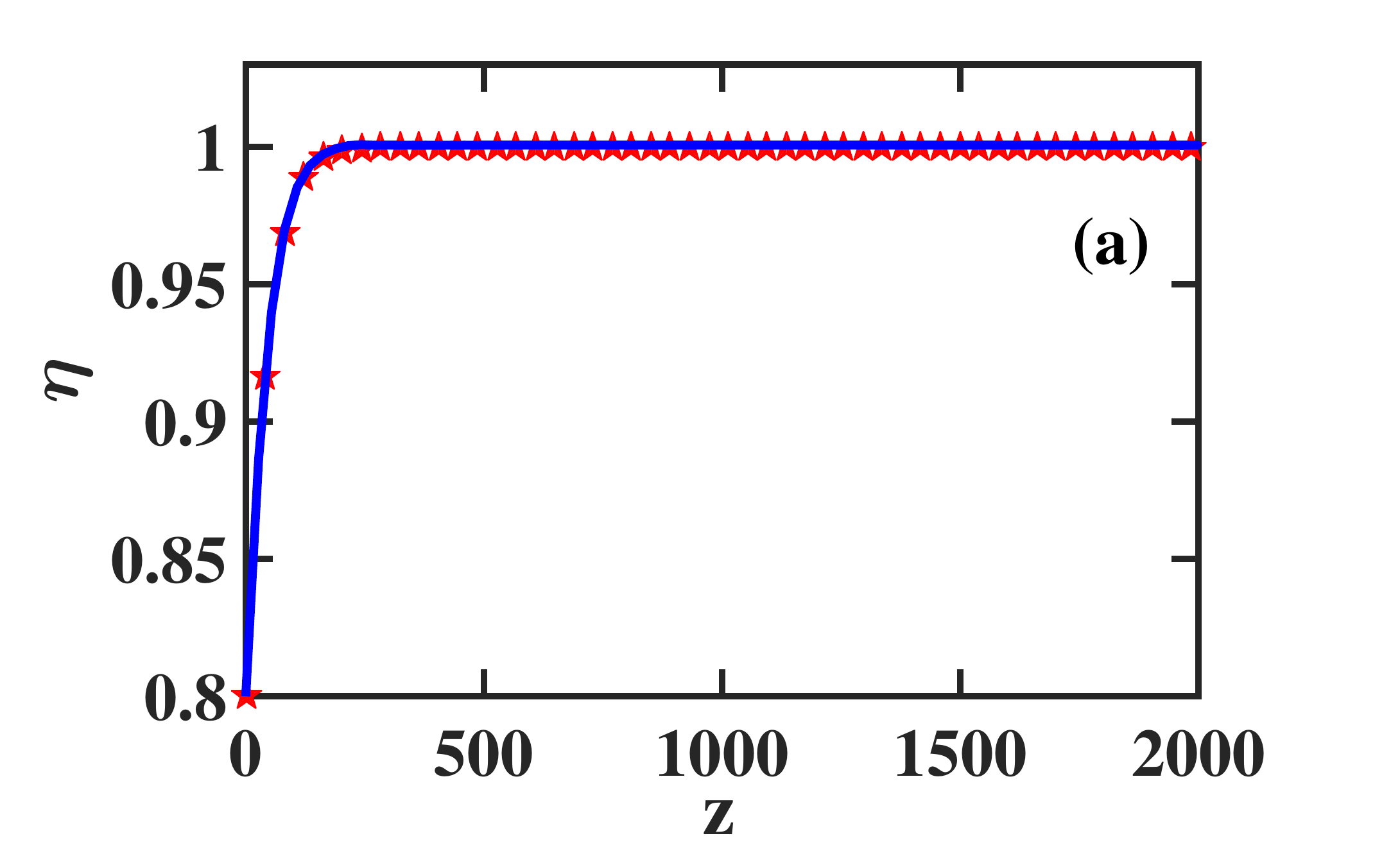} \\
\epsfxsize=8cm  \epsffile{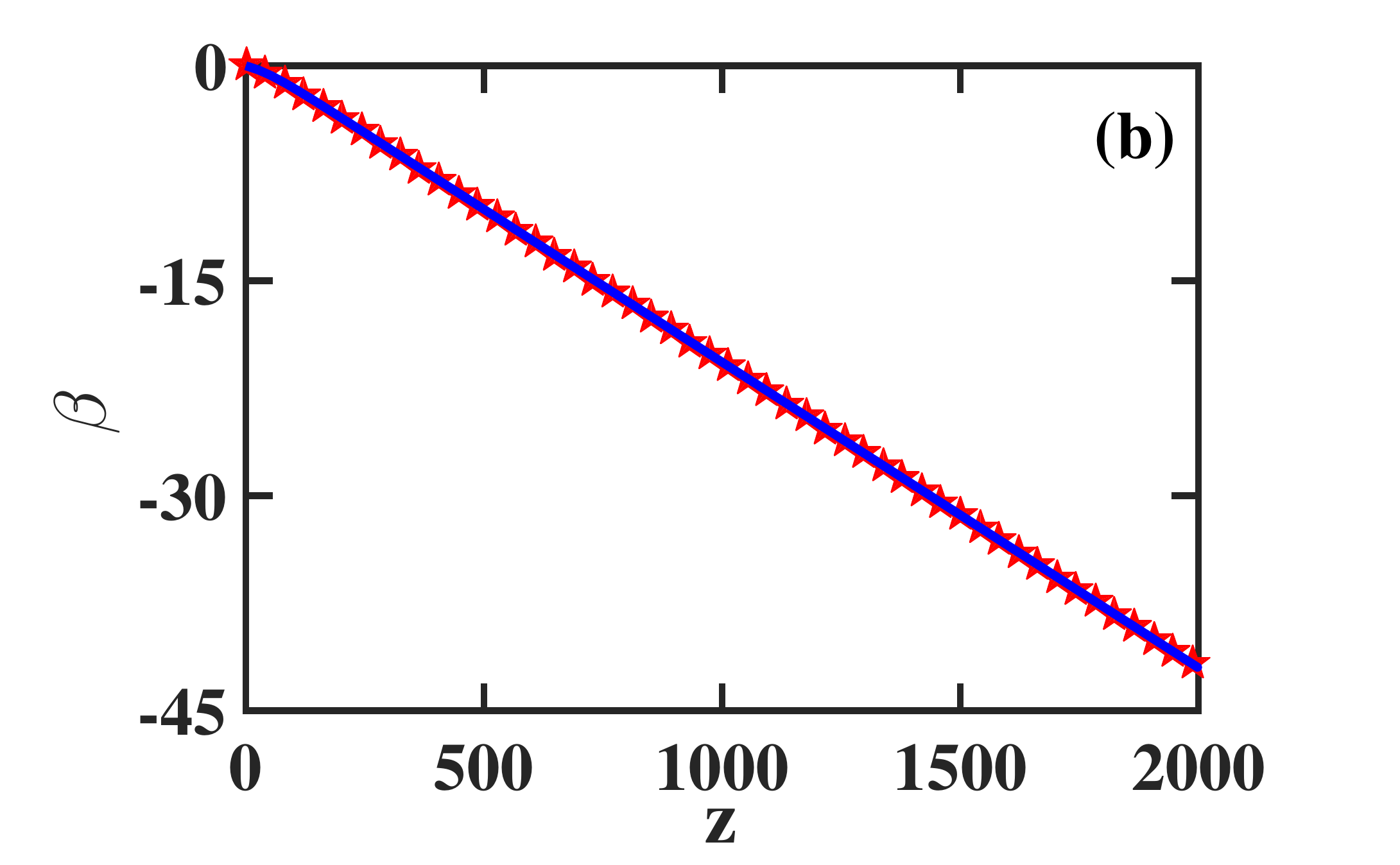}
\end{tabular}
\caption{The $z$ dependence of the soliton amplitude $\eta(z)$ (a) and frequency $\beta(z)$ (b)  
for the closed optical waveguide loop setup considered in Figs. \ref{fig13}-\ref{fig15}. 
The solid blue curves represent the results obtained 
by numerical simulations with Eqs. (\ref{sfs31}) and (\ref{sfs32}).
The red stars correspond to the predictions of the adiabatic perturbation theory, 
obtained with Eq. (\ref{sfs16}) in (a) and with Eqs. (\ref{sfs22}) and (\ref{sfs16}) in (b).}
\label{fig16}
\end{figure}

The enhancement of transmission quality in waveguides with frequency dependent 
linear gain-loss, cubic loss, and delayed Raman response is also manifested in the dynamics of 
the soliton's amplitude and frequency. Figures \ref{fig16}(a) and \ref{fig16}(b) show 
the $z$ dependence of the soliton's amplitude and frequency obtained in the simulations. 
Also shown are the predictions of the adiabatic perturbation theory, obtained with 
Eqs. (\ref{sfs16}) and  (\ref{sfs22}). We observe that the numerically obtained soliton amplitude tends to 
the equilibrium value $\eta_{0}=1$ at short distances and stays close to this value throughout 
the propagation, in excellent agreement with the perturbation theory prediction. 
Furthermore, the value of the soliton frequency obtained in the simulations remains close 
to the $z$ dependent value predicted by the adiabatic perturbation theory throughout the propagation. 
Thus, the efficient suppression of radiation emission in waveguides with frequency dependent linear 
gain-loss, cubic loss, and delayed Raman response enables observation of stable amplitude and frequency dynamics 
along significantly larger distances compared with the distances obtained with  
the closed optical waveguide loop setup considered in section \ref{Raman_sfs1}.  
We also point out that the waveguide setups considered in the current subsection can be used for 
inducing large frequency shifts, which are not accompanied by pulse distortion, 
in soliton-based optical waveguide transmission systems.

\section{Pulse dynamics in waveguides with linear gain, cubic loss, and guiding filters}
\label{filters}
{\it Introduction.} 
The enhancement of transmission quality and stability in waveguides 
with frequency dependent linear gain-loss and delayed Raman response, 
which was demonstrated in subsection \ref{Raman_sfs2}, is somewhat 
similar to transmission stability enhancement in waveguides 
with linear gain and guiding filters with a varying central frequency. 
Indeed, in the latter waveguides, the guiding filters play a role similar to that of the 
frequency dependent linear gain-loss, i.e., their presence leads to suppression of 
radiation emission with frequencies that are significantly different from the soliton's frequency. 
In addition, the variation of the central frequency of the guiding filters with propagation distance 
plays a role similar to that of the Raman self-frequency shift, that is, 
it leads to the separation of the soliton's Fourier spectrum from the radiation's Fourier spectrum. 
For this reason it is useful to compare the dynamics of optical solitons in the two waveguide systems. 
We therefore turn to study soliton propagation in optical waveguide loops 
with frequency independent linear gain, cubic loss, and optical guiding filters. 
We start by considering guiding filters with a constant central frequency in 
subsection \ref{filters1}, and treat the case of guiding filters with a varying 
central frequency in subsection \ref{filters2}. 
We point out that stabilization of soliton-based transmission in optical fibers   
by guiding filters with a varying central frequency was theoretically and experimentally
demonstrated in Refs. \cite{Mollenauer92,Mollenauer97,MM98,MMN96,Mollenauer2006}. 
Since these studies focused on optical fiber transmission, 
the effects of cubic loss were neglected. 
In the current section, we extend the theoretical treatment of Refs. \cite{Mollenauer92,Mollenauer97,Mollenauer2006}   
and take into account the effects of cubic loss in addition to the effects of linear gain 
and guiding filters.

\subsection{Waveguides with linear gain, cubic loss, and guiding filters with a constant central frequency}
\label{filters1}

We consider propagation of pulses of light in nonlinear optical waveguides 
in the presence of weak linear gain, weak cubic loss, and guiding optical filters.  
Following the treatment in Refs. \cite{Mollenauer92,Mollenauer97,Mollenauer2006},  
we assume that the response function of the optical filter can be approximated 
by a Gaussian with a maximum that is equal to 1 and that is located at the frequency 
$\omega_{p}$. Under this assumption, the propagation is described by the following 
perturbed NLS equation \cite{Mollenauer92,Mollenauer97,Mollenauer2006}: 
\begin{eqnarray}&&
i\partial_z\psi+\partial_t^2\psi+2|\psi|^2\psi=
ig_{0}\psi/2 - i\epsilon_{3}|\psi|^2\psi
- i\epsilon_{\omega}\left(i\partial_{t} - \omega_{p}\right)^{2}\psi,
\label{sfs41}
\end{eqnarray}  
where $\epsilon_{\omega}>0$ is the second-order filtering coefficient, 
and $\omega_{p}$ is assumed to be constant in the current subsection \cite{dimensions3}.    
Equation (\ref{sfs41}) can also be written as  
\begin{eqnarray}
i\partial_z\psi+\partial_t^2\psi+2|\psi|^2\psi=
i\left(g_{0}/2-\epsilon_{\omega}\omega_{p}^{2}\right)\psi 
- i\epsilon_{3}|\psi|^2\psi 
- 2\epsilon_{\omega}\omega_{p}\partial_{t}\psi 
+ i\epsilon_{\omega}\partial_{t}^{2}\psi.
\nonumber \\
\label{sfs42}
\end{eqnarray}

Using the adiabatic perturbation theory for the NLS soliton, 
we find that the dynamics of the soliton's amplitude and frequency is given by:   
\begin{eqnarray}&&
\frac{d\eta}{dz} = 
\eta\left\lbrace g_{0} - 2\epsilon_{\omega}\left[\eta^{2}/3 + 
\left(\beta - \omega_{p}\right)^{2}\right] - 4\epsilon_{3}\eta^{2}/3\right\rbrace ,
\label{sfs43}
\end{eqnarray}
and
\begin{eqnarray}&&
\nonumber \\
\frac{d\beta}{dz} = -4\epsilon_{\omega}\left(\beta - \omega_{p}\right)\eta^{2}/3 .
\label{sfs44}
\end{eqnarray}
In the current subsection, we try to realize stable transmission with constant amplitude $\eta=\eta_{0}>0$
and frequency $\beta=\beta_{0}$. We therefore require that $(\eta_{0},\beta_{0})$ is an equilibrium 
point of Eqs. (\ref{sfs43}) and (\ref{sfs44}). We obtain: 
$g_{0} = 2\epsilon_{\omega}\eta_{0}^{2}/3 + 4\epsilon_{3}\eta_{0}^2/3$ 
and $\beta_{0} = \omega_{p}$. As a result, Eq. (\ref{sfs43}) takes the form 
\begin{eqnarray}&&
\frac{d\eta}{dz} = 
2\eta\left[2\epsilon_{3}\left(\eta_{0}^{2}-\eta^{2}\right)\!/3
+ \epsilon_{\omega}\left(\eta_{0}^{2} - \eta^{2}\right)\!/3
- \epsilon_{\omega}\left(\beta - \omega_{p}\right)^{2}\right] .
\label{sfs45}
\end{eqnarray}
Thus, dynamics of the soliton's amplitude and frequency is described by Eqs. (\ref{sfs44}) and (\ref{sfs45}). 
Linear stability analysis shows that $(\eta_{0},\omega_{p})$ is a stable node of  the system (\ref{sfs44})-(\ref{sfs45}).  
In addition to the equilibrium point at $(\eta_{0},\omega_{p})$ there is a line of equilibrium points at $(0,\beta)$.
These additional equilibrium points are asymptotically stable for $\beta > \omega_{p} + r_{p}\eta_{0}$ or 
$\beta < \omega_{p} - r_{p}\eta_{0}$ and are unstable for 
$\omega_{p} - r_{p}\eta_{0} < \beta < \omega_{p} + r_{p}\eta_{0}$, 
where $r_{p}=[(2\epsilon_{3}+\epsilon_{\omega})/(3\epsilon_{\omega})]^{1/2}$. 
Note that similar stability conditions hold for small amplitude wave solutions 
of the form $\psi_{l}(t,z)=\bar C \exp(-ikz+i\omega t)$ of the propagation model 
\begin{eqnarray}&&
i\partial_z\psi+\partial_t^2\psi= ig_{0}\psi/2 
- i\epsilon_{\omega}\left(i\partial_{t} - \omega_{p}\right)^{2}\psi,
\label{sfs46}
\end{eqnarray}  
which is the linear part of Eq. (\ref{sfs41}). 
Indeed, substitution of $\psi_{l}(t,z)$ into Eq. (\ref{sfs46}) yields  
\begin{equation}
k(\omega) = \omega^{2} + i \left[g_{0}/2 - \epsilon_{\omega}(\omega+\omega_{p})^{2}\right].  
\label{sfs46_A}
\end{equation}   
As a result, the small amplitude wave solutions $\psi_{l}(t,z)$ are stable for  
\begin{equation}
\omega < - \omega_{p} - r_{p}\eta_{0}  \;\;\; \mbox{or} \;\;\;    \omega > - \omega_{p} + r_{p}\eta_{0},  
\label{sfs47}
\end{equation}   
and are unstable for 
\begin{equation}
- \omega_{p} - r_{p}\eta_{0} < \omega < - \omega_{p} + r_{p}\eta_{0}  .  
\label{sfs48}
\end{equation}    
Furthermore, Eq. (\ref{sfs46_A}) also indicates that suppression of radiation emission 
by the guiding filters is more efficient at frequencies that are far from the equilibrium value 
of the soliton's frequency $\omega_{p}$ (see also Refs. \cite{Mollenauer92,Mollenauer97,Mollenauer2006}).

{\it Numerical simulations.} 
Equation  (\ref{sfs42}) is numerically solved on a domain $[t_{\mbox{min}},t_{\mbox{max}}]=[-400,400]$ 
with periodic boundary conditions. The initial condition is in the form of a single NLS soliton with 
amplitude $\eta(0)$, frequency $\beta(0)$, position $y(0)=0$, and phase $\alpha(0)=0$. 
As a typical example, we present here the results of the simulations with 
$\epsilon_{3}=0.01$, $\epsilon_{\omega}=0.04$, $\omega_{p}=42.7$, 
$\eta(0)=0.8$, and $\beta(0)=42.5$. This choice of the physical parameter values  
enables comparison with results of numerical simulations in previous sections 
and in section \ref{filters2}. We point out that similar results are obtained 
for other physical parameters values.  Due to the presence of the guiding filters 
and due to the initial nonzero soliton frequency, the soliton experiences a very large 
position shift during the propagation. As a result, the soliton passes through 
the computational domain's boundaries multiple times during the simulation. 
To avoid soliton destruction, we do not employ damping at the boundaries. 
Thus, the simulations describe soliton propagation in a closed optical waveguide loop. 
The values of the transmission quality distance and the final propagation distance  
obtained in the simulations are $z_{q}=96$ and $z_{f}=208$.

\begin{figure}[ptb]
\begin{tabular}{cc}
\epsfxsize=5.8cm  \epsffile{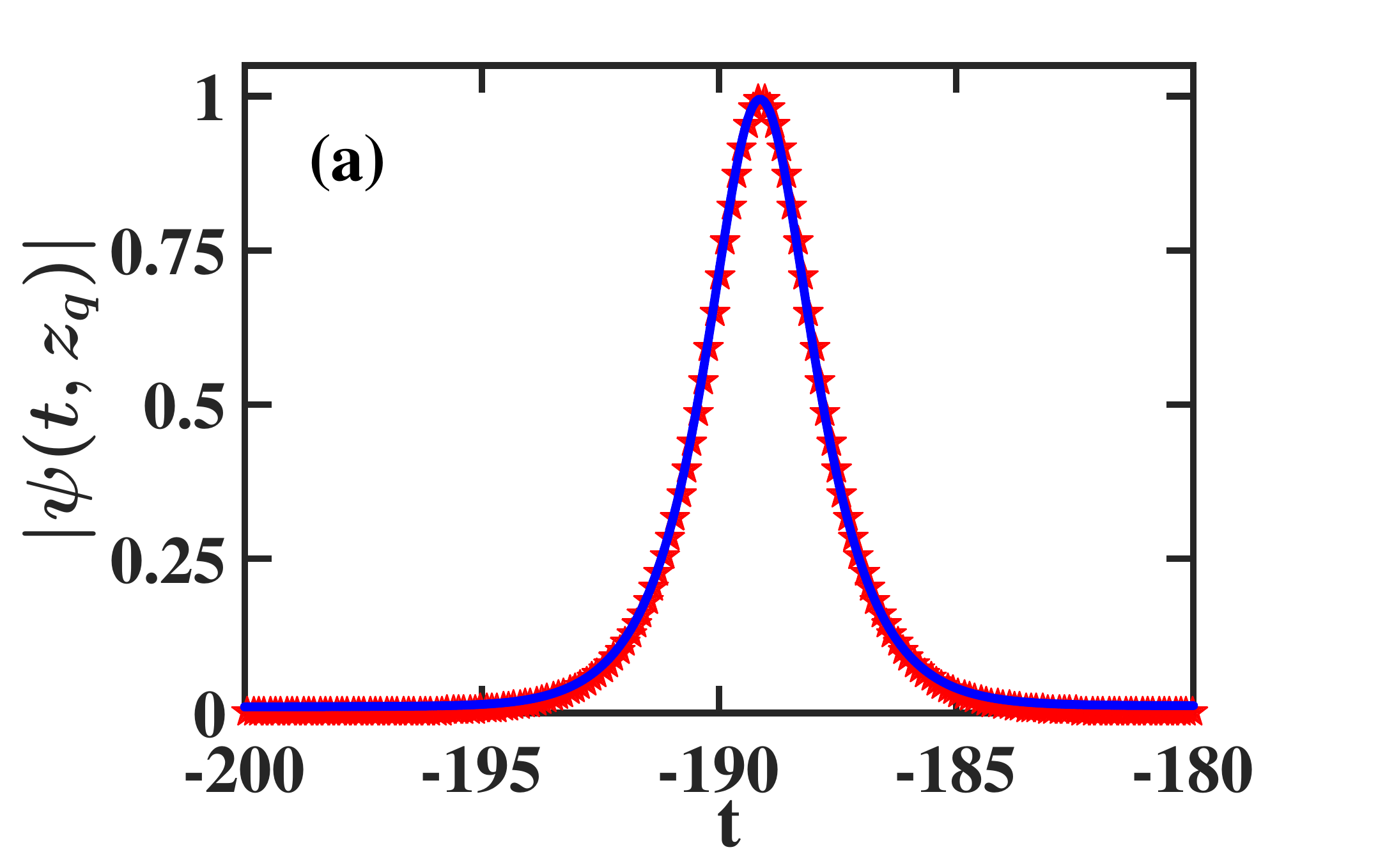} &
\epsfxsize=5.8cm  \epsffile{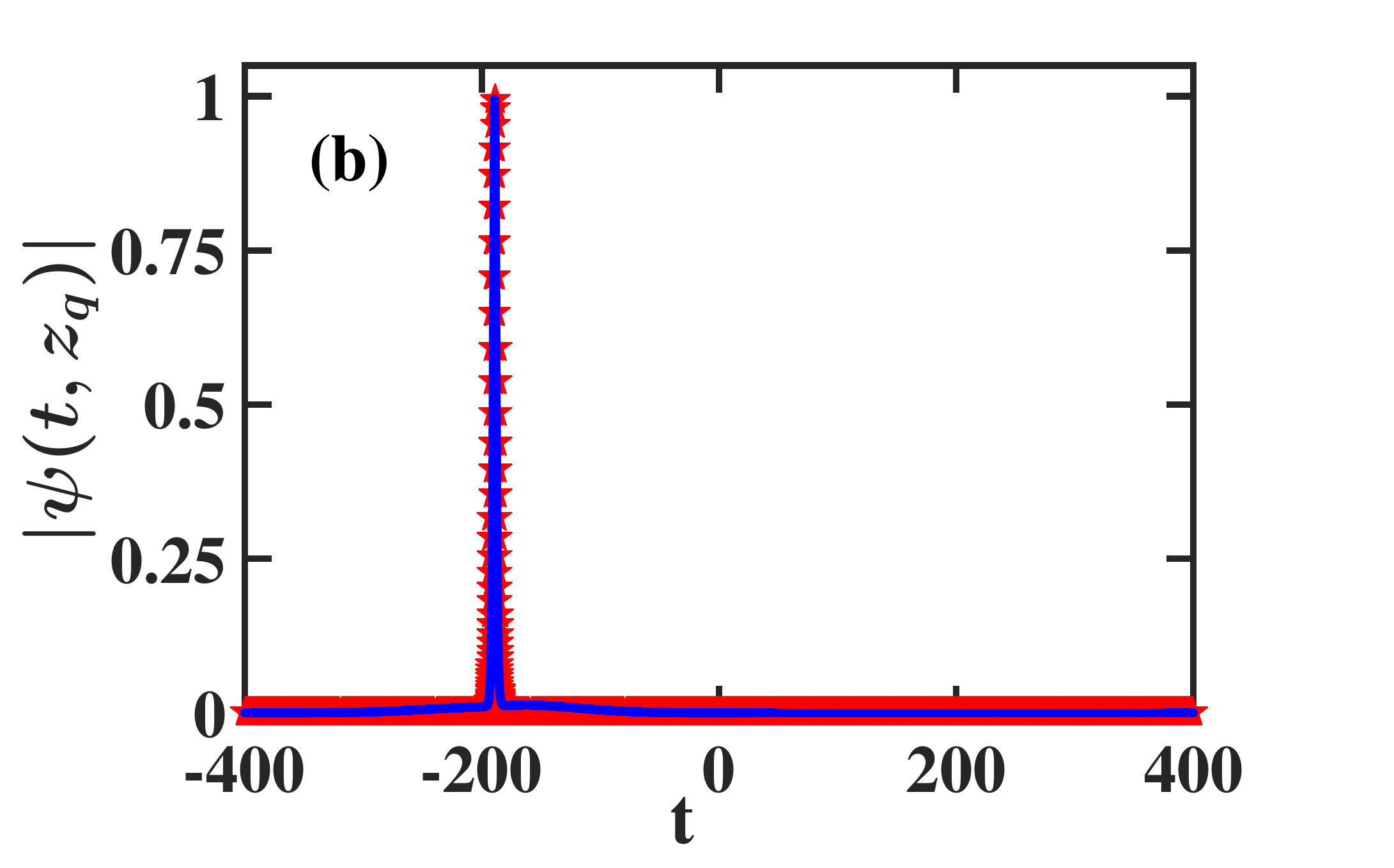} \\
\epsfxsize=5.8cm  \epsffile{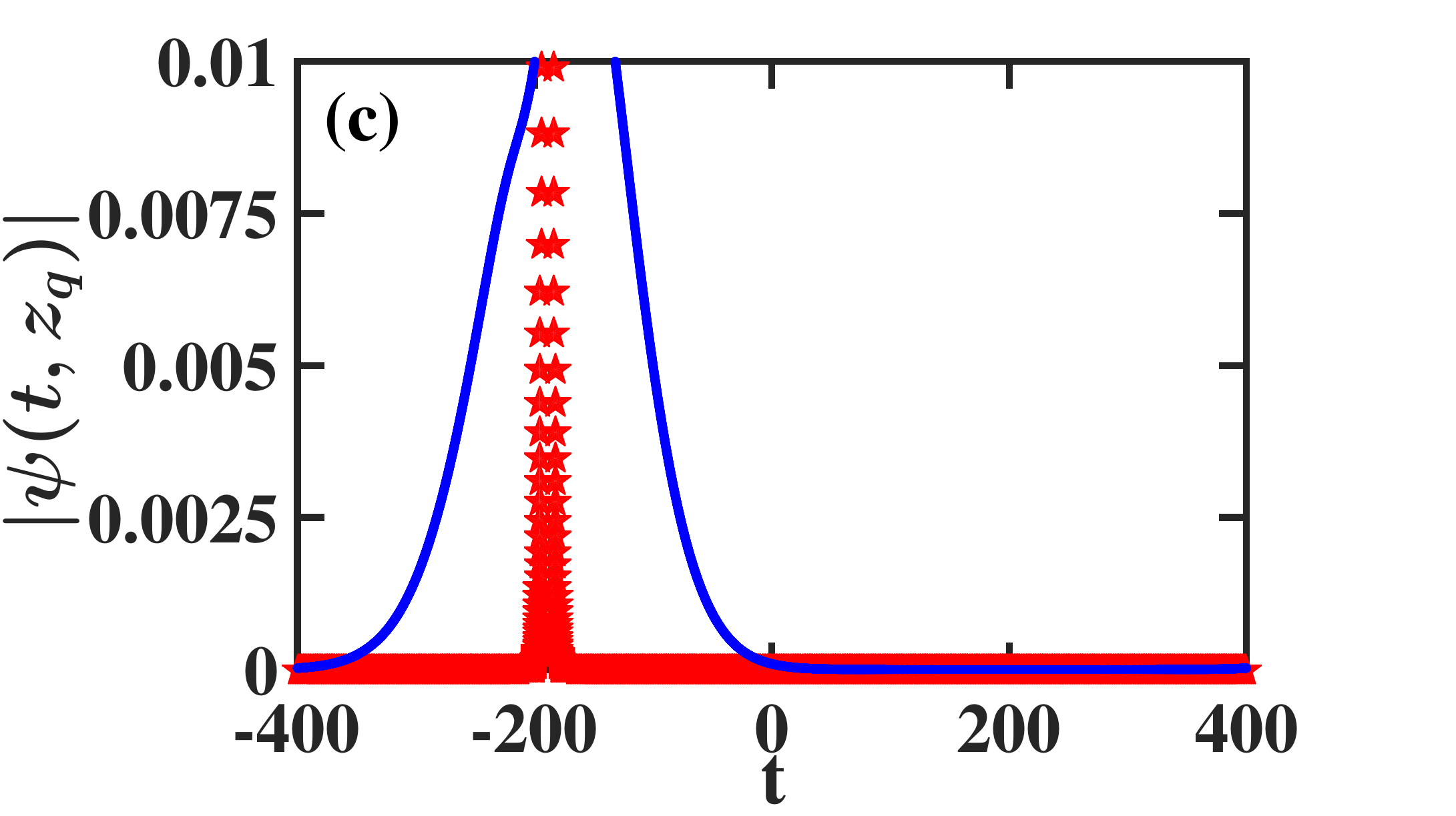} &
\epsfxsize=5.8cm  \epsffile{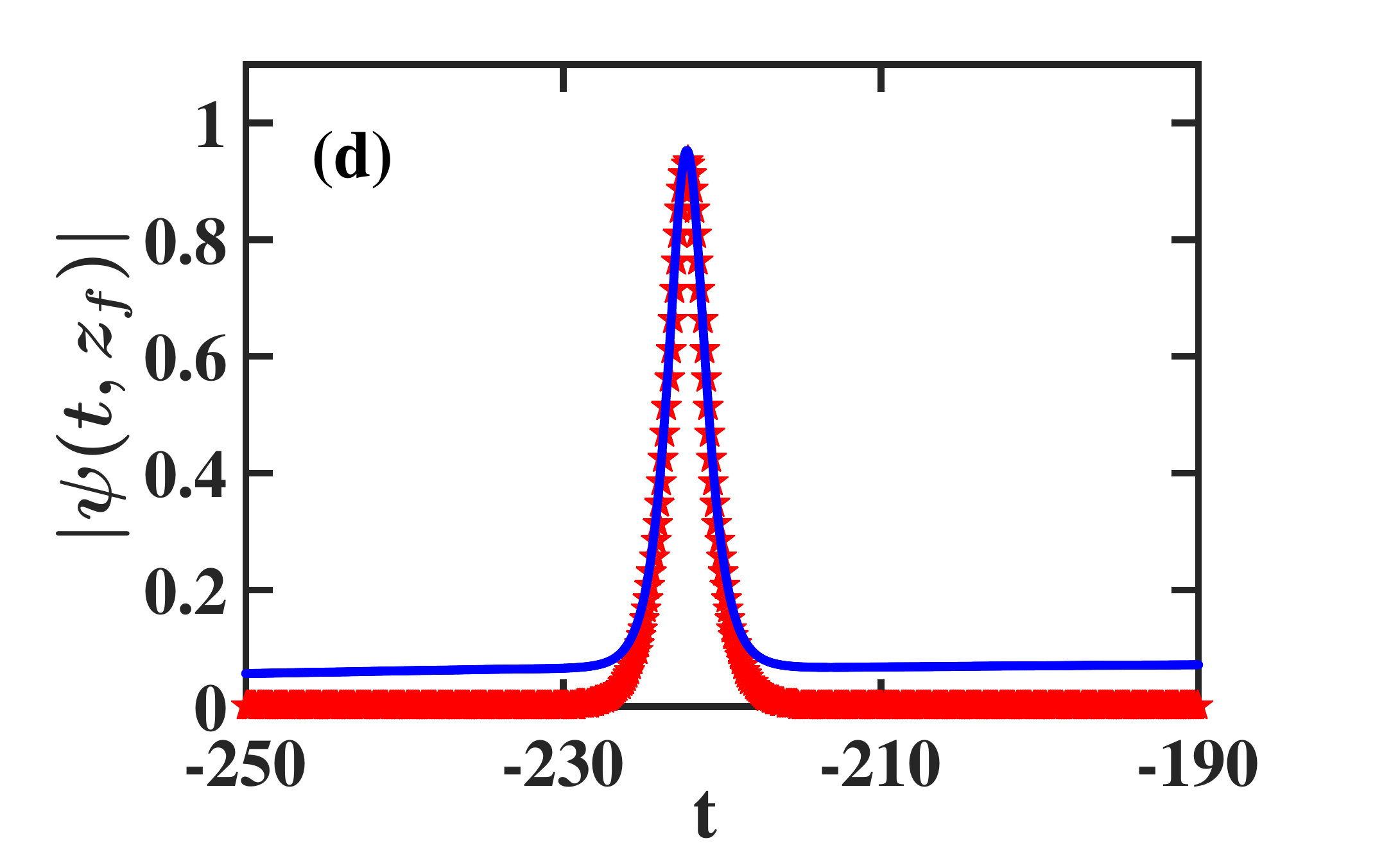} \\ 
\epsfxsize=5.8cm  \epsffile{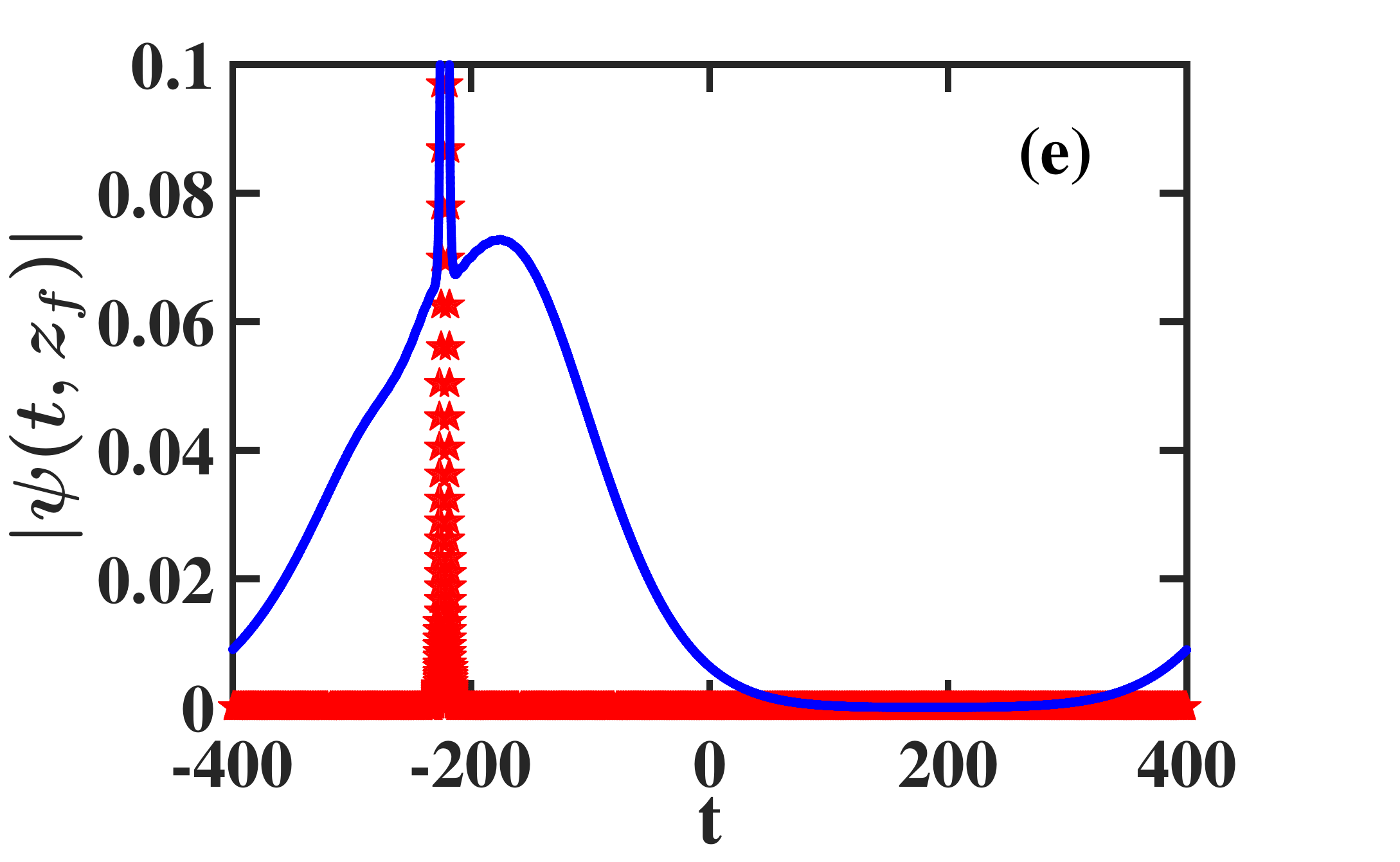}
\end{tabular}
\caption{The pulse shape $|\psi(t,z)|$ at $z_{q}=96$ [(a), (b), and (c)]  
and at $z_{f}=208$ [(d) and (e)] for soliton propagation in a closed optical waveguide loop  
with weak frequency independent linear gain, cubic loss, 
and guiding filters with a constant central frequency. 
The physical parameter values are $\epsilon_{3}=0.01$, $\epsilon_{\omega}=0.04$, 
$\omega_{p}=42.7$, $\eta(0)=0.8$, and $\beta(0)=42.5$.
The solid blue curve represents the result obtained by numerical 
simulations with Eq.  (\ref{sfs42}), while the red stars correspond to 
the perturbation theory prediction of Eqs. (\ref{Iz1}), (\ref{sfs44}),  and  (\ref{sfs45}).}
 \label{fig17}
\end{figure}

\begin{figure}[ptb]
\begin{tabular}{cc}
\epsfxsize=10cm  \epsffile{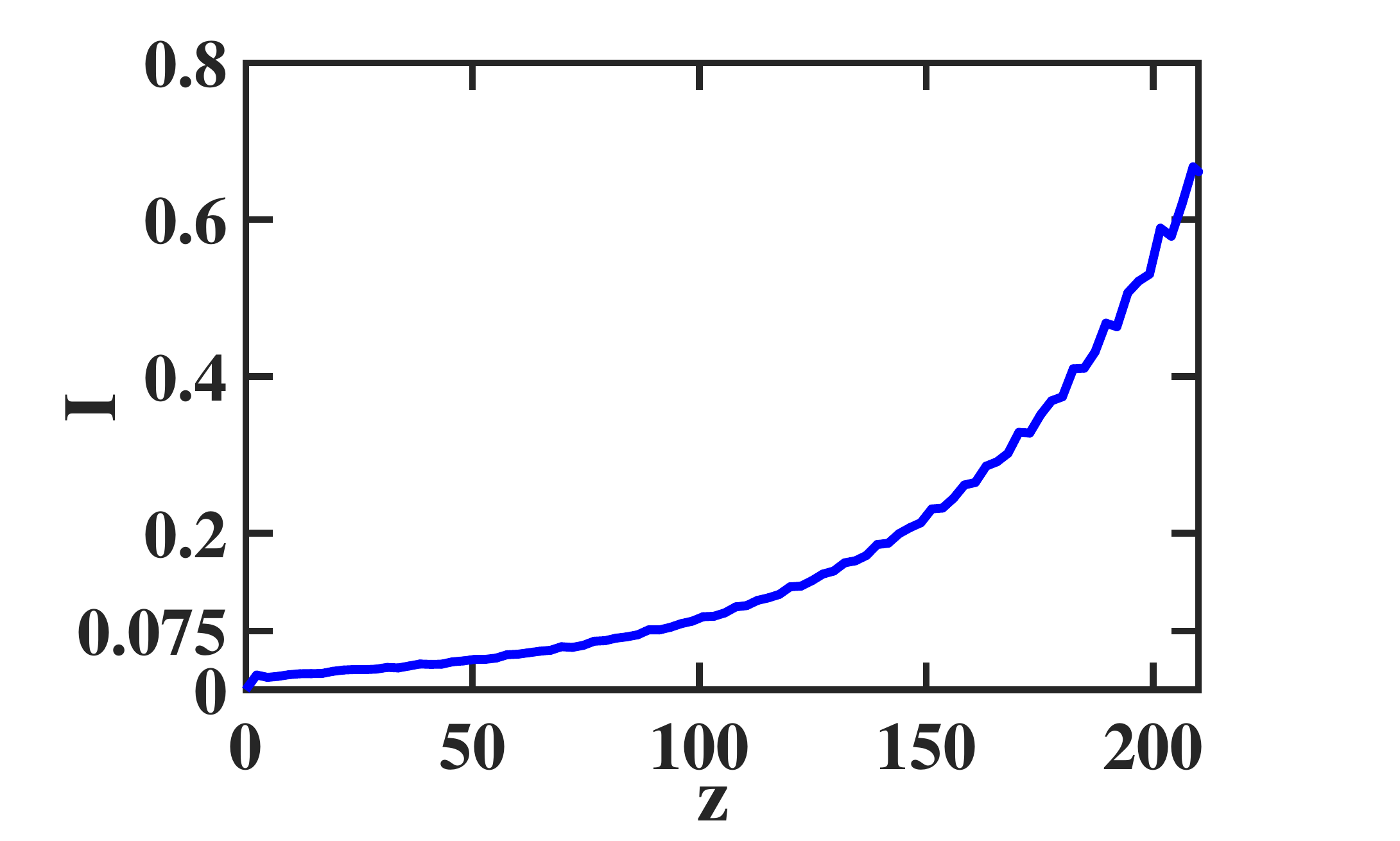} 
\end{tabular}
\caption{The $z$ dependence of the transmission quality integral $I(z)$ obtained 
by numerical simulations with Eq. (\ref{sfs42}) for the same optical waveguide 
setup considered in Fig. \ref{fig17}.}
\label{fig18}
\end{figure}

Figure \ref{fig17} shows the pulse shape $|\psi(t,z)|$ at $z=z_{q}$ and at $z=z_{f}$, 
obtained in the simulations. Also shown is a comparison with the 
prediction of the adiabatic perturbation theory, obtained with Eqs. (\ref{Iz1}), (\ref{sfs44}), and  (\ref{sfs45}). 
As seen in Figs. \ref{fig17}(a) and  \ref{fig17}(b), the pulse shape obtained in the simulations 
at $z=z_{q}$ is close to the analytic prediction. However, the comparison of the analytic 
prediction with the numerical result for small $|\psi(t,z_{q})|$ values in Fig. \ref{fig17}(c) shows 
that an appreciable radiative tail exists at $z=z_{q}$. 
Additionally, as seen in Figs. \ref{fig17}(d) and \ref{fig17}(e), 
the radiative tail continues to grow as the soliton continues to propagate along the waveguide. 
As a result, the value of the transmission quality integral $I(z)$
increases from 0.075 at $z_{q}=96$ to 0.6674 at $z_{f}=208$ [see Fig. \ref{fig18}]. 
We note that the radiative tail observed for the current optical waveguide setup 
is much larger than the radiative tail observed  in section \ref{no_shifting} 
for waveguides with linear gain or loss and cubic loss and with no guiding filters 
[compare Fig. \ref{fig17}(c) with Figs. \ref{fig1}(c) and \ref{fig5}(c)].  
In addition, the $z_{q}$ and $z_{f}$ values for the current waveguide setup are considerably smaller 
compared with the $z_{q}$ and $z_{f}$ values obtained with the waveguide setups of section \ref{no_shifting}. 
Based on these findings we deduce that transmission quality in waveguide loops 
with weak frequency independent linear gain, cubic loss, 
and guiding filers with a constant central frequency is significantly reduced compared 
with the waveguide setups considered in section \ref{no_shifting}.

\begin{figure}[ptb]
\begin{center}
\begin{tabular}{cc}
\epsfxsize=5.8cm  \epsffile{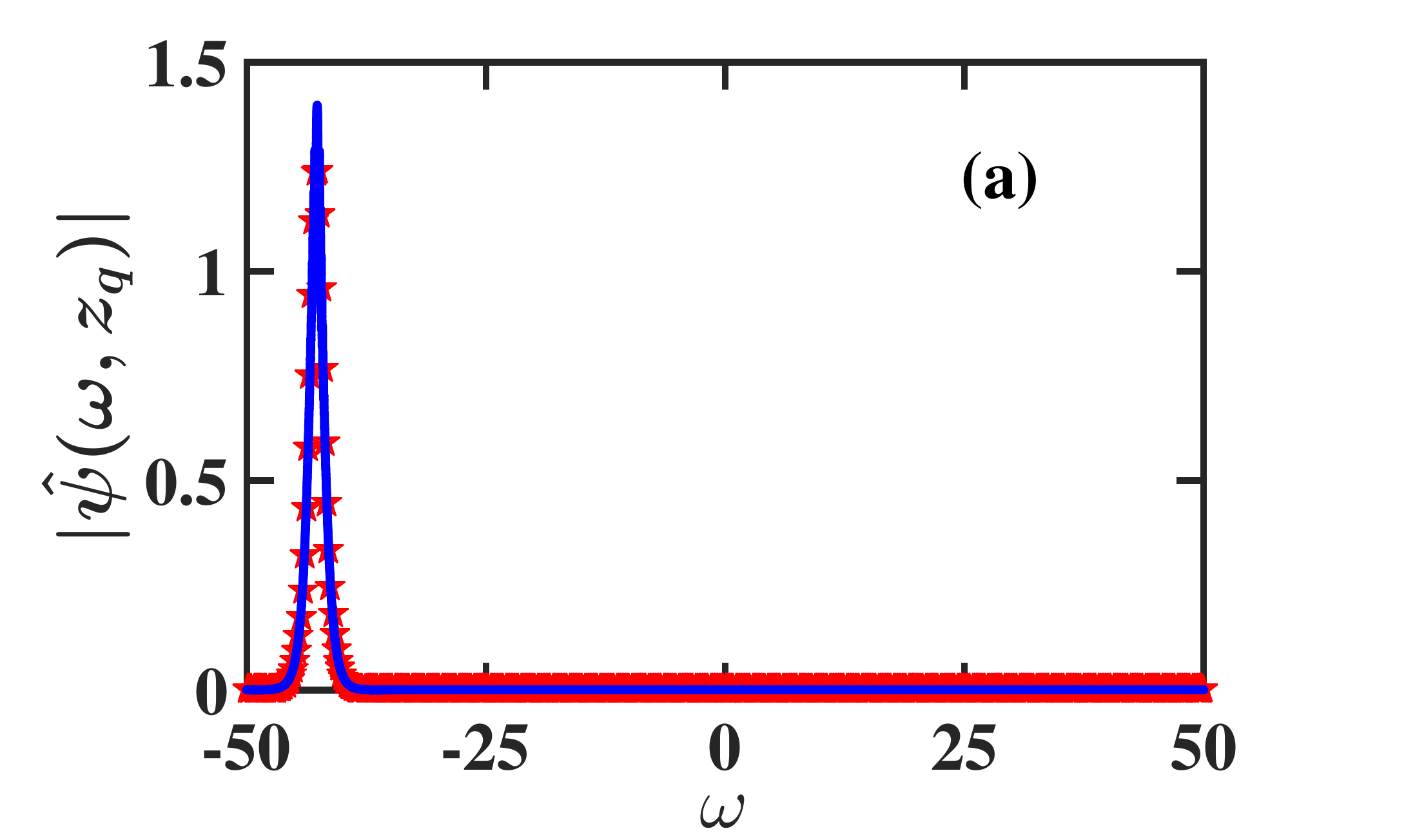} &
\epsfxsize=5.8cm  \epsffile{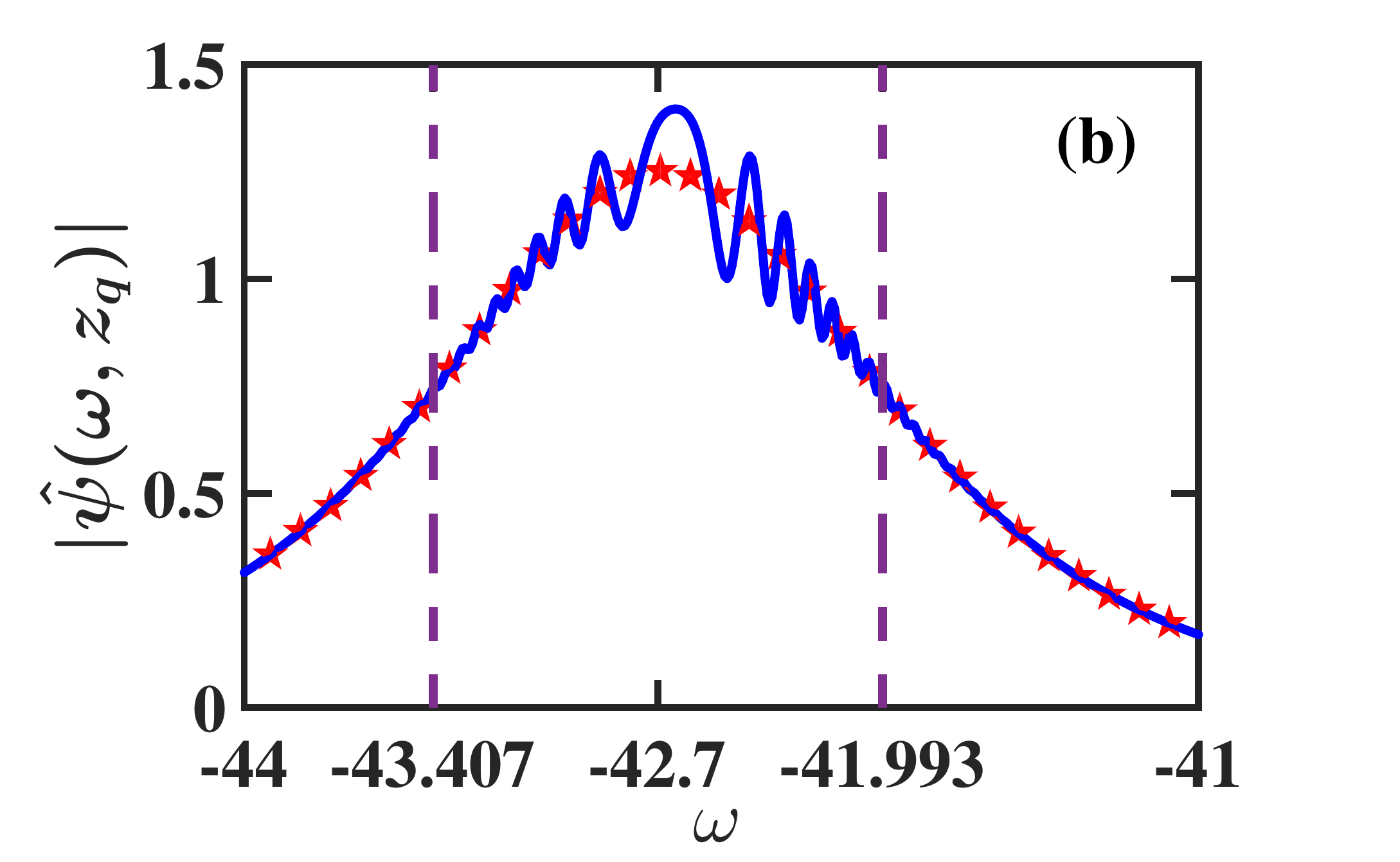} \\
\epsfxsize=5.8cm  \epsffile{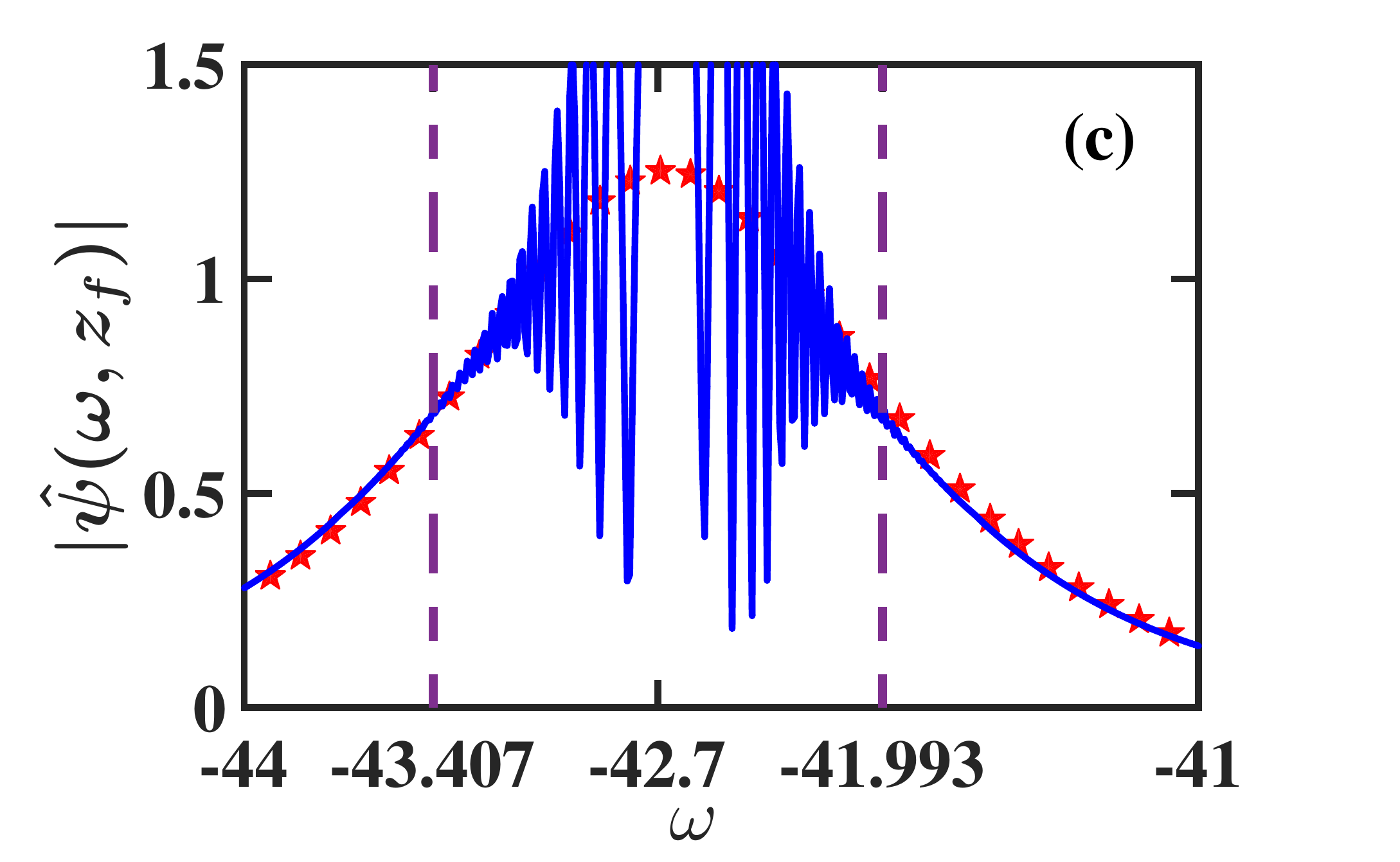} & 
\epsfxsize=5.8cm  \epsffile{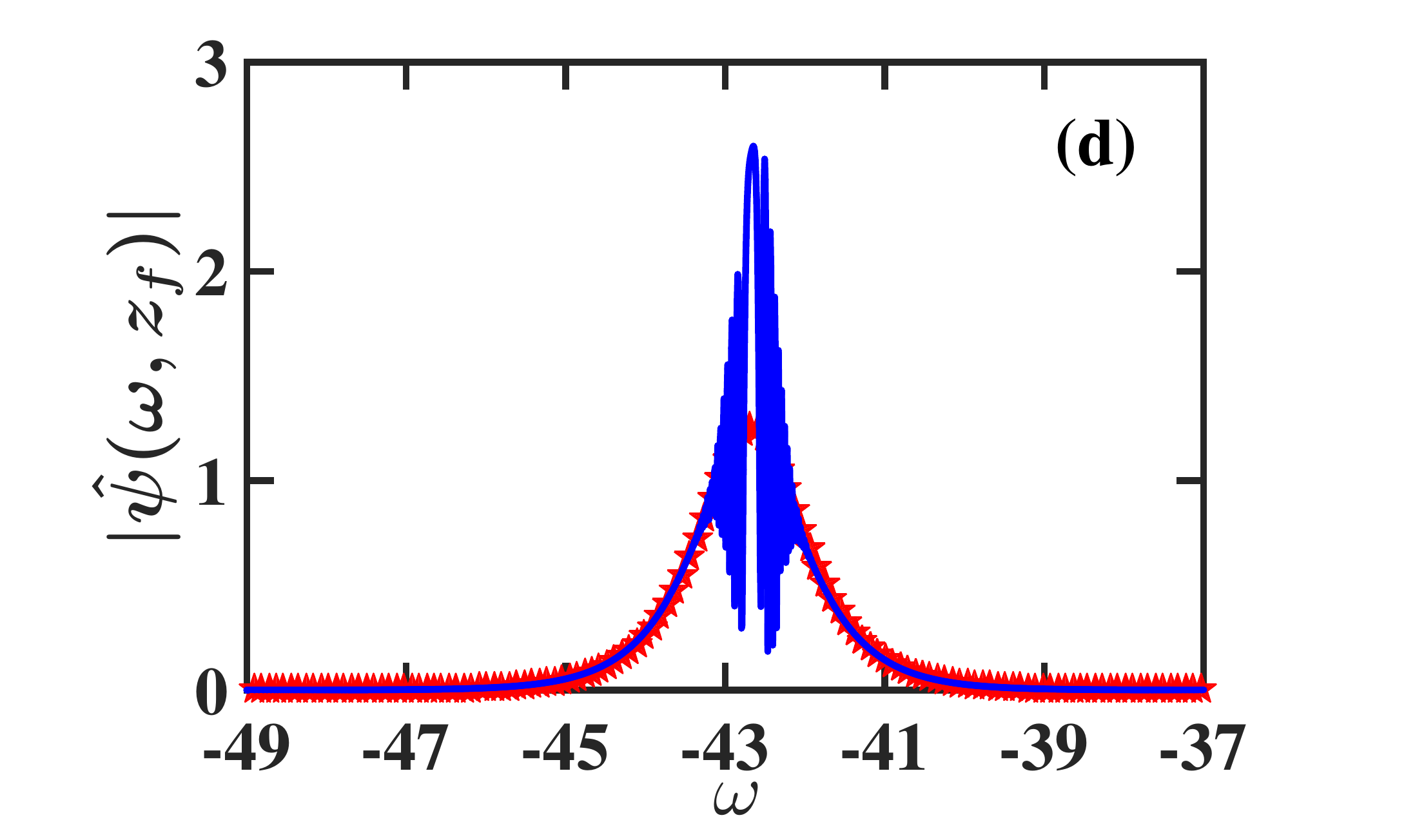} \\
 \epsfxsize=5.8cm  \epsffile{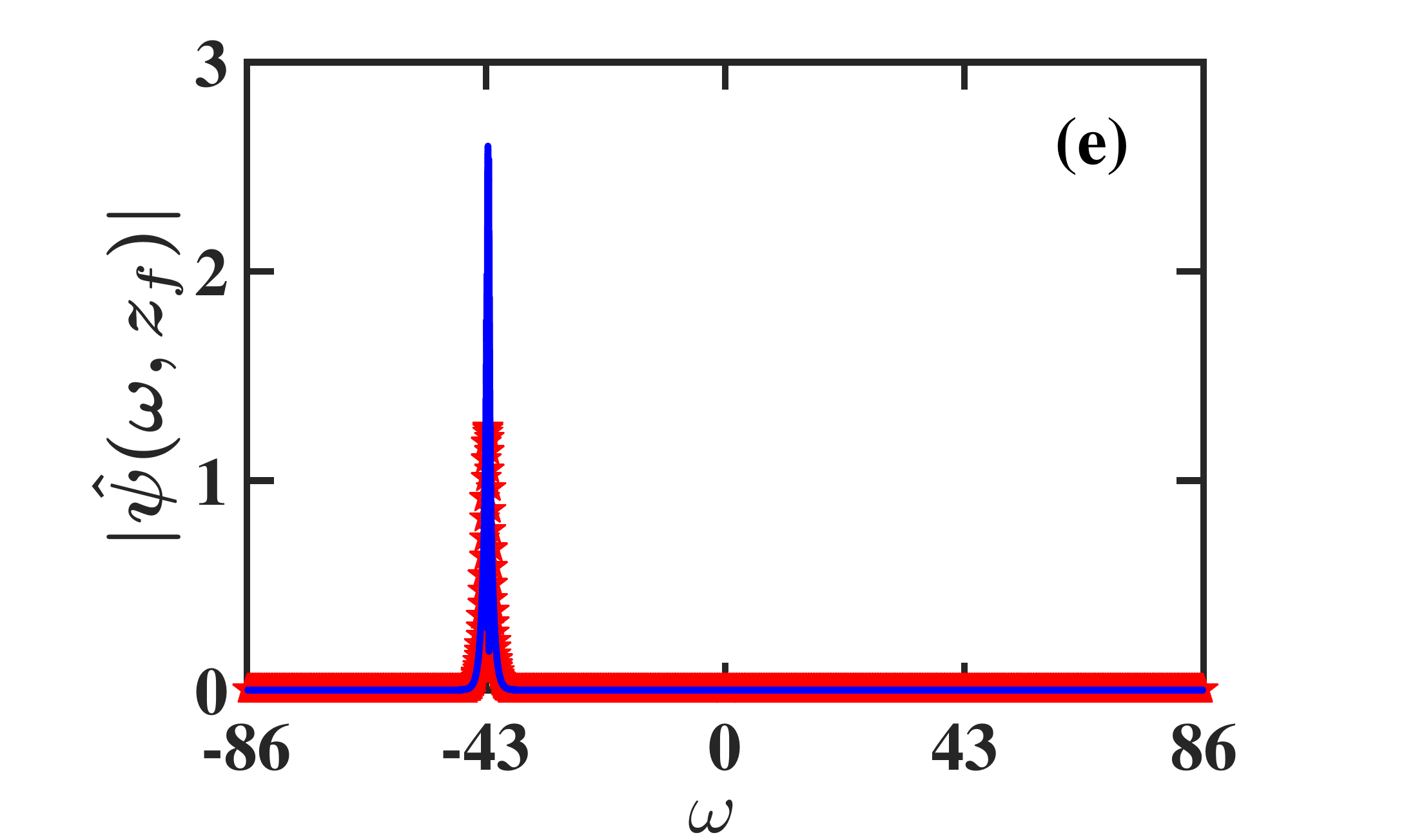}   
\end{tabular}
\end{center}
\caption{The Fourier transform of the pulse shape $|\hat\psi(\omega,z)|$ 
at $z_{q}=96$ [(a) and (b)] and at $z_{f}=208$ [(c), (d), and (e)] 
for the same optical waveguide setup considered in Fig. \ref{fig17}. 
The solid blue curve represents the result obtained by numerical 
simulations with Eq. (\ref{sfs42}). The red stars correspond 
to the prediction of the adiabatic perturbation theory, 
obtained with Eqs. (\ref{Iz3}), (\ref{sfs44}), and  (\ref{sfs45}).
The dashed purple vertical lines in (b) and (c) correspond to the end points 
of the instability interval of Eq. (\ref{sfs48}) for small  amplitude wave solutions 
of Eq. (\ref{sfs46}).}                   
 \label{fig19}
\end{figure}

The reduction in transmission quality of the waveguides considered in the current subsection 
compared with the waveguides considered in section \ref{no_shifting} can be partially attributed 
to the following factors. First, the closed waveguide loop setup, which leads to accumulation of radiation, 
and second, the smaller size of the computational domain used in the simulations in the current subsection.     
The other major factors leading to the reduced transmission quality 
can be explained by analyzing the dynamics of the 
Fourier transform of the optical field $|\hat\psi(\omega,z)|$. 
Figure \ref{fig19} shows the numerically obtained $|\hat\psi(\omega,z)|$ 
at $z=z_{q}$ and at $z=z_{f}$ together with the prediction of the adiabatic perturbation theory, 
obtained with Eqs. (\ref{Iz3}), (\ref{sfs44}), and  (\ref{sfs45}). 
We observe that the graphs of $|\hat\psi(\omega,z_{q})|$ and $|\hat\psi(\omega,z_{f})|$ 
vs $\omega$ are somewhat similar to the graphs obtained in section \ref{no_shifting} for soliton 
propagation in waveguides with linear gain or loss and cubic loss 
[compare Fig. \ref{fig19} with Figs. \ref{fig3} and \ref{fig7}].     
More specifically, the deviation of the numerical result from 
the prediction of the perturbation theory is noticeable already at $z=z_{q}$ 
and is of order 1 at $z=z_{f}$. 
This deviation appears as fast oscillations in the graph of the 
numerically obtained $|\hat\psi(\omega,z)|$ vs $\omega$, 
which are most pronounced near the soliton's central frequency $\beta(z)$.  
Additionally, as seen in Figs. \ref{fig19}(b) and \ref{fig19}(c), 
the frequency interval in which the oscillations are most pronounced coincides 
with the instability interval in Eq. (\ref{sfs48}) for small  amplitude wave solutions 
$\psi_{l}(t,z)$ of the linear propagation model (\ref{sfs46}) 
(for the parameter values used in the simulations, 
the instability interval of Eq. (\ref{sfs48}) is $-43.407 < \omega < -41.993$).      
Furthermore, there is no significant separation between the soliton's spectrum 
and the radiation's spectrum.
Based on these observations we identify three additional factors besides the 
closed waveguide loop setup and the size of the computational domain 
that lead to reduced transmission quality in the current waveguide setup. 
(1) Additional emission of radiation due to the presence of the guiding filters. 
(2) Instability of small amplitude waves with frequencies close to the equilibrium 
value of the soliton's frequency $\omega_{p}$. 
(3) The lack of significant separation between the soliton's spectrum and the radiation's spectrum,  
which makes suppression of radiation emission by the guiding filters inefficient 
(see Eq. (\ref{sfs46_A}) and Refs. \cite{Mollenauer92,Mollenauer97,Mollenauer2006}).   
The combination of the factors (1)-(3) together with the closed waveguide loop setup 
and the smaller size of the computational domain leads to smaller $z_{q}$ and $z_{f}$ 
values in the current waveguide setup compared with the values obtained 
in section \ref{no_shifting} for soliton propagation in the absence of guiding filters.

\begin{figure}[ptb]
\begin{tabular}{cc}
\epsfxsize=8cm  \epsffile{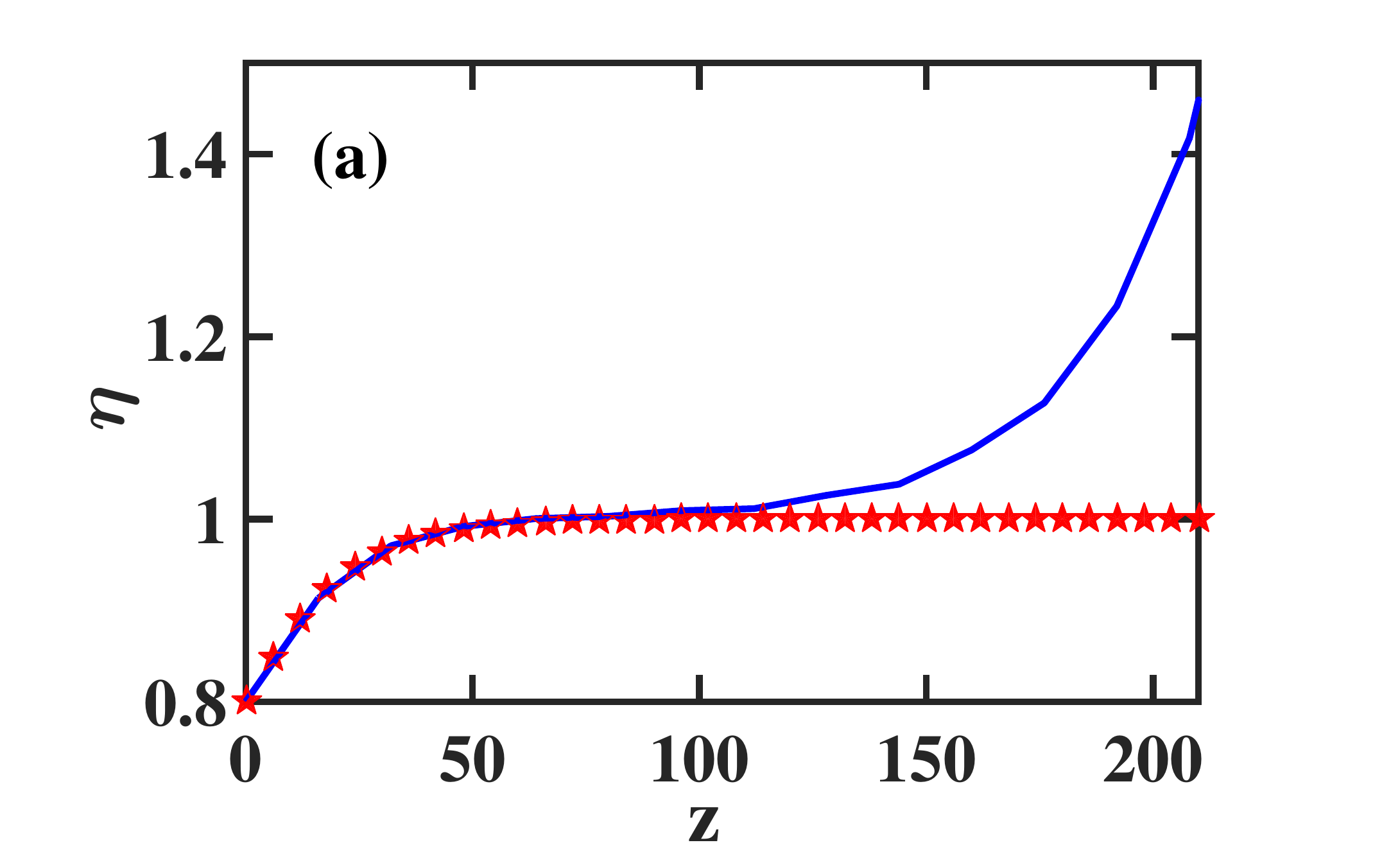} \\
\epsfxsize=8cm  \epsffile{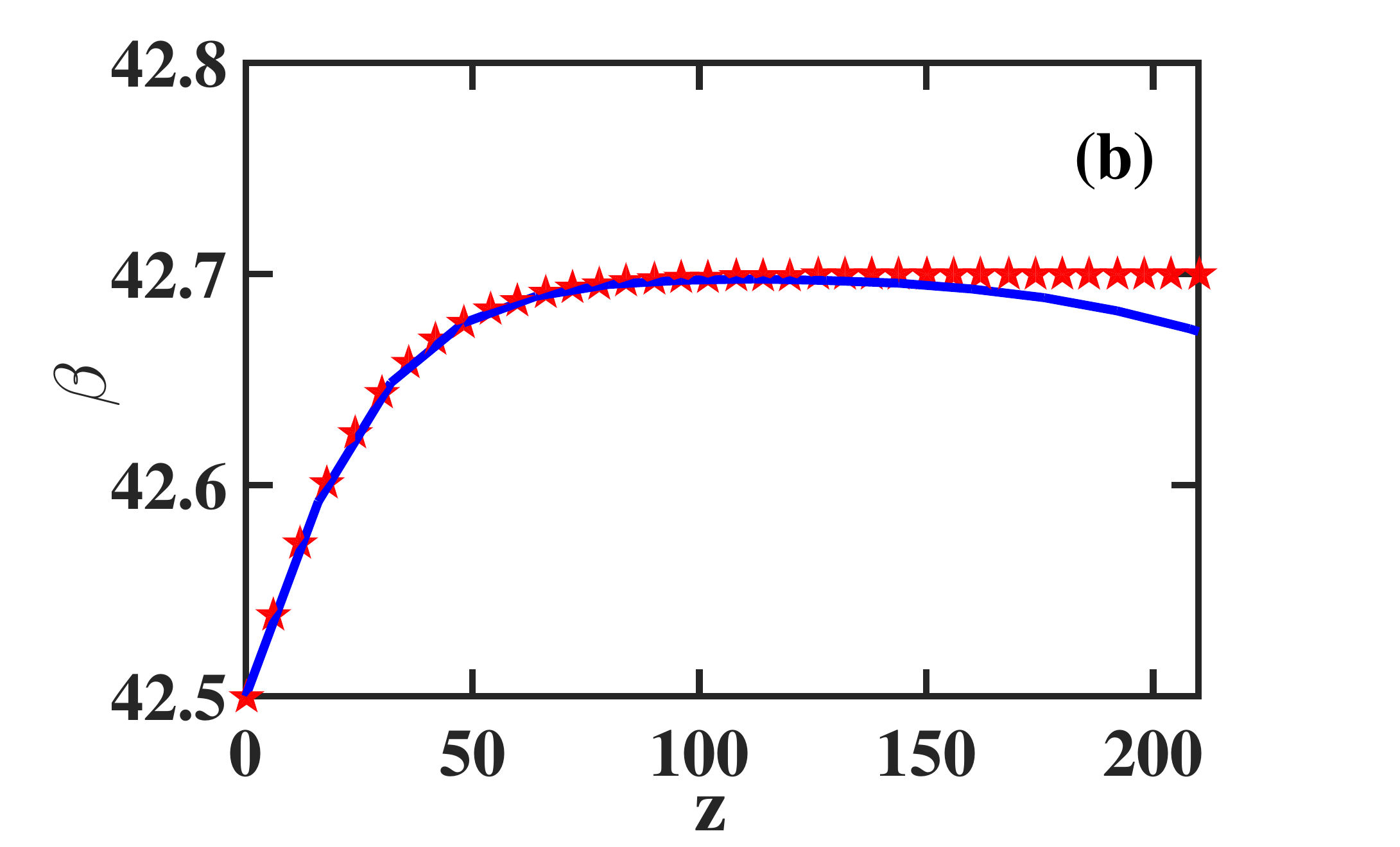}
\end{tabular}
\caption{The $z$ dependence of the soliton amplitude $\eta(z)$ (a) and frequency $\beta(z)$ (b)  
for the closed optical waveguide loop setup considered in Figs. \ref{fig17}-\ref{fig19}. 
The solid blue curves represent the results obtained by numerical simulations with Eq. (\ref{sfs42}).
The red stars correspond to the predictions of the adiabatic perturbation theory, 
obtained with Eqs. (\ref{sfs44}) and (\ref{sfs45}).}
\label{fig20}
\end{figure}

The reduction in transmission quality in waveguides with guiding filters with a constant 
central frequency is also manifested in the dynamics of the soliton's amplitude and frequency. 
Figures \ref{fig20}(a) and \ref{fig20}(b) show 
the $z$ dependence of the soliton's amplitude and frequency obtained in the simulations. 
The predictions of the adiabatic perturbation theory, obtained with Eqs. (\ref{sfs44}) and (\ref{sfs45}), 
are also shown. We observe that for $0 \le z \le 100$, the numerically obtained amplitude and frequency 
tend to the equilibrium values $\eta_{0}=1$ and $\omega_{p}=42.7$, in good agreement with the 
prediction of the adiabatic perturbation theory. However, for $100 < z \le 208$, the numerically 
obtained curves of $\eta(z)$ and $\beta(z)$ deviate significantly from the curves predicted by the  
perturbation theory. These deviations coincide with the increase in the value of $I(z)$ observed in 
Fig. \ref{fig18} and with the deterioration of the pulse shape observed 
in Figs. \ref{fig17} and \ref{fig19}. Based on these observations and on the smaller 
values of $z_{q}$ and $z_{f}$ for the current waveguide setup compared with 
the values obtained for the waveguide setups considered in sections \ref{no_shifting} and \ref{Raman_sfs}, 
we conclude that the introduction of guiding filters with a {\it constant} central frequency 
does not lead to improvement of transmission quality.

\subsection{Waveguides with linear gain, cubic loss, and guiding filters with a varying central frequency}
\label{filters2}

As we saw in subsection \ref{filters1}, suppression of radiation emission in 
waveguides with guiding filters with a constant central frequency is inefficient 
due to the lack of significant separation between the soliton's spectrum and the 
radiation's spectrum. This leads to reduced transmission quality for these waveguides. 
However, as shown in Refs. \cite{Mollenauer92,Mollenauer97,Mollenauer2006} (for optical fibers), 
this drawback can be circumvented by using guiding filters with a varying 
central frequency $\omega_p(z)$, which is a monotonous function of $z$. 
In this case at large distances, the soliton's spectrum is centered around 
a $z$ dependent frequency $\tilde\beta_{0} + \omega_p(z)$, while the 
radiation's spectrum is centered near the constant frequency $\tilde\beta_{0}$. 
Since $\omega_p(z)$ is a monotonous function of $z$,  
at sufficiently large $z$ $|\omega_p(z)| \gg 1$, 
and therefore the radiation's spectrum is well-separated from the soliton's spectrum. 
As a result, in this case suppression of radiation emission by the guiding filters 
becomes very efficient at intermediate and large distances.

We therefore turn to study soliton propagation in the presence
of weak linear gain, weak cubic loss, and guiding filters with a varying central frequency. 
Similar to the treatment in Refs. \cite{Mollenauer92,Mollenauer97,Mollenauer2006}, 
we assume that the response function of the guiding filters can be approximated 
by a Gaussian with a maximum that is equal to 1 and that is located at the frequency 
$\omega_{p}(z)$. Thus, the propagation is described by Eq. (\ref{sfs41}) or by Eq. (\ref{sfs42}), 
where $\omega_{p}$ is now $z$ dependent. 
In addition, the dynamics of the soliton's amplitude and frequency is described 
by Eqs. (\ref{sfs43}) and (\ref{sfs44}) in first-order in $\epsilon_{\omega}$ and $\epsilon_{3}$. 
Similar to the treatment in Refs. \cite{Mollenauer92,Mollenauer97,Mollenauer2006}, 
we assume that $\omega_{p}$ changes linearly with $z$, that is, $\omega_{p}=\omega_{p}' z$, 
where $\omega_{p}' \equiv d \omega_{p}/dz = C_{1}$, and $C_{1}$ is a constant. 
We define a new frequency $\tilde\beta$ by: $\tilde\beta(z)=\beta(z)-\omega_{p}(z)$. 
The new system of equations for the dynamics of $\eta$ and $\tilde\beta$ is: 
\begin{eqnarray}&&
\frac{d\eta}{dz} = 
\eta\left[ g_{0} - 2\epsilon_{\omega}\left(\eta^{2} \!/3 + \tilde\beta^{2}\right) 
-4\epsilon_{3}\eta^{2} \!/3\right] ,
\label{sfs51}
\end{eqnarray}
and
\begin{eqnarray}&&
\nonumber \\
\frac{d\tilde\beta}{dz} = -C_{1} - 4\epsilon_{\omega}\tilde\beta\eta^{2} \!/3 .
\label{sfs52}
\end{eqnarray} 
We are interested in realizing stable transmission with constant amplitude $\eta=\eta_{0}>0$
and frequency $\tilde\beta=\tilde\beta_{0}\ne 0$.      
We therefore require that $(\eta_{0},\tilde\beta_{0})$ is an equilibrium 
point of Eqs. (\ref{sfs51}) and (\ref{sfs52}). We obtain: 
$g_{0} = 2\epsilon_{\omega}\eta_{0}^{2}/3 +  
2\epsilon_{\omega}\tilde\beta_{0}^{2} + 4\epsilon_{3}\eta_{0}^2/3$ 
and $\tilde\beta_{0} = -3\omega_{p}'/(4\epsilon_{\omega}\eta_{0}^2)$.         
Thus, Eq. (\ref{sfs51}) takes the form
\begin{eqnarray}&&
\frac{d\eta}{dz} = 
2\eta\left[2\epsilon_{3}\left(\eta_{0}^{2}-\eta^{2}\right)\!/3
+ \epsilon_{\omega}\left(\eta_{0}^{2} - \eta^{2}\right)\!/3
+ \epsilon_{\omega}\left(\tilde\beta_{0}^{2} - \tilde\beta^{2} \right)\right] .
\label{sfs53}
\end{eqnarray}     
Dynamics of the soliton's amplitude and frequency is therefore described 
by Eqs. (\ref{sfs52}) and (\ref{sfs53}). Linear stability analysis shows 
that $(\eta_{0},\tilde\beta_{0})$ is a stable equilibrium point of 
the system (\ref{sfs52})-(\ref{sfs53}) [a stable node], 
provided that $\omega_{p}'$ satisfies the condition 
\begin{equation}
|\omega_{p}'| < \left(\frac{8}{27}\right)^{1/2}\epsilon_{\omega}
\left(1+\frac{2\epsilon_{3}}{\epsilon_{\omega}}\right)^{1/2}\eta_{0}^{3} .
\label{sfs54}
\end{equation}

{\it Numerical simulations}.                  
Equation (\ref{sfs42}) is numerically integrated on a domain 
$[t_{\mbox{min}},t_{\mbox{max}}]=[-400,400]$ with periodic boundary conditions. 
The initial condition is in the form of an NLS soliton with amplitude $\eta(0)$, 
frequency $\beta(0)=0$, position $y(0)=0$, and phase $\alpha(0)=0$. 
To enable comparison with the results of the numerical simulations in subsection \ref{filters1}, 
we use parameter values that are similar to the ones used in this subsection. 
In particular, we carry out the simulations with $\epsilon_{3}=0.01$, 
$\epsilon_{\omega}=0.04$, $\omega_{p}(0)=0$, and $\eta(0)=0.8$. 
We realize efficient separation between the soliton's spectrum and the radiation's 
spectrum by choosing $\omega_{p}'=0.0218$, which is close to the largest value 
allowed by inequality (\ref{sfs54}).   
We emphasize, however, that similar results are obtained for other values of the physical parameters. 
Similar to the simulations in sections \ref{Raman_sfs} and \ref{filters1}, the soliton passes 
multiple times through the computational domain's boundaries during the simulation   
and therefore the simulation describes soliton propagation in a closed waveguide loop.  
To avoid soliton destruction, we do not employ damping at the boundaries. 
The simulation is run up to a final propagation distance $z_{f}=2000$, 
at which the value of the transmission quality integral $I(z)$ is still smaller than 0.075.

\begin{figure}[ptb]
\begin{tabular}{cc}
\epsfxsize=5.8cm  \epsffile{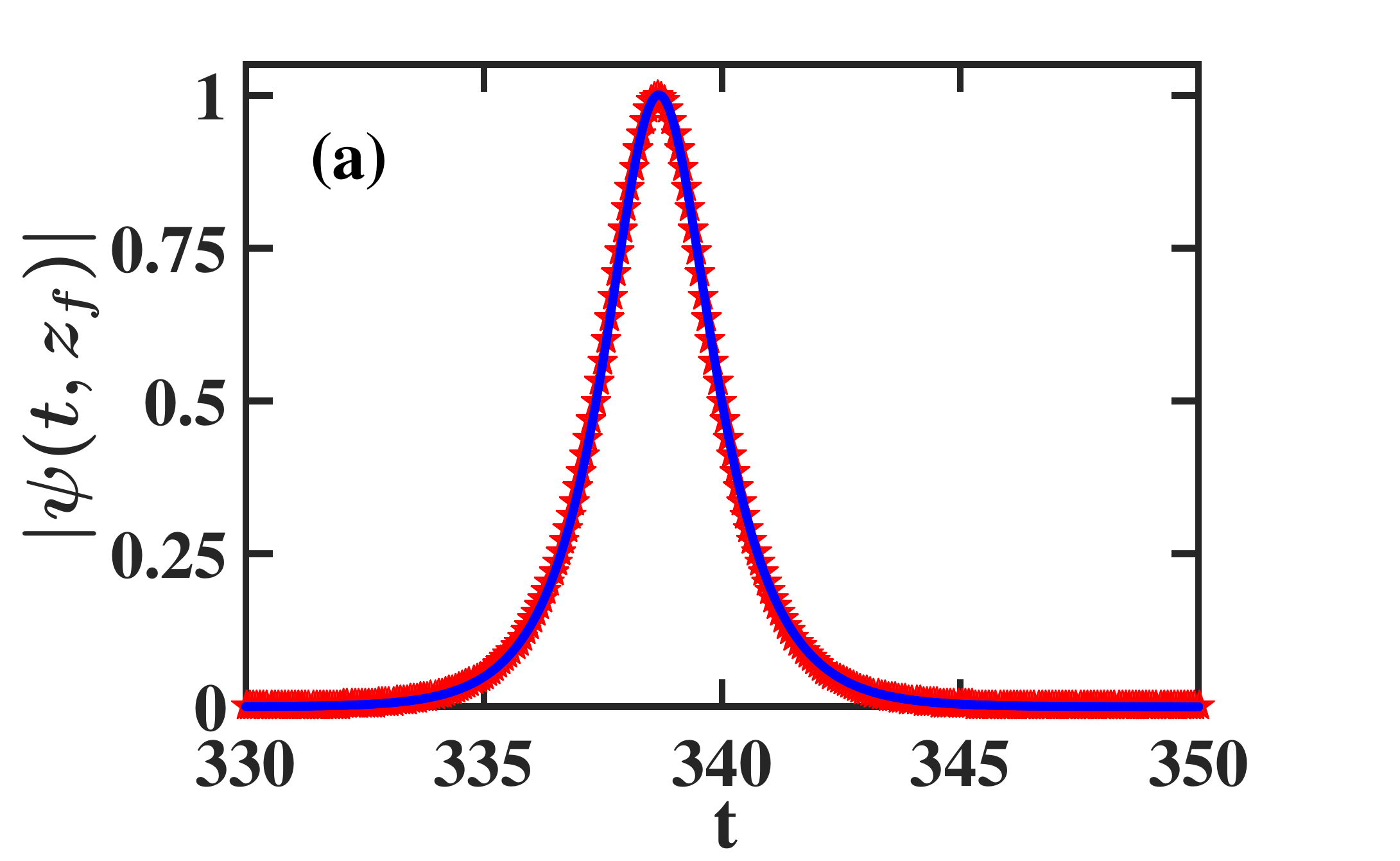} &
\epsfxsize=5.8cm  \epsffile{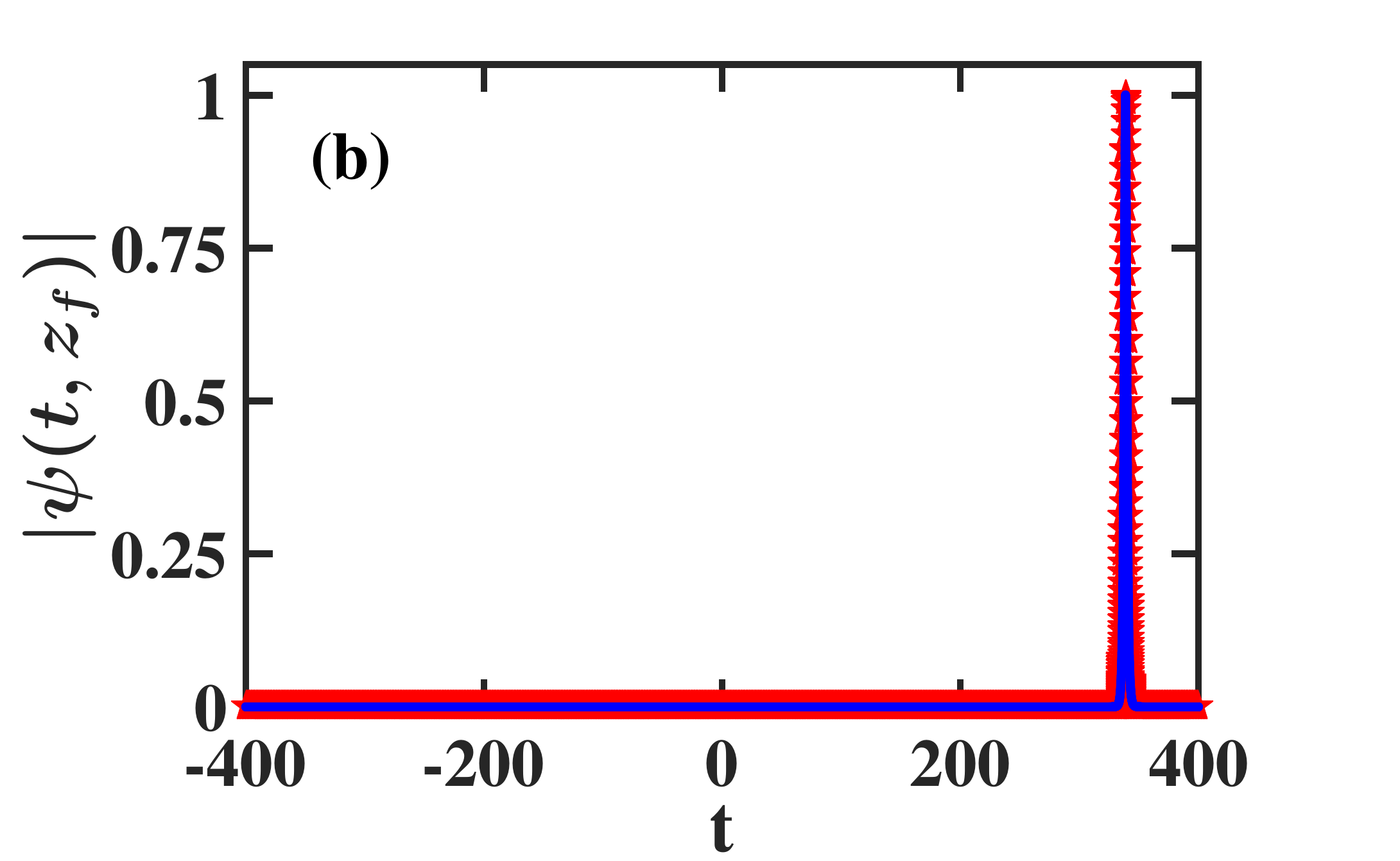} \\
\epsfxsize=5.8cm  \epsffile{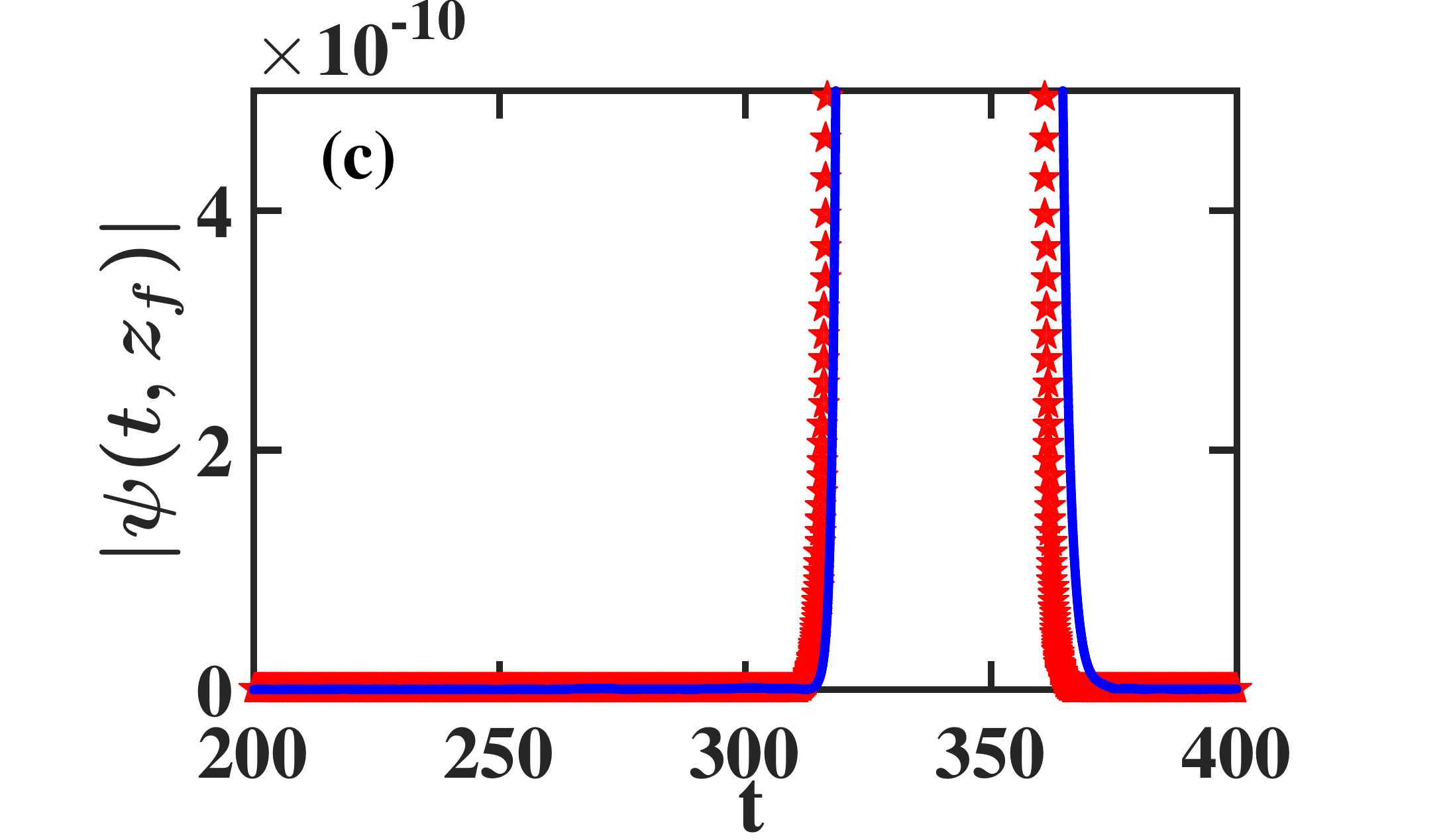} 
\end{tabular}
\caption{The pulse shape $|\psi(t,z_{f})|$,  where $z_{f}=2000$, 
for soliton propagation in a closed optical waveguide loop 
with weak frequency independent linear gain, cubic loss, 
and guiding filters with a varying central frequency. 
The physical parameter values are $\epsilon_{3}=0.01$, 
$\epsilon_{\omega}=0.04$, $\omega_{p}(0)=0$, $\omega_{p}'=0.0218$, 
$\eta(0)=0.8$, and $\beta(0)=0$.
The solid blue curve represents the result obtained by numerical 
simulations with Eq.  (\ref{sfs42}), while the red stars correspond to the 
prediction of the perturbation theory, obtained with 
Eqs. (\ref{Iz1}), (\ref{sfs52}), and (\ref{sfs53}).}
 \label{fig21}
\end{figure}

\begin{figure}[ptb]
\begin{tabular}{cc}
\epsfxsize=10cm  \epsffile{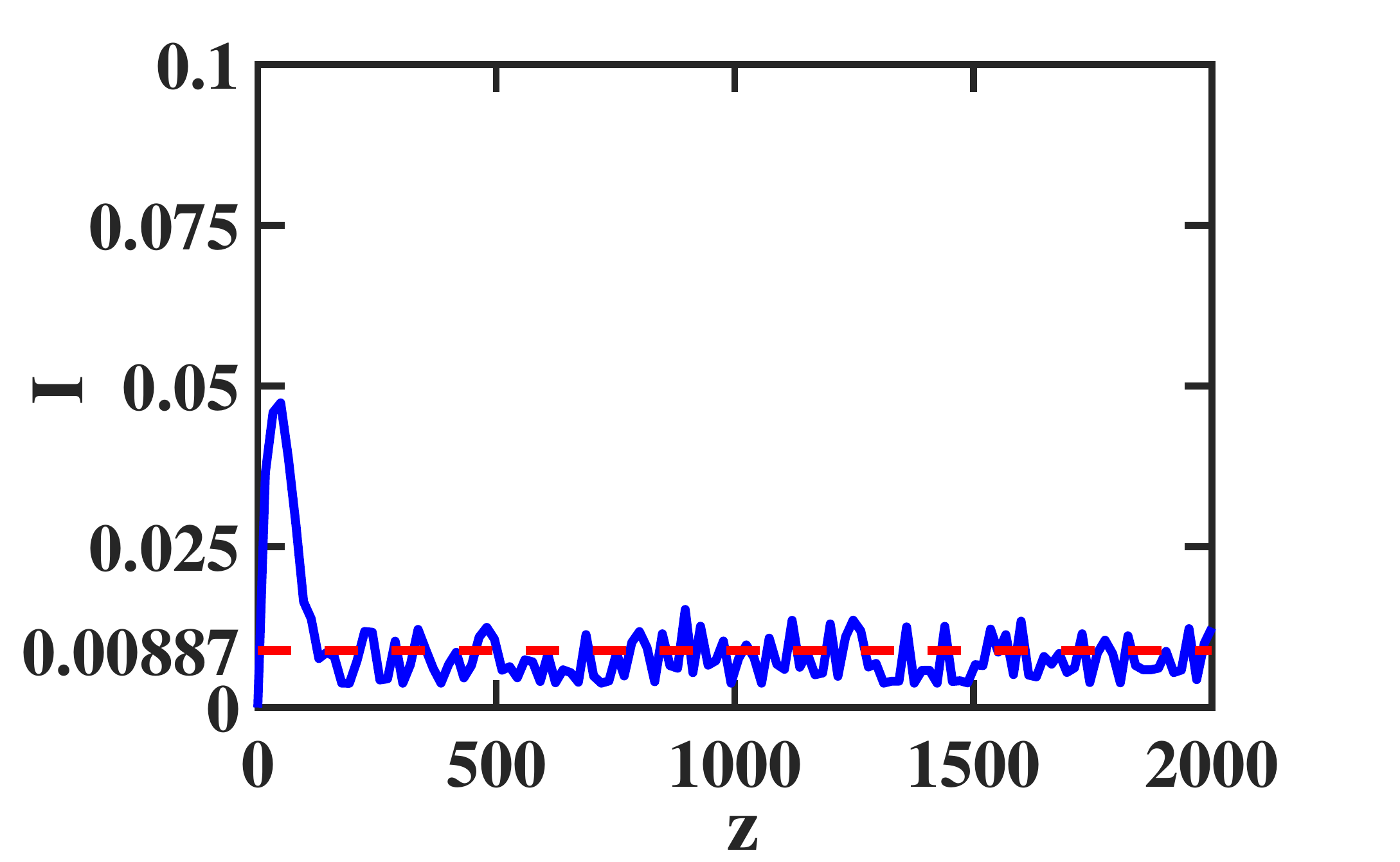} 
\end{tabular}
\caption{The $z$ dependence of the transmission quality integral $I(z)$ obtained 
by numerical simulations with Eq. (\ref{sfs42}) for the same optical waveguide 
setup considered in Fig. \ref{fig21}. The solid blue curve represents $I(z)$ 
and the dashed red horizontal line corresponds to $\langle I(z) \rangle$.}
\label{fig22}
\end{figure}

Figure \ref{fig21} shows the pulse shape $|\psi(t,z)|$ at $z=z_{f}$, as obtained in the simulations.  
Also shown is the prediction of the adiabatic perturbation theory, obtained with Eqs. (\ref{Iz1}), (\ref{sfs52}),   
and (\ref{sfs53}). As seen in Figs. \ref{fig21}(a) and  \ref{fig21}(b),   
the numerically obtained pulse shape at $z=z_{f}$ is very close to the analytic prediction 
and no significant radiative tail is observed. Furthermore, as seen in Fig. \ref{fig21}(c), 
the deviation of the numerical result for $|\psi(t,z_{f})|$ from the theoretical one is smaller than 
$10^{-9}$ for all $t$ values. Therefore, the introduction of guiding filters with a central frequency 
that changes linearly with propagation distance leads to significant enhancement of transmission quality 
compared with the waveguide setups considered in sections \ref{no_shifting}, \ref{Raman_sfs1}, and \ref{filters1}. 
The enhancement of transmission quality is also demonstrated in Fig. \ref{fig22}, which shows the 
$z$ dependence of the transmission quality integral $I$ obtained in the simulations 
along with the average $\langle I(z) \rangle$. As seen in this figure, 
the value of $I(z)$ is smaller than 0.05 throughout the propagation 
and is smaller than 0.02 for $96 \le z \le 2000$. In addition, $\langle I(z) \rangle=0.00887$

\begin{figure}[ptb]
\begin{center}
\begin{tabular}{cc}
\epsfxsize=8.0cm  \epsffile{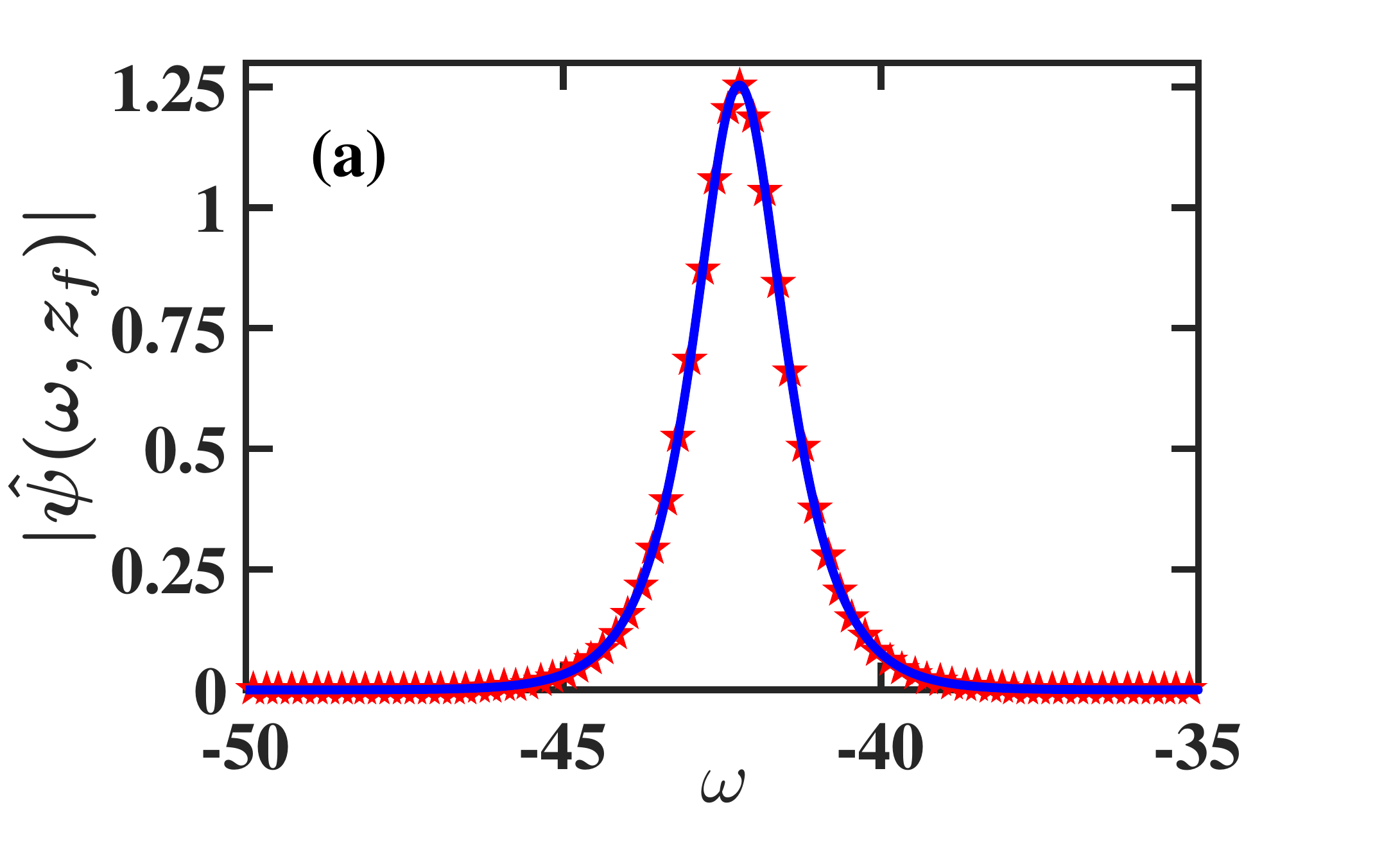} \\
\epsfxsize=8.0cm  \epsffile{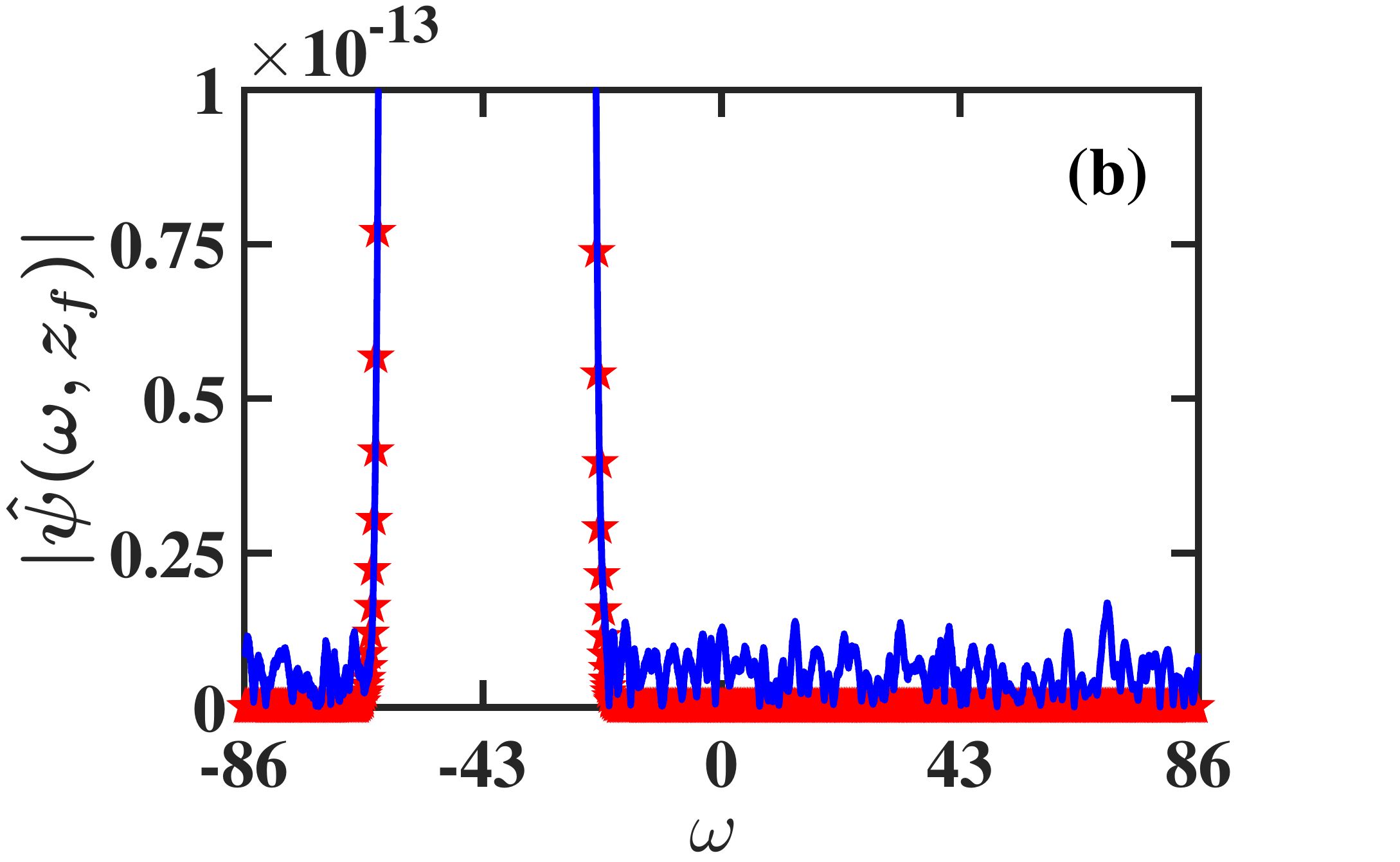} 
\end{tabular}
\end{center}
\caption{The Fourier transform of the pulse shape $|\hat\psi(\omega,z)|$ at $z_{f}=2000$ 
for soliton propagation in a closed optical waveguide loop  
with weak frequency independent linear gain, cubic loss, 
and guiding filters with a varying central frequency. 
The physical parameter values are the same as in Fig. \ref{fig21}. 
The solid blue curve represents the result obtained by numerical 
simulations with Eq.  (\ref{sfs42}). The red stars correspond 
to the prediction of the adiabatic perturbation theory, 
obtained with Eqs. (\ref{Iz3}), (\ref{sfs52}), and (\ref{sfs53}).}                        
 \label{fig23}
\end{figure}

Further insight into the enhanced transmission quality can be gained from  
the Fourier transform of the pulse $|\hat\psi(\omega,z)|$.                          
Figure \ref{fig23} shows the numerically obtained Fourier transform $|\hat\psi(\omega,z)|$ 
at $z=z_{f}$ together with the prediction of the adiabatic perturbation theory, 
obtained with Eqs. (\ref{Iz3}), (\ref{sfs52}), and (\ref{sfs53}). 
The agreement between the two results is excellent. 
In particular, the Fourier transform $|\hat\psi(\omega,z_{f})|$ obtained in the simulation 
does not contain any fast oscillations in the main peak such as the oscillations seen in 
Figs. \ref{fig3} and \ref{fig7} in section \ref{no_shifting}, and in Fig. \ref{fig19} in section \ref{filters1}. 
Additionally, $|\hat\psi(\omega,z_{f})|$ does not contain any peaks associated with radiation emission 
such as the one seen in Fig. \ref{fig11} in section \ref{Raman_sfs1}. 
Based on these findings and based on the comparison with the results obtained in section \ref{filters1}, 
we deduce that the introduction of a varying central frequency of the guiding filters leads to significant 
enhancement of transmission quality. Similar to the situation in waveguides with delayed Raman response, 
the monotonous increase of the central filtering frequency $\omega_{p}$ leads to separation 
of the soliton's spectrum from the radiation's spectrum. 
The separation of the two spectra  enables efficient suppression of radiation emission 
with frequencies that are significantly different from the soliton's frequency 
due to the presence of the guiding filters.

\begin{figure}[ptb]
\begin{tabular}{cc}
\epsfxsize=8cm  \epsffile{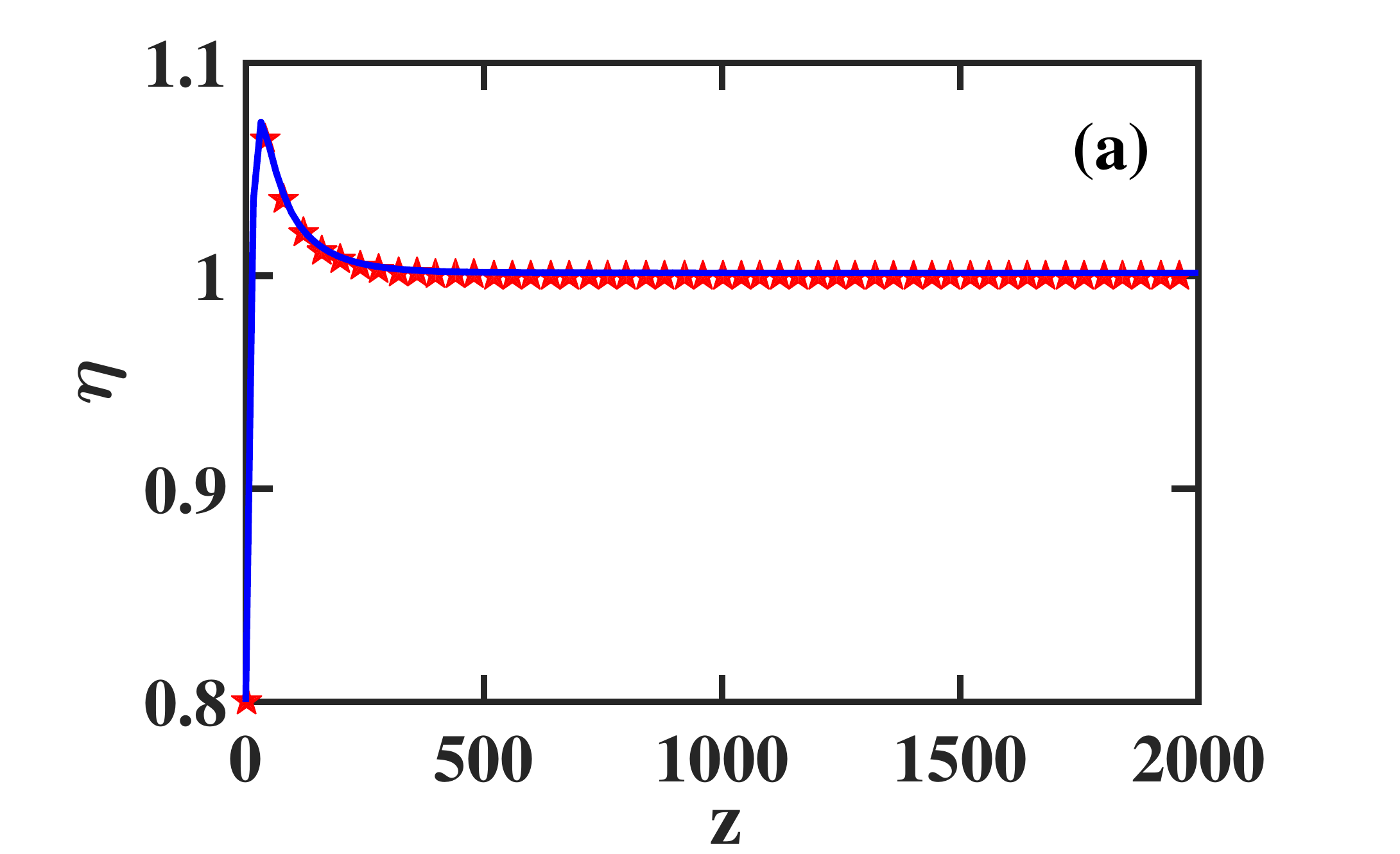} \\
\epsfxsize=8cm  \epsffile{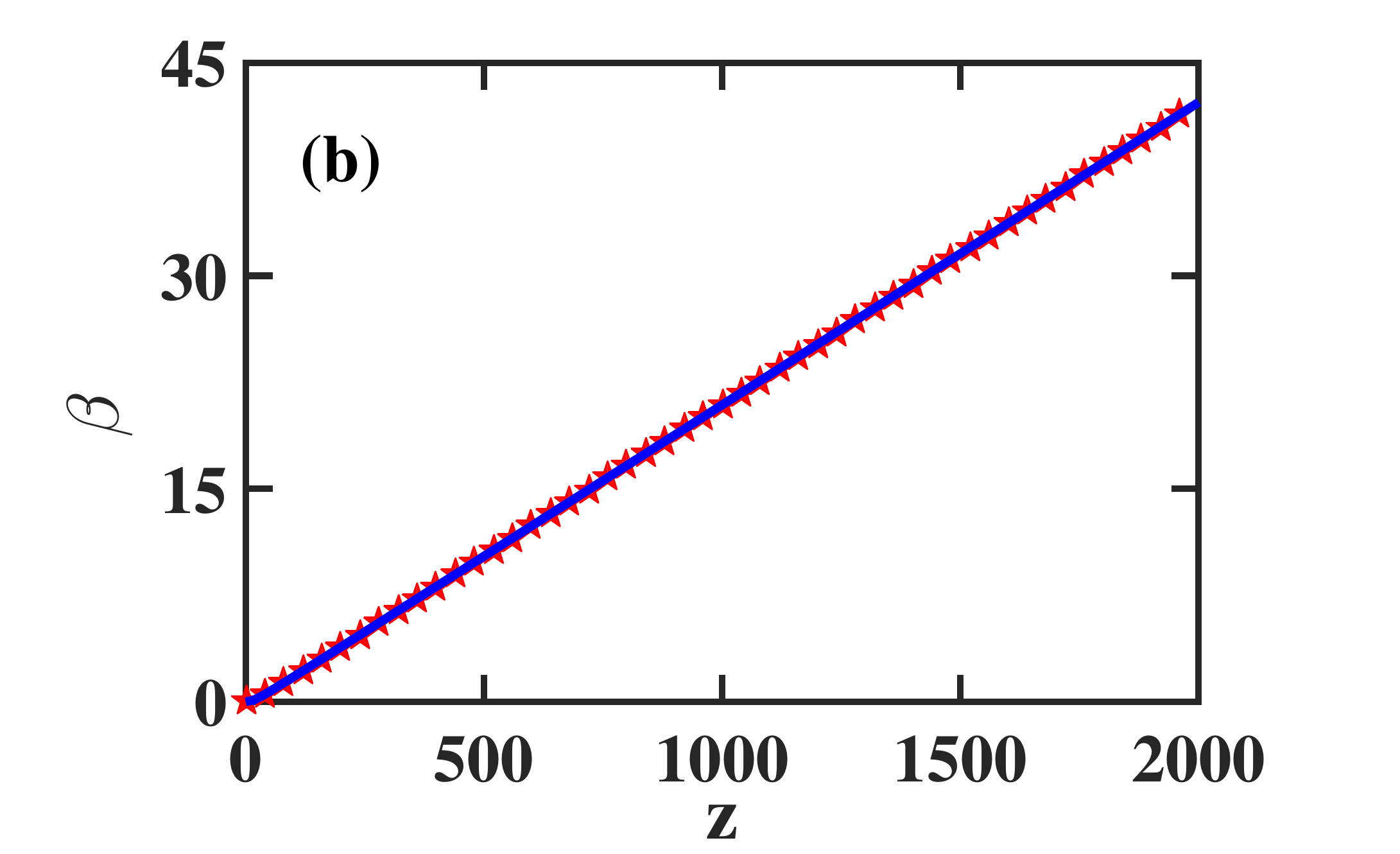}
\end{tabular}
\caption{The $z$ dependence of the soliton amplitude $\eta(z)$ (a) and frequency $\beta(z)$ (b)  
for the closed optical waveguide loop setup considered in Figs. \ref{fig21}-\ref{fig23}. 
The solid blue curves represent the results obtained by numerical simulations with Eq. (\ref{sfs42}).
The red stars correspond to the predictions of the adiabatic perturbation theory, 
obtained with Eqs. (\ref{sfs52}) and (\ref{sfs53}).}
\label{fig24}
\end{figure}

Figures \ref{fig24}(a) and \ref{fig24}(b) show the $z$ dependence 
of the soliton's amplitude and frequency obtained in numerical simulations with Eq. (\ref{sfs42}). 
Also shown are the predictions of the adiabatic perturbation theory, obtained with 
Eqs. (\ref{sfs52}) and  (\ref{sfs53}). It is seen that the numerically obtained soliton amplitude tends to 
the equilibrium value $\eta_{0}=1$ at short distances and stays close to this value throughout 
the propagation, in excellent agreement with the perturbation theory prediction. 
Additionally, the value of the soliton frequency obtained in the simulations remains close 
to the $z$ dependent value predicted by the adiabatic perturbation theory throughout the propagation.    
Based on these findings and on similar results obtained for other values of the physical parameters 
we conclude that the efficient suppression of radiation emission in waveguides with 
frequency independent linear gain, cubic loss, and guiding filters with a varying central frequency 
enables observation of stable amplitude and frequency dynamics along significantly larger distances 
compared with the distances obtained with the closed optical waveguide loop setups 
considered in sections \ref{Raman_sfs1} and \ref{filters1}. 
In this sense, stabilization of amplitude and frequency dynamics in waveguides with guiding filters with a varying 
central frequency is similar to the stabilization observed in Fig. \ref{fig16}, for waveguides with 
frequency dependent linear gain-loss, cubic loss, and delayed Raman response.

\section{Conclusions}
\label{conclusions}

We studied transmission stabilization against radiation emission for single-soliton 
propagation in nonlinear optical waveguides with weak linear gain-loss, cubic loss, 
and delayed Raman response. The value of the linear gain coefficient for waveguides 
with frequency independent linear gain was chosen such that stable soliton transmission 
with a constant amplitude can be realized. However, the presence of the linear gain 
can lead to an unstable growth of small amplitude waves (radiation) emitted by the 
soliton. We therefore looked for ways for stabilizing the transmission by frequency dependent 
linear gain-loss and delayed Raman response. 
We characterized transmission quality and stability by calculating the transmission quality integral, 
which measures the deviation of the pulse shape obtained in numerical simulations 
with perturbed NLS equations from the shape expected by the adiabatic perturbation theory 
for the NLS soliton. Additionally, we characterized stability of amplitude and frequency dynamics by comparing 
the numerically obtained $z$ dependence of the soliton's amplitude and frequency 
with the $z$ dependence expected by the adiabatic perturbation theory.

We first studied soliton propagation in the absence of delayed Raman response. 
Our numerical simulations with the perturbed NLS propagation models showed that 
transmission quality in waveguides with frequency independent linear gain and cubic loss 
is comparable to transmission quality in waveguides with frequency dependent linear gain-loss 
and cubic loss. Furthermore, we found that in the absence of delayed 
Raman response, the presence of frequency dependent linear gain-loss does not lead 
to enhancement of transmission quality due to the lack of significant separation between 
the soliton's Fourier spectrum and the radiation's Fourier spectrum.

We then included the effects of delayed Raman response in the perturbed NLS model. 
Our numerical simulations showed that in waveguides with frequency independent linear gain, 
cubic loss, and delayed Raman response, the soliton's spectrum becomes separated 
from the radiation's spectrum due to the Raman self-frequency shift experienced by the soliton. 
However, in this case transmission quality was not improved compared with transmission quality 
in the absence of delayed Raman response due to the lack of an efficient mechanism 
for suppression of radiation emission. For the same reason, dynamics of the soliton's amplitude 
and frequency became unstable at intermediate propagation distances.

Drastic enhancement of transmission quality was demonstrated in waveguides  
with weak frequency dependent linear gain-loss, cubic loss, and delayed Raman response. 
In this case, our numerical simulations showed that the presence of delayed Raman response 
leads to separation of the soliton's spectrum from the radiation's spectrum, 
while the presence of frequency dependent linear gain-loss with relatively 
strong loss far from the soliton's frequency leads to efficient suppression of radiation emission. 
This enabled the observation of distortion-free soliton propagation and stable 
amplitude and frequency dynamics along significantly larger distances compared 
with the distances obtained in the absence of delayed Raman response 
and compared with the distances obtained in waveguides with 
frequency independent linear gain, cubic loss, and delayed Raman response.
Further numerical simulations showed that enhancement of transmission quality  
in waveguides with weak frequency dependent linear gain-loss, cubic loss, and delayed Raman response 
is similar to transmission quality enhancement in waveguides with weak frequency independent linear gain, 
cubic loss, and guiding filters with a varying central frequency. 
More specifically, the simulations demonstrated that the variation of the central filtering frequency leads 
to separation of the soliton's spectrum from the radiation's spectrum, 
while the presence of the guiding filters leads to efficient suppression of radiation emission.

\appendix
\section{The adiabatic perturbation theory for the fundamental NLS soliton} 
\label{appendA}
In this appendix we give a brief summary of the adiabatic perturbation theory 
for the fundamental NLS soliton, which was developed by Kaup \cite{Kaup90,Kaup91,Kaup76}. 
The theory was used for analyzing soliton dynamics in a variety of optical waveguide systems, 
see, e.g., Refs. \cite{CCDG2003,Hasegawa95} and references therein.

To illustrate the approach, consider the perturbed NLS equation 
\begin{eqnarray}
i\partial_z\psi+\partial_t^2\psi+2|\psi|^2\psi = \epsilon h(t,z),
\label{perturbation1}
\end{eqnarray}
where $0 < |\epsilon| \ll 1$. 
We look for a solution of Eq. (\ref{perturbation1}) in the form: 
\begin{eqnarray} 
\psi(t,z)=\psi_{s}(t,z)+\psi_{rad}(t,z)=
\eta(z)\frac{\exp[i\chi(t,z)]}{\cosh(x)}
+v(t,z)\exp[i\chi(t,z)],
\label{perturbation2}
\end{eqnarray}
where $x=\eta(z)\left[t-y(z)\right]$, $\chi(t,z)=\alpha(z)-\beta(z)\left[t-y(z)\right]$, 
$y(z)=y(0)-2\int_{0}^{z} dz' \beta(z')$, 
and $\alpha(z)=\alpha(0)+\int_{0}^{z} dz' \left[\eta^{2}(z')+\beta^{2}(z')\right]$. 
The first term on the right hand side of Eq. (\ref{perturbation2}) 
is the soliton solution with slow varying parameters, while the second term, 
which is of $O(\epsilon)$, is the radiation part.   
We now substitute Eq. (\ref{perturbation2}) into (\ref{perturbation1}) 
and keep terms up to $O(\epsilon)$. The resulting equation and its complex conjugate can 
be written in the following vector form:     
\begin{eqnarray} &&
\frac{i}{\cosh(x)} {1 \choose -1}\eta \left(\frac{d\alpha}{dz}+\beta\frac{dy}{dz}
-\eta^{2}+\beta^{2}\right)
+\frac{\tanh(x)}{\cosh(x)} {1 \choose 1}\eta^{2} \left(\frac{dy}{dz}+2\beta\right)
\nonumber \\&&
-\frac{ix}{\cosh(x)} {1 \choose -1} \frac{d\beta}{dz}
-\frac{\left[x\tanh(x)-1\right]}{\cosh(x)} {1 \choose 1}\frac{d\eta}{dz} 
+ \partial _{z} {v  \choose v^{\ast }} -i\eta^{2}{\cal L} {v  \choose v^{\ast }} 
\nonumber \\&&
-2\beta \partial _{t} {v  \choose v^{\ast }}= 
-i \epsilon{h(t,z) e^{-i\chi} \choose - h^{\ast}(t,z) e^{i\chi}} .
\label{perturbation3} 
\end{eqnarray}
The linear operator ${\cal L}$ in Eq. (\ref{perturbation3}) is defined by: 
\begin{eqnarray}
{\cal L}=\left(\partial _{x}^{2}-1\right) \pmb{\sigma_{3}}
+\frac{2}{\cosh^{2}(x)} \left(2\pmb{\sigma_{3}}+i\pmb{\sigma_{2}}\right) ,
\label{perturbation4} 
\end{eqnarray}
where $\pmb{\sigma_{j}}$ with $1 \le j \le 3$ are the Pauli spin matrices.

The complete set of orthogonal eigenfunctions of ${\cal L}$ was found in Refs. \cite{Kaup90,Kaup91,Kaup76}. 
It includes four localized eigenfunctions, which appear in the first four terms on the 
left hand side of Eq. (\ref{perturbation3}):  
\begin{eqnarray} && 
f_{0}(x)=\frac{1}{\cosh(x)} {1 \choose -1}, \;\;\;\;
f_{1}(x)=\frac{\tanh(x)}{\cosh(x)} {1 \choose 1},  
\nonumber \\&&
f_{2}(x)=\frac{x}{\cosh(x)} {1 \choose -1}, \;\;\;\;
f_{3}(x)=\frac{x\tanh(x)-1}{\cosh(x)}{1 \choose 1} .
\label{perturbation5} 
\end{eqnarray}
The eigenfunctions $f_{0}(x)$ and $f_{1}(x)$ have a zero eigenvalue, 
while $f_{2}(x)$ and $f_{3}(x)$ satisfy ${\cal L}f_{2}=-2f_{1}$ and ${\cal L}f_{3}=-2f_{0}$ \cite{Kaup90,Kaup91,Kaup76}. 
The left localized eigenfunctions of ${\cal L}$, which are given by $f_{m}^{T}\pmb{\sigma_{3}}$ for $0 \le m \le 3$, 
satisfy the following relations \cite{Kaup90,Kaup91,Kaup76}:  
\begin{eqnarray} && 
\int\limits_{-\infty }^{+\infty } dx f_{2}^{T}(x) \pmb{\sigma_{3}}f_{1}(x)=2, \;\;\;\;
\int\limits_{-\infty }^{+\infty } dx f_{0}^{T}(x) \pmb{\sigma_{3}}f_{3}(x)=-2.
\label{perturbation6} 
\end{eqnarray}
In addition, the set of eigenfunctions of ${\cal L}$ contains an infinite set of unlocalized eigenfunctions, 
which are characterized by a continuous index $q$, where $-\infty < q < \infty$. 
We obtain the dynamic equations for the four soliton parameters by projecting 
both sides of Eq. (\ref{perturbation3}) on the four left localized eigenfunctions of ${\cal L}$. 
In particular, the equations for amplitude and frequency dynamics are obtained by projecting 
both sides of Eq. (\ref{perturbation3}) on the left eigenfunctions 
$f_{0}^{T}(x)\pmb{\sigma_{3}}=\mbox{sech}(x)(1,1)$ and 
$f_{1}^{T}(x)\pmb{\sigma_{3}}=\mbox{sech}(x)\tanh(x)(1,-1)$, respectively.

\section{Calculation of the transmission quality integral $I(z)$}
\label{appendB}                        
In this appendix we present the method used for calculating the 
transmission quality integral $I(z)$ and the transmission quality distance $z_{q}$ 
from the results of the numerical simulations. 
In addition, we present the theoretical predictions for the soliton's shape 
and its Fourier transform, which were used in the analysis of transmission quality.

The theoretical prediction for the soliton's shape and the calculation of $I(z)$ 
are based on the adiabatic perturbation theory for the NLS soliton 
(see Refs. \cite{Hasegawa95,Kaup90,Kaup91,CCDG2003} and \ref{appendA}). 
According to the theory, the total optical field can be written as a sum of the soliton part $\psi_{s}$ 
and the radiation part $\psi_{rad}$, where the soliton part is given by 
the expression for the soliton solution to the unperturbed 
NLS equation with slowly varying parameters [see Eq. (\ref{perturbation2})]. 
We therefore take $\psi_{s}(t,z)$ as the theoretical prediction for the soliton part, 
i.e., $\psi^{(th)}(t,z) \equiv \psi_{s}(t,z)=\eta(z)\mbox{sech}(x)\exp(i\chi)$, 
where $x$ and $\chi$ were defined in \ref{appendA}. Therefore, the theoretical 
prediction for the soliton's shape is given by 
\begin{eqnarray} 
|\psi^{(th)}(t,z)|=
\eta(z)\mbox{sech}\left[\eta(z)\left(t-y(z)\right)\right],
\label{Iz1}
\end{eqnarray}  
where $\eta(z)$ and $y(z)$ can be calculated by solving the equations 
for $d\eta/dz$ and $dy/dz$, which are obtained within the framework 
of the adiabatic perturbation theory. We point out that the value of $y(z)$ 
is not changed by linear gain-loss and by cubic loss. 
In addition, the value of $y(z)$ is affected by the Raman perturbation 
in first-order in $\epsilon_{R}$ only via the $z$ dependence of the soliton's frequency. 
Therefore, in the current paper, we calculate the value of $\eta(z)$ in Eq. (\ref{Iz1}) 
by solving the perturbation theory's equation for $d\eta/dz$, while the value of 
$y(z)$ is measured from the results of the numerical simulations.      
The Fourier transform of $\psi_{s}(t,z)$ with respect to time is 
\begin{eqnarray} 
\hat\psi_{s}(\omega,z)=
\left(\frac{\pi}{2}\right)^{1/2}
\frac{\exp[i\alpha(z)-i\omega y(z)]}
{\cosh\left[\pi\left(\omega+\beta(z)\right)/\left(2\eta(z)\right)\right]}.
\label{Iz2}
\end{eqnarray}                         
Thus, the theoretical prediction for the Fourier transform of the soliton's shape is 
given by: 
\begin{eqnarray} 
|\hat\psi^{(th)}(\omega,z)|=
\left(\frac{\pi}{2}\right)^{1/2}
\mbox{sech}\left[\pi\left(\omega+\beta(z)\right)/\left(2\eta(z)\right)\right], 
\label{Iz3}
\end{eqnarray}   
where $\eta(z)$ and $\beta(z)$ are calculated by solving the equations 
for $d\eta/dz$ and $d\beta/dz$ that are obtained with the adiabatic perturbation theory.

The transmission quality integral $I(z)$ measures the deviation of the pulse shape  
obtained in the numerical simulations $|\psi^{(num)}(t,z)|$ from the 
soliton's shape predicted by the adiabatic perturbation theory $|\psi^{(th)}(t,z)|$.  
We use the same definition of $I(z)$ that was used in Ref. \cite{PNT2016} 
for characterizing transmission stability in multisequence soliton-based 
optical waveguide systems. Thus, $I(z)$ is defined by the relation 
\begin{eqnarray} &&
I(z)=\tilde I^{(dif)}(z)/\tilde I(z), 
\label{Iz4}
\end{eqnarray}   
where $\tilde I^{(dif)}(z)$ and $\tilde I(z)$ are defined by
\begin{eqnarray} &&
\tilde I^{(dif)}(z)=     
\left\{\int_{t_{min}}^{t_{max}} dt\,
\left[\;\left|\psi^{(th)}(t,z) \right| - 
\left|\psi^{(num)}(t,z) \right| \; \right]^2 \right\}^{1/2},   
\label{Iz5}
\end{eqnarray}                      
and
\begin{eqnarray} &&
\tilde I(z)=     
\left[ \int_{t_{min}}^{t_{max}} dt\,
\left| \psi^{(th)}(t,z) \right|^2 \right]^{1/2}.   
\label{Iz6}
\end{eqnarray}       
From this definition it is clear that $I(z)$ measures both distortion in the pulse shape due 
to radiation emission and deviations of the numerically obtained values of the soliton's parameters 
from the values predicted by the adiabatic perturbation theory. 
The transmission quality distance $z_{q}$ is defined as the distance at which the value of $I(z)$ 
first exceeds a constant value $C$. In the current paper we used $C=0.075$. 
We emphasize, however, that the values of the transmission quality 
distance obtained by using this definition are not very sensitive to the value of 
the constant $C$. That is, we found that small changes in the value of $C$ 
lead to small changes in the measured $z_{q}$ values.

\section{Amplitude dynamics in the presence of frequency dependent linear gain-loss}
\label{appendC}

In the current appendix we derive Eq. (\ref{sfs14}) for the dynamics of the soliton's 
amplitude in waveguides with frequency dependent linear gain-loss and cubic loss. 
The calculation of the effects of cubic loss on amplitude dynamics is straightforward 
and has been presented in earlier works (see, e.g., Refs. \cite{PNC2010,Silberberg90}). 
We therefore concentrate mainly on calculating the effects of frequency dependent 
linear gain-loss on amplitude dynamics.

We introduce the following notations for the two perturbation terms 
on the right hand side of Eq. (\ref{sfs11}): 
$h_{1}(t,z)=i{\cal F}^{-1}(\hat g(\omega) \hat\psi)/2$  
and $h_{2}(t,z)=-i\epsilon_{3}|\psi|^2\psi$, and assume that 
$\hat g(\omega)$ can be approximated by Eq. (\ref{sfs13}). 
In the leading order of the perturbation theory, we approximate $\psi$ and $\hat\psi$ 
by the soliton parts $\psi_{s}$ and $\hat\psi_{s}$, which are given 
by Eqs. (\ref{perturbation2}) and (\ref{Iz2}), respectively. 
Therefore \cite{eta_equal_0}: 
\begin{eqnarray}
h_{1}(t,z) \simeq
i{\cal F}^{-1}(\hat g(\omega) \hat\psi_{s})/2 ,
\label{loss1}
\end{eqnarray}          
and 
\begin{eqnarray}
h_{2}(t,z) \simeq - i\epsilon_{3}|\psi_{s}|^2\psi_{s}.
\label{loss2}
\end{eqnarray}

We first calculate the contribution of $h_{1}(t,z)$ to the right hand side of Eq. (\ref{sfs14}).
Using the convolution theorem, we obtain: 
\begin{eqnarray}
{\cal F}^{-1}(\hat g(\omega) \hat\psi_{s}) =
(2\pi)^{-1/2}\int_{-\infty}^{\infty} ds g(s)\psi_{s}(t-s,z) .
\label{loss3}
\end{eqnarray}     
Calculation of the inverse Fourier transform of $\hat g(\omega)$ yields 
\begin{eqnarray}
g(t)=-(2\pi)^{1/2}g_{L}\delta(t) 
+\left(\frac{2}{\pi}\right)^{1/2}(g_{0}+g_{L})
\exp[-i\beta(0)t]\sin(Wt/2)/t,
\label{loss4}
\end{eqnarray}       
where $\delta(t)$ is the Dirac delta function. 
Substituting Eq. (\ref{loss4}) into Eq. (\ref{loss3}) while using the expression 
for $\psi_{s}(t,z)$ in Eq. (\ref{perturbation2}), we obtain the following equation 
for the leading order approximation for $-ih_{1}(t,z)$:
\begin{eqnarray}
-ih_{1}(t,z) \simeq
\frac{-g_{L}\eta e^{i\chi}}{2\cosh(x)}
+\frac{(g_{0}+g_{L})}{2\pi}\eta e^{i\chi} 
\int_{-\infty}^{\infty} ds \frac{\sin(Ws/2)}{s\cosh(x-\eta s)} .
\label{loss5}
\end{eqnarray}       
From Eq. (\ref{loss5}) it follows that                               
\begin{eqnarray} &&
-i {h_{1}(t,z) e^{-i\chi} \choose - h_{1}^{\ast}(t,z) e^{i\chi}} \simeq
\frac{-g_{L}\eta}{2\cosh(x)} {1 \choose 1}
+\frac{(g_{0}+g_{L})}{2\pi}\eta {1 \choose 1}  
\int_{-\infty}^{\infty} ds \frac{\sin(Ws/2)}{s\cosh(x-\eta s)} . 
\nonumber \\&&
\label{loss6} 
\end{eqnarray}
A straightforward calculation for the contribution of the cubic loss term yields: 
\begin{eqnarray} 
-i {h_{2}(t,z) e^{-i\chi} \choose - h_{2}^{\ast}(t,z) e^{i\chi}} \simeq
-\frac{\epsilon_{3}\eta^{3}}{\cosh^{3}(x)} {1 \choose 1} .
\label{loss7} 
\end{eqnarray}

Substituting Eqs. (\ref{loss6}) and (\ref{loss7}) into Eq. (\ref{perturbation3}) 
and projecting  both sides of the resulting equation on the left eigenfunction 
$f_{0}^{T}(x)\pmb{\sigma_{3}}=\mbox{sech}(x)(1,1)$ 
of the linear operator ${\cal L}$, we obtain: 
\begin{eqnarray} 
\frac{d\eta}{dz} =
\left[ -g_{L}  + \left(g_{0} + g_{L}\right)\tilde J(\eta;W)/(2\pi)
-4\epsilon_{3}\eta^{2}/3\right]\eta.
\label{loss8}
\end{eqnarray}
The function $\tilde J(\eta;W)$ in Eq. (\ref{loss8}) is given by: 
\begin{eqnarray} &&
\tilde J(\eta;W) \!=\!
\int_{-\infty}^{\infty} \!\!\! ds \frac{\sin(Ws/2)}{s}
\int_{-\infty}^{\infty} \frac{dx}{\cosh(x)\cosh(x-\eta s)} 
\nonumber \\&&
= 2\pi \,\mbox{sgn}(\eta) \tanh \! \left(\frac{\pi W}{4\eta}\right)  
\label{loss9}
\end{eqnarray}
for $\eta \ne 0$, where $W>0$ is used \cite{eta_equal_0}.  
Substituting Eq. (\ref{loss9}) into Eq. (\ref{loss8}) and using the notation 
$V=\pi W/(4\eta)$, we obtain \cite{eta_tends_to_0}: 
\begin{eqnarray}
\frac{d\eta}{dz} =
\left[ -g_{L}  + \left( g_{0} + g_{L} \right)\mbox{sgn}(\eta)\tanh(V) - 4\epsilon_{3}\eta^{2}/3\right]\eta.
\label{loss10}
\end{eqnarray}
Since in the physical problem $\eta \ge 0$, we arrive at: 
\begin{eqnarray}
\frac{d\eta}{dz} =
\left[ -g_{L}  + \left( g_{0} + g_{L} \right) \tanh(V) - 4\epsilon_{3}\eta^{2}/3\right]\eta,
\label{loss11}
\end{eqnarray}
which is Eq. (\ref{sfs14}).

We now discuss stability properties of the equilibrium points $\eta=\eta_{0}$ and $\eta=0$ 
of Eq. (\ref{sfs16}). Stability of the equilibrium point $\eta=0$ is established in a more convenient 
manner with the help of Eq. (\ref{loss10}). We therefore use Eq. (\ref{loss10}) in the following analysis. 
Substituting Eq. (\ref{sfs15}) for $g_{0}$ into Eq. (\ref{loss10}), we obtain:       
\begin{equation}
\frac{d\eta}{dz} = \eta\left\lbrace g_{L}\left[\frac{\mbox{sgn}(\eta)\tanh(V)}{\tanh\left(V_{0}\right)} - 1 \right]
+\frac{4}{3}\epsilon_{3}\left[\eta_{0}^{2}\frac{\mbox{sgn}(\eta)\tanh(V)}{\tanh\left(V_{0}\right)} - \eta^{2} \right]
\right\rbrace . 
\label{loss12}
\end{equation}  
Denote the right hand side of Eq. (\ref{loss12}) by $H(\eta)$. It is straightforward to show that 
$H(\eta)<0$ for $\eta>\eta_{0}$, $H(\eta)>0$ for $0<\eta<\eta_{0}$, and $H(\eta)<0$ 
for $-\eta_{0}<\eta<0$. It follows that there are no additional equilibrium points with $\eta>0$, 
and that $\eta=\eta_{0}$ is a stable equilibrium point, while $\eta=0$ is an unstable equilibrium point. 
Thus, the number, locations, and stability properties of the equilibrium points of Eq. (\ref{sfs16}) 
and Eq. (\ref{sfs4}) are the same.

\section*{Acknowledgments}
D.C. is grateful to the Mathematics Department of NJCU 
for providing technological support for the computations.


\begin{thebibliography}{}
\bibitem{dimensions1} The dimensionless distance $z$ in Eq. (\ref{sfs1})  is 
$z=X/(2L_{D})$, where $X$ is the dimensional distance, 
$L_{D}=\tau_{0}^{2}/|\tilde\beta_{2}|$ is the dispersion length,
$\tau_{0}$ is the soliton width, and $\tilde\beta_{2}$ is the second-order
dispersion coefficient. The dimensionless time is
$t=\tau/\tau_{0}$, where $\tau$ is time. 
$\psi=(\gamma \tau_{0}^{2}/|\tilde\beta_{2}|)^{1/2}E$, 
where $E$ is the electric field and $\gamma$ is the Kerr nonlinearity coefficient.
The coefficients $g_{0}$ and $\epsilon_{3}$ are related 
to the dimensional linear gain and cubic loss coefficients $\rho_{1}$ and $\rho_{3}$ 
by $g_{0}=2\tau_{0}^{2}\rho_{1}/|\tilde\beta_{2}|$  
and $\epsilon_{3}=2\rho_{3}/\gamma$.
\bibitem{dimensions3} The dimensionless second-order filtering coefficient $\epsilon_{\omega}$ 
in Eq. (\ref{sfs41}) is $\epsilon_{\omega}=2\tilde\epsilon_{\omega}/|\tilde\beta_{2}|$, 
where $\tilde\epsilon_{\omega}$ is the dimensional second-order filtering coefficient. 
The dimensionless central filtering frequency is $\omega_{p}=\tilde\omega_{p}\tau_{0}$, 
where $\tilde\omega_{p}$ is the dimensional central filtering frequency.
\bibitem{eta_equal_0} Note that for $\eta(z)=0$, $\psi_{s}(t,z) \equiv 0$ 
and $\hat\psi_{s}(\omega,z) \equiv 0$. Therefore, in this case, $h_{1}(t,z)=0$
and $d\eta/dz=0$ in the leading order of the perturbation theory.
\bibitem{eta_tends_to_0} Note that similar to the right hand side of Eq. (\ref{sfs3}), 
the right hand side of Eq. (\ref{loss10}) is a continuous function of $\eta$ for any $\eta$, 
including $\eta=0$. Additionally, the right hand side of Eq. (\ref{loss10}) tends to 0 as 
$\eta$ tends to 0, in accordance with the statements in Ref. \cite{eta_equal_0}.
\end{thebibliography}
\end{document}